\documentclass[a4paper,11pt]{article}
\pdfoutput=1 

\usepackage{jheppub} 

\usepackage[T1]{fontenc} 
\usepackage[utf8]{inputenc}

\preprint{SNUTP19-004}

\title{\boldmath Universal 3d Cardy Block and Black Hole Entropy}

\author[a]{Sunjin Choi}
\author[b]{and Chiung Hwang}

\affiliation[a]{Department of Physics and Astronomy \& Center for
Theoretical Physics,\\Seoul National University, Seoul 08826, Korea}
\affiliation[b]{Dipartimento di Fisica, Universit\`{a} di Milano-Bicocca
\& INFN, Sezione di Milano-Bicocca,\\I-20126 Milano, Italy}

\emailAdd{csj37100@snu.ac.kr}
\emailAdd{chiung.hwang@unimib.it}

\abstract{We discuss the Cardy limit of 3d supersymmetric partition functions which allow the factorization into the hemisphere indices: the generalized superconformal index, the refined topologically twisted index and the squashed sphere partition function. In the Cardy limit, the hemisphere index can be evaluated by the saddle point approximation where there exists a dominant saddle point contribution, which we call the \emph{Cardy block}. The Cardy block turns out to be a simple but powerful object as it is a building block of other partition functions in the Cardy limit. The factorization to the Cardy block allows us to find universal relations among the partition functions, which we formulate as index theorems. Furthermore, if we consider a holographic 3d SCFT and its large $N$ limit, those partition functions relate to various entropic quantities of the dual gravity theory in AdS$_4$. As a result, our result provides the microscopic derivation of the universal relations among those entropic quantities of the gravity theory. We also discuss explicit examples, which confirm our general index theorems.}

\begin{document} 
\maketitle
\flushbottom

\section{Introduction}

Last few years, the localization has played an important role in understanding supersymmetric theories on compact manifolds. Thanks to this technique, one can compute various exact BPS observables such as partition functions and Wilson loops, which turn out very useful to test the non-perturbative phenomena in strongly coupled systems.

One of those observables is the holomorphic block, which can be defined as a partition function on $D^2 \times S^1$ in 3d \cite{Beem:2012mb} and as that on $D^2 \times S^2$ in 4d \cite{Nieri:2015yia}. The holomorphic block is of particular interest because it is a basic building block of various other supersymmetric partition functions. A partition function defined on a circle fibered, either trivially or non-trivially, over a sphere is written in terms of holomorphic blocks as follows:
\begin{align}
\label{eq:fact}
    \mathcal Z_{\mathcal M} = \sum_{\alpha} ||\mathcal B^\alpha||_{\mathcal M}^2
\end{align}
where the fusion rule $||\cdots||^2$ depends on the manifold $\mathcal M$. $\alpha$ specifies the boundary condition defining the holomorphic block, and one can obtain a new supersymmetric partition function $\mathcal Z_{\mathcal M}$ by summing the square of the holomorphic block over those boundary conditions. The holomorphic block also relates to the open topological string amplitude and the vortex partition function on $\Omega$-deformed $\mathbb R^2 \times S^1$ \cite{Pasquetti:2011fj,Beem:2012mb,Hwang:2012jh,Taki:2013opa}. The holomorphic block appears in various contexts such as the AGT-like correspondence \cite{Alday:2009aq} and supersymmetric dualities. See, e.g., \cite{Aganagic:2013tta,Aganagic:2014oia} and \cite{Kim:2012uz,Beem:2012mb,Hwang:2012jh,Hwang:2015wna,Hwang:2017kmk,Hwang:2018uyj}.
\\

While the holomorphic block turns out to be a useful quantity in various contexts, its matrix integral expression needs some care as one should specify the integration contour depending on the boundary condition $\alpha$. See \cite{Beem:2012mb} for explicit examples. On the other hand, if one is interested in the Cardy limit, which is usually called the semi-classical limit, of the blocks and especially the dominant contribution, one may perform a saddle point approximation of the integral and pay particular attention only to the leading saddle point. In this way, one can circumvent the explicit determination of the integration contour and can extract the dominant contribution in the Cardy limit. In this paper, we give a particular name for such dominant contribution, \emph{Cardy block}, as it plays an important role in our discussion.

Such Cardy block has been somehow overlooked due to its simplicity. However, recently it was found that the 3d Cardy block is indeed extremely useful, especially for the microstate counting of AdS$_4$ black holes. The counting of black hole microstates is one of the most interesting problems in quantum gravity. Since the seminal work by Strominger and Vafa \cite{Strominger:1996sh}, there have been various attempts and notable success such as the microstate counting of static dyonic black holes\footnote{For example, see \cite{Benini:2015eyy,Hosseini:2016tor,Hosseini:2016ume,Benini:2016rke,Azzurli:2017kxo,Hosseini:2017fjo,Benini:2017oxt} for AdS$_4$. See \cite{Zaffaroni:2019dhb} for a more comprehensive list of references, including higher dimensional generalizations.} and, rather recently, that of rotating electric black holes in $d > 3$ dimensions \cite{Choi:2019zpz, Nian:2019pxj,Cabo-Bizet:2018ehj,Choi:2018hmj,Choi:2018vbz,Benini:2018ywd,Honda:2019cio,ArabiArdehali:2019tdm,Kim:2019yrz,Cabo-Bizet:2019osg,Larsen:2019oll,Lezcano:2019pae,Choi:2019miv,Nahmgoong:2019hko}. The former is achieved by the large $N$ computation of the topologically twisted index of the dual field theory while the latter relates to the large $N$ limit of the superconformal index. In contrast to $d > 3$, the large $N$ computation of the $d = 3$ superconformal index becomes more challenging due to the existence of supersymmetric monopole operators. The localization saddles of the 3d superconformal index include infinite monopole configurations, whose individual contribution has to be summed to obtain the exact superconformal index.

This problem can be rephrased in a slightly different way using the holomorphic block. According to \eqref{eq:fact}, the superconformal index is also written in terms of the holomorphic blocks where the \emph{Cardy block} gives the dominant contribution. Furthermore, the Cardy block itself can be independently computed using the localization, which provides another way to compute the superconformal index in the Cardy limit. In that case, the infinite monopole summation is already encoded in the Cardy block. Indeed, it was shown in \cite{Choi:2019zpz} that the large $N$ limit of the Cardy block, more precisely its square, successfully reproduces the known entropy function of the rotating BPS black hole in AdS$_4 \times S^7$ \cite{Choi:2018fdc} using the ABJM theory and the mirror dual of the maximal SYM theory \cite{Aharony:2008ug,Kapustin:2010xq}. The Legendre transformation of such entropy function yields the Bekenstein-Hawking entropy of the dual black hole with the large angular momentum.
\\

Inspired by this, in this paper, we extend this analysis to more general theories and partition functions. We will consider the Cardy limit of the hemisphere index and various other partition functions allowing the factorization into the holomorphic blocks: the generalized superconformal index, the refined topologically twisted index\footnote{While the topologically twisted index can be defined on a generic Riemann surface $\Sigma_g$ of arbitrary genus $g$, the refined version is only available for $S^2$, which has $U(1)$ isometry \cite{Benini:2015noa,Benini:2016hjo,Closset:2016arn}. In this paper, we focus on the refined topologically twisted index on $S^2$ because such refinement is essential for the factorization and the Cardy limit, which play crucial roles in our discussion.} and the squashed sphere partition function. Especially in the Cardy limit, the free energies of those partition functions, defined by $F_{\mathcal M} = -\log Z_{\mathcal M}$, are written in terms of the Cardy block $\mathcal C$ in a much simpler manner than \eqref{eq:fact}:
\begin{align}
\begin{aligned}
F_{D^2}(t;\beta) &\approx -\log \mathcal C(t;\beta) \,, \\
F_{S^2}(\mathfrak t,n;\beta) &\approx N \log \beta - \log \mathcal C(t;\beta) - \log \overline{\mathcal C}(\bar t;-\beta) \,, \\
F_{S^2}^\text{twisted}(\mathfrak t,n;\beta) &\approx N \log \beta - \log \mathcal C(t;\beta) - \log \mathcal C(\bar t^{-1};-\beta) \,, \\
F_{S^3_b}(\Delta;\beta) &\approx \frac{N}{2} \log \beta - \log \mathcal C(e^{\Delta};\beta) \,,
\end{aligned}
\end{align}
which only refer to the dominant saddle. We may call those formulae \emph{Cardy factorizations}. We should mention that the Cardy factorization of the topologically twisted index happens only in special circumstances such as the large $N$ limit. See section \ref{sec:fact} for more detailed discussions including the notation. Also note that our analysis will be applied on generic 3d $\mathcal N = 2$ supersymmetric theories having UV Lagrangian description.

Surprisingly, the Cardy factorization allows us to find various universal relations among those partition functions in the Cardy limit, some of which, in particular, are phrased as index theorems in section \ref{sec:universal}. For example, we formulate an index theorem relating the generalized superconformal index and the squashed sphere partition function:
\begin{align}
    F_{S^2}(e^\Delta,n;\beta) = F_{S^3_b}(\Delta+\beta n;\beta) + \overline F_{S^3_b}(-\Delta+\beta n;-\beta)+\mathcal O(\beta) \,
\end{align}
where $\Delta$ denotes flavor chemical potentials while $n$ denotes flavor magnetic flux collectively. For the superconformal index, $\beta$ is the chemical potential for the angular momentum on $S^2$, with a shift by the $R$-charge due to the supersymmetry condition, while, for the squashed sphere partition function, it is related to the squashing parameter $b$ of the sphere by $\beta = \pi i b^2$. In our Cardy limit, $\beta$ is taken to zero: $\beta \rightarrow 0^+$. 

In addition to the Cardy limit, the large $N$ limit of 3d superconformal field theories is of particular interest because a large class of 3d SCFTs are known to have holographic dual gravity theories in AdS$_4$ in the large $N$ limit. Accordingly, if we further assume large $N$, we find another index theorem relating the superconformal index and the topologically twisted index to the round sphere partition function:
\begin{align}
\begin{aligned}
\label{eq:theorem2_intro}
    &F_{S^2}(\Delta,n;\beta) = \frac{(\pi i+\beta)^2}{4 \pi i \beta} F_{S^3} \left(-\frac{\Delta+\beta n}{\pi i+\beta}\right) + \frac{(\pi i+\beta)^2}{4 \pi i \beta} \overline F_{S^3} \left(-\frac{\Delta-\beta n}{\pi i+\beta}\right) \,, \\
    \\
    &F_{S^2}^\text{twisted}(\Delta,n;\beta) = \frac{(\pi i+\beta)^2}{4 \pi i \beta} F_{S^3} \left(-\frac{\Delta+\beta n}{\pi i+\beta}\right) - \frac{(\pi i-\beta)^2}{4 \pi i \beta} F_{S^3} \left(-\frac{\Delta-\beta n}{\pi i-\beta}\right) \,,
\end{aligned}
\end{align}
which relies on the large $N$ relation between the squashed and round sphere partition functions found in \cite{Martelli:2011fu}:
\begin{align}
   F_{S^3_b}\left(-\pi i b Q \delta;\pi i b^2\right) = \frac{Q^2}{4} F_{S^3} \left(\delta\right)
\end{align}
where $Q = b+1/b$ and $\delta$ parametrizes trial $R$-charges. While our derivation of \eqref{eq:theorem2_intro} is valid up to $\mathcal O(\beta)$, there is strong evidence that it is indeed exact even for finite $\beta$ in the large $N$ limit, at least for the known large $N$ saddle point capturing the dual black hole microstates. Especially, those indices in the large $N$ limit statistically account for the microstates of rotating dyonic BPS black holes in AdS$_4$. While the superconformal index should have the vanishing magnetic flux for the $R$-symmetry, the topologically twisted index has the non-zero $R$-symmetry flux, which leads to a particular asymptotically locally AdS$_4$ spacetime, dubbed mAdS$_4$ \cite{Hristov:2011ye}, on the dual gravity side. We expect that those indices give entropy functions of dual black holes for arbitrary $\beta$, which is confirmed for some examples in section \ref{sec:ex1} and \ref{sec:ex2}. In addition, the squashed sphere partition function relates to the supersymmetric R\'{e}nyi entropy \cite{Nishioka:2013haa}, which accounts for the Bekenstein-Hawking entropy of a charged topological black hole in AdS$_4$ \cite{Huang:2014gca,Nishioka:2014mwa}. Also note that the right hand side is written in terms of the round sphere partition function, which, with the superconformal $R$-charge, is identified with the entanglement entropy for a spherical entangling surface \cite{Dowker:2010yj,Casini:2011kv}. By the AdS/CFT dictionary, it corresponds to the Euclidean on-shell action in AdS$_4$ \cite{Maldacena:1997re}. In that regard, our index theorem shows that various entropic quantities in dual AdS are not independent and indeed mutually related. While we provide a field theoretic derivation of such relations, interestingly, similar relations are discussed in the gravity context recently \cite{Hosseini:2019iad}, using the gravitational blocks, which are supposed to be dual to our Cardy blocks in the large $N$ limit.

Moreover, if we turn off all the magnetic flux for the flavor symmetry for the generalized superconformal index, it reduces to the ordinary superconformal index, which satisfies
\begin{align}
    F_{S^2}(\Delta,n = 0;\beta) = \frac{(\pi i+\beta)^2}{2 \pi i \beta} \mathrm{Re}\left[F_{S^3} \left(-\frac{\Delta}{\pi i+\beta}\right)\right]
\end{align}
where $\mathrm{Re}[\dots]$ should be understood with the conjugation defined by \eqref{eq:conjugate}. If we further turn off the flavor chemical potentials while restore the superconformal $R$-charge, the right hand side is simply written as
\begin{align}
\label{eq:unrefined_intro}
    F_{S^2}(\beta) = \frac{\Delta_R^2}{2 \pi i \beta} \mathrm{Re} \left[F_{S^3} \left(\delta_*\right)\right]
\end{align}
where $\Delta_R = \pi i+ \beta$ is the chemical potential for the superconformal $R$-symmetry and $F_{S^3} \left(\delta_*\right)$ is the round sphere free energy at the superconformal $R$-charge, which is determined by the $F$-maximization \cite{Jafferis:2010un}. This is reminiscent of the Cardy formula for 2d CFTs \cite{Cardy:1986ie} or 4d $\mathcal N = 1$ SCFTs \cite{Kim:2019yrz} in the large $N$ limit, where the real part of the round sphere free energy, $\mathrm{Re} \left[F_{S^3} \left(\delta_*\right)\right]$, plays the role of central charges in 2d and in 4d. Recently the same formula has been obtained both on the gravity side and on the field theory side for a particular class of theories called class $\mathcal R$, which is obtained from M5-branes wrapped on hyperbolic 3-manifolds, using the 3d-3d correspondence \cite{Bobev:2019zmz,Benini:2019dyp}. Our result validates this formula for generic 3d $\mathcal N = 2$ SCFTs with UV Lagrangian.

Lastly, one can also find interesting relations from the leading term of the right hand side of \eqref{eq:theorem2_intro}. If we expand the right hand side with respect to $\beta$ and take the leading term, we find
\begin{align}
\begin{aligned}
    &F_{S^2}(\mathfrak t e^{-\beta \delta},n;\beta) \approx -\frac{\pi}{2 \beta} \mathrm{Re}\left[F_{S^3}\left(-\frac{\Delta}{\pi i}\right)\right], \\
    &F_{S^2}^\text{twisted}(\mathfrak t,n;\beta) \approx \frac{\pi i}{2} \sum_i \left(n_i-\frac{\Delta_i}{\pi i}\right) \frac{\partial}{\partial \Delta_i} F_{S^3}\left(-\frac{\Delta}{\pi i}\right) + F_{S^3}\left(-\frac{\Delta}{\pi i}\right).
\end{aligned}
\end{align}
Especially, the strict Cardy limit of the topologically twisted index is essentially the unrefined limit of the index, which accounts for the static dyonic BPS black holes in AdS$_4$. Indeed, the second relation is a rederivation of the index theorem for the unrefined topologically twisted indices, and therefore the entropy functions of static black holes, discovered by Hosseini and Zaffaroni \cite{Hosseini:2016tor}. For such unrefined indices with $\beta = 0$, the index theorem was later generalized for those on generic Riemann surfaces with arbitrary genera \cite{Azzurli:2017kxo,Bobev:2017uzs}.

Note that the interpretation of our results as black hole entropy functions should be understood only when they give rise to positive macroscopic entropy, which is not guaranteed for an arbitrary choice of the background, such as the existence of the topological twist or the magnetic fluxes for flavor symmetries. For instance, recently \cite{Benini:2019dyp} have made use of the 3d-3d correspondence to show that the refined topologically twisted index is exponentially suppressed in the large $N$ limit, and also exactly vanishes for some finite $N$, for a particular class of theories with the universal twist, the twist along the exact superconformal $R$-symmetry. Such universal twist corresponds to the so-called universal black holes \cite{Azzurli:2017kxo}, and the vanishing of the index is consistent with the fact that there is no universal black hole solution with near-horizon geometry AdS$_2 \times S^2$.

This shows that macroscopic entropy and dual black hole solutions are not always guaranteed for an arbitrary choice of background. Nonetheless, as long as we have macroscopic entropy from the index, we expect this entropy captures the microstates of some black hole, regardless of whether or not such a black hole solution has been found already. This suggests that our index computation can be a probe for a new black hole solution. Indeed, recently the Macdonald index of 4d $\mathcal N = 4$ SYM theory has been used to predict a new black hole in AdS$_5$ \cite{Choi:2018hmj}. It will be also interesting if such predictions can be made for black holes in AdS$_4$ using our results for 3d field theories.
\\

The paper is organized as follows. In section \ref{sec:fact} we first review the localization results of 3d $\mathcal N = 2$ supersymmetric partition functions and examine their Cardy limits, especially focusing on their factorization properties. In section \ref{sec:universal}, we derive universal relations among the partition functions by combining the results of section \ref{sec:fact}. In particular, we formulate two index theorems: one relating the generalized superconformal index and the squashed sphere partition function in the Cardy limit and the other relating the generalized superconformal index, the refined topologically twisted index and the round sphere partition function in the large $N$ limit. In section \ref{sec:ex1}, we address the $\mathcal N = 4$ $U(N)$ SYM with one fundamental and one adjoint matters as an explicit example. We demonstrate how to obtain four different partition functions in the large $N$ limit using the factorization. Lastly, in section \ref{sec:ex2}, we provide more examples of 3d $\mathcal N \geq 2$ SCFTs. We discuss the large $N$ Cardy limit of the generalized superconformal indices for those examples. We also examine the finite $N$ Cardy formulae for some examples, which provide nontrivial tests for known supersymmetric dualities.
\\

\section{The 3d Cardy Block and Factorization}
\label{sec:fact}

In this section, we first review the localization results of 3d $\mathcal N = 2$ supersymmetric partition functions and examine their Cardy limits, especially focusing on their factorization properties. Since the partition functions are 1-loop exact in the context of the supersymmetric localization, the results are given by finite dimensional matrix integrals whose integrands consist of the classical action contributions as well as the 1-loop determinants. Also the factorization of 3d partition functions have been extensively discussed in the literature; e.g., see \cite{Pasquetti:2011fj,Beem:2012mb,Hwang:2012jh,Taki:2013opa,Fujitsuka:2013fga,Benini:2013yva,Nieri:2015yia,Hwang:2015wna,Hwang:2017kmk,Hwang:2018uyj}. Here we revisit them in our notation, which is chosen to be convenient for our Cardy limit analysis. In particular, we start with the hemisphere index on $D^2 \times S^1$ defined in \cite{Yoshida:2014ssa}, which is closely related to the holomorphic block discussed in \cite{Beem:2012mb}. This is a building block of the other supersymmetric partition functions we will discuss. We then move on to those partition functions on a circle fibered over a sphere, which are known to be factorized, and examine their Cardy limits.
\\

\subsection{The Hemisphere Index}

The first example we discuss is the hemisphere index on $D^2 \times S^1$ \cite{Yoshida:2014ssa}, which is defined by
\begin{align}
    I_{D^2} = \mathrm{tr}_{\mathcal H(D^2;\alpha)} \left[(-1)^F e^{-\beta_1 (\mathsf D-\mathsf R-\mathsf J_3)} e^{-\beta_2 (\mathsf D+\mathsf J_3)} e^{-F_l M_l}\right]
\end{align}
where $\mathsf D$ is the translation generator along $S^1$; $\mathsf R$ is the $\mathcal N=2$ $U(1)$ $R$-charge; $\mathsf J_3$ is the angular momentum; and $F_l$'s are the Cartan charges of the flavor symmetry. The trace is taken over the Hilbert space on $D^2$ with the boundary condition $\alpha$. As usual, this index counts the BPS states saturating $\mathsf D-\mathsf R-\mathsf J_3 \geq 0$ and is thus independent of $\beta_1$ unless there is a flat direction appearing while $\beta_1$ changes. While the hemisphere index itself does not have a factorized structure mentioned in the introduction, it plays a role of a building block of the other partition functions we will consider.

$F$, a fermion number, is typically chosen to be $F = 2 \mathsf J_3$. On the other hand, to define an index, one can also use other choices of $F$; one useful alternative, especially for the factorization, is $F = \mathsf R$. Recall that, for 3d $\mathcal N = 2$ supersymmetric theories, the IR superconformal $R$-charge is determined by the $F$-maximization \cite{Jafferis:2010un}. However, to define an index, one can use a trial, or UV, value of the $R$-charge, from which the IR superconformal value can be achieved as a mixture of the UV $R$-charge and various $U(1)$ flavor charges. For convenience, we take the integer quantized UV $R$-charge and use it to define the index. For the integer quantized $R$-charge, $(-1)^F$ is merely a sign while it would be a nontrivial phase otherwise. In fact, the integer $R$-charge will be eventually required for the comparison with the topologically twisted index, which demands the integer $R$-charge due to twisting. This choice of $F$ will make the comparison between the hemisphere index and other partition functions, especially the squashed sphere partition function, more clear as noted in \cite{Beem:2012mb}.
\\

From the definition of the index, one can see that those two choices are related by the shift of $\beta_1, \, \beta_2$:
\begin{align}
\begin{aligned}
    \beta_1 &\rightarrow \beta_1-\pi i, \\
    \beta_2 &\rightarrow \beta_2+\pi i,
\end{aligned}
\end{align}
which yields extra sign $(-1)^{\mathsf R+2 \mathsf J_3}$. Note that if $\mathsf R$ is even integer quantized, the two choices are identical. Using this shift, one can easily obtain the formula for $F = \mathsf R$ from the localization result for $F = 2 \mathsf J_3$ in \cite{Yoshida:2014ssa}. Setting $\beta_2 = \beta+\pi i$, the hemisphere index with $F = \mathsf R$ is given by
\begin{align}
\label{eq:D2}
    I_{D^2} = \frac{1}{|W_G|} \oint \left(\prod_{a = 1}^{\mathrm{rk}(G)} \frac{d z_a}{2 \pi i z_a}\right) \, \mathcal Z_\text{classical} \, \mathcal Z_\text{vector} \, \mathcal Z_\text{chiral}^{N/D} \, \mathcal Z_\text{vector}^{2d} \, \mathcal Z_\text{chiral/Fermi}^{2d}
\end{align}
where $W(G)$ is the Weyl group of the gauge group $G$.
The nontrivial classical action contribution $\mathcal Z_\text{classical}$ consists of various Chern-Simons terms:
\begin{align}
    e^{\frac{k}{4 \beta} \mathrm{tr} u^2}
\end{align}
for a canonical Chern-Simons term with level $k$ and
\begin{align}
    e^{\frac{1}{2 \beta} \mathrm{tr} (u_A u_B)}
\end{align}
for a mixed Chern-Simons term between $U(1)_A \times U(1)_B$, each of which is either gauge or global $U(1)$. Each $u$ is defined by 
$u = \log z$ where $z$ is the holonomy for the corresponding (either gauge or global) symmetry.

The 1-loop determinants of the 3d bulk fields are as follows:
\begin{align}
    \mathcal Z_\text{vector} &= \prod_{\alpha \in \Delta} e^{-\frac{1}{8 \beta} (\alpha(u)\pm\pi i)^2} \left(z^\alpha;x^2\right), \\
    \mathcal Z_\text{chiral}^N &= \prod_{\rho \, \otimes \, \sigma \in \mathfrak R^N} e^{\frac{\beta}{8} \left((r_\sigma-1)^2-\frac{1}{3}\right)-\frac{r_\sigma-1}{4} (\rho(u)+\sigma(v)-\pi i (r_\sigma\mp1))+\frac{1}{8 \beta} (\rho(u)+\sigma(v)-\pi i (r_\sigma\mp1))^2} \nonumber \\
    &\qquad \times \left(z^\rho t^\sigma x^{r_\sigma} e^{-\pi i r_\sigma};x^2\right)^{-1} \,, \label{eq:Neumann} \\
    \mathcal Z_\text{chiral}^D &= \prod_{\rho \, \otimes \, \sigma \in \mathfrak R^D} e^{-\frac{\beta}{8} \left((r_\sigma-1)^2-\frac{1}{3}\right)+\frac{r_\sigma-1}{4} (\rho(u)+\sigma(v)-\pi i (r_\sigma\mp1))-\frac{1}{8 \beta} (\rho(u)+\sigma(v)-\pi i (r_\sigma\mp1))^2} \nonumber \\
    &\qquad \times \left(z^{-\rho} t^{-\sigma} x^{2-r_\sigma} e^{\pi i r_\sigma};x^2\right) \label{eq:Dirichlet}
\end{align}
with
\begin{align}
    z^\rho = e^{\rho(u)} \,, \qquad t^{\sigma} = e^{\sigma(v)} \,, \qquad x = e^{-\beta}
\end{align}
where $\Delta$ is the set of non-zero roots of the gauge group; $\mathfrak R^{N/D}$ are the representations of the chiral multiplets with the Neumann/Dirichlet boundary conditions respectively, with gauge weight $\rho$ and global weight $\sigma$, which includes $U(1)_R$ charge $r_\sigma$. Here we use the shorthand expression $(a;x^2)$ for q-Pochhammer symbol $(a;x^2)_\infty$:
\begin{align}
    (a;x^2)_\infty = \prod_{k = 0}^\infty \left(1-a x^{2 k}\right).
\end{align}
Due to our choice $F = \mathsf R$, there are extra $e^{-\pi i r_\sigma}$ in contrast to the result of \cite{Yoshida:2014ssa}.

Note that the exponential factor in each determinant spoils the invariance of the determinant under the large gauge transformation. Such exponential factors are remnant of the gauge non-invariant regularization, which should disappear if one regularizes the determinant in a gauge invariant way with the appropriate definitions of the UV Chern-Simons levels understood. However, we here stick to the above definition of the determinants for easy comparisons with the earlier literature. Indeed, the exponential factors should be completely canceled out once the boundary matter contributions are taken into account. Thus, the entire integrand is again invariant under the large gauge transformation. Nevertheless, since each determinant is not invariant under the large gauge transformation, we have to fix the ambiguity; namely, we take the above definition for the arguments in chambers $-2 \pi < \pm \mathrm{Im}(\alpha(u)), \, \pm \mathrm{Im}(\rho(u)+\sigma(v)-\pi i r_\sigma) < 0$. If we don't specify the chamber explicitly, we take the upper plus sign of $\pm$.

The exponential factor in each determinant makes an effective shift of Chern-Simons levels. The effective Chern-Simons terms, dictated by such exponential factors as well as the classical action contribution, yield anomalies at the boundary, which should be canceled by extra boundary degrees of freedom, at least up to a $u$-independent constant. The 1-loop determinants of the boundary matters are
\begin{align}
    \mathcal Z_\text{vector}^{2d} &= \prod_{\alpha \in \Delta^{2d}} e^{-\frac{1}{4 \beta} (\alpha(u)\pm\pi i)^2} \theta\left(z^\alpha;x^2\right), \\
    \mathcal Z_\text{chiral}^{2d} &= \prod_{\rho \, \otimes \, \sigma \in \mathfrak R_\text{chiral}^{2d}} e^{\frac{\beta}{4} \left((r_\sigma-1)^2-\frac{1}{3}\right)-\frac{r_\sigma-1}{2} (\rho(u)+\sigma(v)-\pi i (r_\sigma\mp1))+\frac{1}{4 \beta} (\rho(u)+\sigma(v)-\pi i (r_\sigma\mp1))^2} \nonumber \\
    &\quad \qquad \times \theta \left(z^\rho t^\sigma x^{r_\sigma} e^{-\pi i r_\sigma};x^2\right)^{-1} \,, \\
    \mathcal Z_\text{Fermi}^{2d} &= \prod_{\rho \, \otimes \, \sigma \in \mathfrak R_\text{Fermi}^{2d}} e^{-\frac{\beta}{4} \left((r_\sigma-1)^2-\frac{1}{3}\right)+\frac{r_\sigma-1}{2} (\rho(u)+\sigma(v)-\pi i (r_\sigma\mp1))-\frac{1}{4 \beta} (\rho(u)+\sigma(v)-\pi i (r_\sigma\mp1))^2} \nonumber \\
    &\quad \qquad \times \theta \left(z^{\rho} t^{\sigma} x^{r_\sigma} e^{-\pi i r_\sigma};x^2\right)
\end{align}
where $\theta \left(a;x^2\right)$ is defined by
\begin{align}
    \theta \left(a;x^2\right) = \left(a;x^2\right)_\infty \left(a^{-1} x^2;x^2\right)_\infty.
\end{align}
The 3d determinants and 2d determinants satisfy
\begin{align}
    \frac{\mathcal Z_\text{chiral}^N}{\mathcal Z_\text{chiral}^D} = Z_\text{chiral}^{2d} = \frac{1}{Z_\text{Fermi}^{2d}}
\end{align}
once we assign the same representation and the $R$-charge.
Note that those 2d degrees of freedom are engineered so that the net exponential factor of the integrand is completely canceled.
\\

In the Cardy limit, i.e., $\beta \rightarrow 0^+$ while the other variables kept finite, the 3d determinants are given by
\begin{align}
\begin{aligned}
\label{eq:Cardy}
    &\lim_{\beta \rightarrow 0} \log \mathcal Z_\text{vector}(z;\beta) = -\frac{1}{2 \beta} \sum_\alpha \left[\frac{1}{4} (\alpha(u)\pm\pi i)^2+\mathrm{Li}_2 (z^\alpha x^{-1})\right], \\
    &\lim_{\beta \rightarrow 0} \log \mathcal Z_\text{chiral}^N(z,t;\beta) \\
    &= \frac{1}{2 \beta} \sum_{\rho \, \otimes \, \sigma} \left[\frac{1}{4} (\rho(u)+\sigma(v)-\pi i (r_\sigma\mp1)-\beta (r_\sigma-1))^2+\mathrm{Li}_2 (z^\rho t^\sigma x^{r_\sigma-1} e^{-\pi i r_\sigma})\right], \\
    &\lim_{\beta \rightarrow 0} \log \mathcal Z_\text{chiral}^D(z,t;\beta) \\
    &= -\frac{1}{2 \beta} \sum_{\rho \, \otimes \, \sigma} \left[\frac{1}{4} (\rho(u)+\sigma(v)-\pi i (r_\sigma\mp1)-\beta (r_\sigma-1))^2+\mathrm{Li}_2(z^{-\rho} t^{-\sigma} x^{1-r_\sigma} e^{\pi i r_\sigma})\right]
\end{aligned}
\end{align}
up to $\mathcal O(\beta)$.
On the other hand, the 2d determinants become
\begin{align}
\begin{aligned}
\label{eq:2d}
    \lim_{\beta \rightarrow 0} \log \mathcal Z_\text{vector}^{2d}(z;\beta)
    &= \frac{\pi^2}{6 \beta} \\
    \lim_{\beta \rightarrow 0} \log \mathcal Z_\text{chiral}^{2d}(z,t;\beta)
    &= -\frac{\pi^2}{12 \beta} \\
    \lim_{\beta \rightarrow 0} \log \mathcal Z_\text{Fermi}^{2d}(z,t;\beta)
    &= \frac{\pi^2}{12 \beta}.
\end{aligned}
\end{align}
for each pair of $(\alpha,-\alpha)$ and for each $\rho \, \otimes \, \sigma$. While those do not vanish in the Cardy limit, we will see that their role is rather minimal when we construct other 3d partition functions upon the hemisphere index. Since $\mathcal Z_\text{chiral}^N$ and $\mathcal Z_\text{chiral}^D$ are related by $\mathcal Z_\text{chiral/Fermi}^{2d}$, also their distinction will not be very significant in such situations. Thus, we mostly call them $\mathcal Z_\text{chiral}$ unless their distinction is necessary.
\\

Note that, in each determinant, $\beta$ always appears in a combination with $v$: $\sigma(v)-\pi i r_\sigma-\beta (r_\sigma-1)$ except $\beta^{-1}$ in front. From now on, for convenience, let us distinguish $\beta$ in a combination with $v$ and $\beta$ in front by denoting the former by $\hat \beta$. And we also introduce $\mathsf t = (t,-\hat x e^{-\pi i})$, on which $\sigma$ act as follows:
\begin{align}
    \mathsf t^\sigma = t^\sigma \hat x^{r_\sigma-1} e^{-\pi i r_\sigma}
\end{align}
where $\hat x = e^{-\hat \beta}$.
In this way, the explicit $\beta$-dependence of the above determinants is only $\beta^{-1}$ in front. At the end, we should restore $\hat \beta$ to $\beta$. Also note that although we have set $r_\sigma$ to be the UV integer value, one can obtain the index with the non-integer $R$-charge, such as the IR superconformal one, by shifting the flavor chemical potentials $v$ by $v \rightarrow v-(\beta+\pi i) \delta$. Then the deformed value of the $R$-charge is $r_\sigma+\sigma(\delta)$.

In the small $\beta$ limit, one can evaluate the integral \eqref{eq:D2} using the saddle point approximation as follows:
\begin{align}
I_{D^2} = \left( \frac{\beta}{\pi}\right)^{\frac{N}{2}} \sum_{*} \exp \left(-\frac{1}{2\beta} \mathcal{W}^* \right) \left( \det ( - \partial_z^2 \mathcal{W})^*\right)^{-\frac{1}{2}} \left(\prod_{a=1}^{N} \frac{1}{z_a^*} +\mathcal O(\beta) \right),
\end{align}
where $N$ is the gauge group rank. Our effective potential is defined by
\begin{align}
    -\frac{1}{2 \beta} \mathcal W = \sum \lim_{\beta \rightarrow 0} \log\mathcal Z,
\end{align}
where the summation is taken over the collection of the above determinants. The $\beta$-dependence of $\mathcal W$ is hidden in $\mathsf t$ as $\hat \beta$, which will be taken back to be $\beta$ at the end. ${}^*$ denotes the value at each saddle point, which is a solution to the equation
\begin{align}
    \partial_z \mathcal W = 0 \,.
\end{align}
As long as there is a saddle satisfying $\mathrm{Re} (\mathcal W^*/\beta) < 0$, the index exponentially grows in the Cardy limit. In such cases, there is a dominant saddle such that
\begin{align}
I_{D^2}(t;\beta) \approx \mathcal C(t;\beta)
\end{align}
where $\mathcal C$ is the contribution at the dominant saddle, which we call the \emph{Cardy block}.
Thus, the free energy, $F_{D^2} = -\log I_{D^2}$, is written as
\begin{align}
\label{eq:Cardy block}
    F_{D^2}(t;\beta) = -\log \mathcal C(t;\beta) + \mathcal O\left(e^{-\beta^{-1}}\right) = \frac{1}{\beta} G^{(0)} (\mathsf t) - \frac{N}{2} \log \beta + G^{(1)}(\mathsf t)+\mathcal O(\beta)
\end{align}
where $G^{(0)} = \frac{1}{2} \mathcal W^*$ is the dominant saddle value of the effective potential while the other saddle point contributions are suppressed exponentially. $G^{(1)}$ is the collection of the remaining contributions at the dominant saddle, e,g,. that of the Hessian. One should remember that $\mathsf t$ includes $\hat \beta = \beta$ in such way that $\mathsf t^\sigma = t^\sigma \hat x^{r_\sigma-1} e^{-\pi i r_\sigma}$.
\\

\subsection{The Generalized Superconformal Index}

Our next example is the generalized superconformal index \cite{Bhattacharya:2008zy,Bhattacharya:2008bja,Kapustin:2011jm}, which is defined by
\begin{align}
    I_{S^2} = \mathrm{tr}_{\mathcal H_{BPS}(S^2;n)} \left[(-1)^F e^{-\beta (\mathsf R+2 \mathsf J_3)} e^{-F_l M_l}\right]
\end{align}
where the BPS condition $\mathsf D-\mathsf R-\mathsf J_3 = 0$ is understood. $n$ denotes the external magnetic flux for the flavor symmetry collectively.
Again, while a typical choice of $F$ is $F = 2 \mathsf J_3$, our choice here is $F = \mathsf R$. Those two are related by shift of $\beta$: $\beta \rightarrow \beta-\pi i$. Also $\mathsf R$ will be taken to be integer quantized. The superconformal $\mathsf R$ can be restored by shifting $M_l$.

One can evaluate such superconformal index using the supersymmetric localization. The localization result for $F = \mathsf J_3$ was obtained in \cite{Kim:2009wb,Imamura:2011su}. Similarly, the superconformal index with $F = \mathsf R$ is written as
\begin{align}
    I_{S^2} = \frac{1}{|W_G|} \sum_{m \in \Gamma_G^\vee} \oint \left(\prod_{a = 1}^{\mathrm{rk}(G)} \frac{d \mathfrak z_a}{2 \pi i \mathfrak z_a}\right) \, Z_\text{classical} \, Z_\text{vector} \, Z_\text{chiral} \,.
\end{align}
$m$ runs over the GNO charges including the Weyl equivalent ones. Thus, the symmetry factor is just the order of the Weyl group of the gauge group $G$. The 1-loop determinants of the vector multiplet and the chiral multiplet are
\begin{align}
    Z_\text{vector} &= \prod_{\alpha \in \Delta} \left(x e^{-\pi i}\right)^{-\frac{|\alpha(m)|}{2}} \left(1-\mathfrak z^\alpha x^{|\alpha(m)|}\right), \\
    Z_\text{chiral} &= \prod_{\rho \, \otimes \, \sigma \in \mathfrak R} \left(\mathfrak z^\rho \mathfrak t^\sigma x^{r_\sigma-1} e^{-\pi i (r_\sigma-1)}\right)^{-\frac{|\rho(m)+\sigma(n)|}{2}} \frac{\left(\mathfrak z^{-\rho} \mathfrak t^{-\sigma} x^{2-r_\sigma+|\rho(m)+\sigma(n)|} e^{\pi i r_\sigma};x^2\right)}{\left(\mathfrak z^\rho \mathfrak t^\sigma x^{r_\sigma+|\rho(m)+\sigma(n)|} e^{-\pi i r_\sigma};x^2\right)}
\end{align}
with
\begin{align}
    \mathfrak z^\rho = e^{i \rho(a)} \,, \qquad \mathfrak t^\sigma = e^{i \sigma(b)} \,, \qquad x = e^{-\beta} \,.
\end{align}
$\Delta$ is the set of non-zero roots. $\mathfrak R$ is the representation of the chiral multiplets under the gauge and global symmetry groups, including the $R$-symmetry, with weights $\rho$ and $\sigma$ respectively. Note that there are extra $e^{-\pi i r_\sigma}$ compared to the usual localization result due to the choice $F = \mathsf R$. The nontrivial contribution of the classical action again comes from various Chern-Simons terms. A canonical Chern-Simons term with level $k$ gives rise to
\begin{align}
    e^{i k \, \mathrm{tr} (a m)}
\end{align}
while a mixed Chern-Simons term between $U(1)_A \times U(1)_B$ gives
\begin{align}
    \mathfrak z_A^{m_B} \mathfrak z_B^{m_A}
\end{align}
where $U(1)_{A,B}$ are either gauge or global $U(1)$. Each pair of $\mathfrak z = e^{i a}$ and $m$ denote the holonomy and the magnetic flux for the corresponding symmetry, either gauge or global.
\\

One can massage the 1-loop determinants to be written in terms of \emph{formal} holomorphic and anti-holomorphic variables. Using the identity \cite{Dimofte:2011py}
\begin{align}
\label{eq:ident}
\left(-z x^{-1}\right)^{\frac{|m|+m}{2}} \frac{\left(z^{-1} x^{2-m};x^2\right)_{\frac{|m|+m}{2}}}{\left(z x^{-m};x^2\right)_{\frac{|m|+m}{2}}} = 1,
\end{align}
we have
\begin{align}
    Z_\text{vector}
    &= \prod_{\alpha} e^{-\frac{1}{8 \beta} \left(\alpha(u)^2-\alpha(\bar u)^2\right)} \frac{\left(z^{\alpha};x^2\right)}{\left(\bar z^\alpha x^{2};x^2\right)} \,, \\
    Z_\text{chiral}
    &= \prod_{\rho \, \otimes \, \sigma} e^{-\frac{r_\sigma-1}{4} (\rho(u)+\sigma(v)+\rho(\bar u)+\sigma(\bar v))+\frac{1}{8 \beta} \left((\rho(u)+\sigma(v)-\pi i (r_\sigma-1))^2-(\rho(\bar u)+\sigma(\bar v)+\pi i (r_\sigma-1))^2\right)} \nonumber \\
    &\qquad \times \frac{\left(\bar z^{\rho} \bar t^{\sigma} x^{2-r_\sigma} e^{\pi i r_\sigma};x^2\right)}{\left(z^\rho t^\sigma x^{r_\sigma} e^{-\pi i r_\sigma};x^2\right)}
\end{align}
where we have defined that
\begin{gather}
\begin{gathered}
    z^\rho = e^{\rho(u)} = \mathfrak z^\rho x^{-\rho(m)} \,, \qquad t^\sigma = e^{\sigma(v)} = \mathfrak t^\sigma x^{-\sigma(m)} \,, \\
    \bar z^\rho = e^{\rho(\bar u)} = \mathfrak z^{-\rho} x^{-\rho(m)} \,, \qquad \bar t^\sigma = e^{\sigma(\bar v)} = \mathfrak t^{-\sigma} x^{-\sigma(m)} \,;
\end{gathered}
\end{gather}
i.e.,
\begin{gather}
\begin{gathered}
\label{eq:u}
    u = \log z = i a+\beta m \,, \qquad v = \log t = i b+\beta n \,, \\
    \bar u = \log \bar z = -i a+\beta m \,, \qquad \bar v = \log \bar t = -i b+\beta n \,.
\end{gathered}
\end{gather}
When $a, \, b$ and $\beta$ are real, the barred variables are the complex conjugates of the unbarred variables. On the other hand, if we relax such reality conditions, the barred variables should be understood as the formal conjugates defined by the following map:
\begin{align}
\begin{aligned}
\label{eq:conjugate}
    a \, &\rightarrow \, -a \,, \\
    m \, &\rightarrow \, -m \,, \\
    \beta \, &\rightarrow \, -\beta
\end{aligned}
\end{align}
and similarly for $b$ and $n$. This formal conjugate will be understood throughout the paper.

Note that each 1-loop determinant is completely factorized into a holomorphic piece and an anti-holomorphic piece:
\begin{align}
\begin{aligned}
\label{eq:fusion_SCI}
    Z_\text{vector} &= \mathcal Z_\text{vector}(z;\beta) \times \overline{\mathcal Z}_\text{vector}(\bar z;-\beta) \,, \\
    Z_\text{chiral} &= \mathcal Z_\text{chiral}(z,t;\beta) \times \overline{\mathcal Z}_\text{chiral}(\bar z,\bar t;-\beta)
\end{aligned}
\end{align}
where $\mathcal Z_\text{vector}$ and $\mathcal Z_\text{chiral}$ are defined by
\begin{align}
\begin{aligned}
\label{eq:fact1}
    \mathcal Z_\text{vector}(z;\beta) &= \prod_{\alpha} e^{-\frac{1}{8 \beta} (\alpha(u)+\pi i)^2} \left(z^{\alpha};x^2\right) \,, \\
    \mathcal Z_\text{chiral}(z,t;\beta) &= \prod_{\rho \, \otimes \, \sigma} e^{\frac{\beta}{8} \left((r_\sigma-1)^2-\frac{1}{3}\right)-\frac{r_\sigma-1}{4} (\rho(u)+\sigma(v)-\pi i (r_\sigma-1))+\frac{1}{8 \beta} (\rho(u)+\sigma(v)-\pi i (r_\sigma-1))^2} \nonumber \\
    &\qquad \times \left(z^\rho t^\sigma x^{r_\sigma} e^{-\pi i r_\sigma};x^2\right)^{-1}.
\end{aligned}
\end{align}
Here $\overline{\mathcal Z}$ is defined such that any imaginary coefficient flips its sign. Also the following definition of $(a;x^2)$ for $|x| \neq 1$ is understood:
\begin{align}
    \left(a;x^2\right) = \left\{\begin{array}{ll}
    \left(a;x^2\right)_\infty, \qquad & |x| < 1 \,, \\
    \left(a x^{-2};x^{-2}\right)_\infty^{-1}, \qquad & |x| > 1 \,.
    \end{array}\right.
\end{align}
While we have introduced the exponential factor $e^{\frac{\beta}{8} \left((r_\sigma-1)^2-\frac{1}{3}\right)+\frac{\pi i}{4} (r_\sigma-1)^2}$ in $\mathcal Z_\text{chiral}(z,t;\beta)$ to be matched with the hemisphere determinant, it is irrelevant because it is canceled by the same factor of $\overline{\mathcal Z}_\text{chiral}(\bar z,\bar t;-\beta)$.
Their Cardy limits are given by
\begin{align}
\begin{aligned}
\label{eq:Cardy1}
    \lim_{\beta \rightarrow 0} \log \mathcal Z_\text{vector}(z;\beta) &= -\frac{1}{2 \beta} \sum_\alpha \left[\frac{1}{4} (\alpha(u)+\pi i)^2+\mathrm{Li}_2(z^\alpha x^{-1})\right], \\
    \lim_{\beta \rightarrow 0} \log \mathcal Z_\text{chiral}(z,t;\beta) &= \frac{1}{2 \beta} \sum_{\rho \, \otimes \, \sigma} \left[\frac{1}{4} (\rho(u)+\sigma(v)-(\beta+\pi i) (r_\sigma-1))^2+\mathrm{Li}_2 (z^\rho t^\sigma x^{r_\sigma-1} e^{-\pi i r_\sigma})\right]
\end{aligned}
\end{align}
up to $\mathcal O(\beta)$.

We should comment that there is another way of factorizing the 1-loop determinant of the chiral multiplet:
\begin{align}
    Z_\text{chiral} &= \mathcal Z_\text{chiral}(z,t;\beta) \times \overline{\mathcal Z}_\text{chiral}(\bar z,\bar t;-\beta) \,, \\
    \mathcal Z_\text{chiral}(z,t;\beta) &= \prod_{\rho \, \otimes \, \sigma} e^{-\frac{\beta}{8} \left((r_\sigma-1)^2-\frac{1}{3}\right)+\frac{r_\sigma-1}{4} (\rho(u)+\sigma(v)-\pi i r_\sigma)} e^{-\frac{1}{8 \beta} (\rho(u)+\sigma(v)-\pi i r_\sigma)^2} \nonumber \\
    &\qquad \times \left(z^{-\rho} t^{-\sigma} x^{2-r_\sigma} e^{\pi i r_\sigma}\right), \label{eq:fact2}
\end{align}
which can be obtained by replacing $m$ by $-m$ in the identity \eqref{eq:ident}.
The Cardy limit is given by
\begin{align}
\label{eq:Cardy2}
    & \lim_{\beta \rightarrow 0} \log \mathcal Z_\text{chiral}(z,t;\beta) \nonumber \\
    &= -\frac{1}{2 \beta} \sum_{\rho \, \otimes \, \sigma} \left[\frac{1}{4} (\rho(u)+\sigma(v)-(\beta+\pi i) (r_\sigma-1))^2+\mathrm{Li}_2(z^{-\rho} t^{-\sigma} x^{1-r_\sigma} e^{\pi i r_\sigma})\right].
\end{align}
One should note that the holomorphic part of the first factorization \eqref{eq:fact1} is identified with the determinant on $D^2 \times S^1$ with the Neumann boundary condition \eqref{eq:Neumann} while the holomorphic part of the second factorization \eqref{eq:fact2} is identified with the determinant on $D^2 \times S^1$ with the Dirichlet boundary condition \eqref{eq:Dirichlet}. As we have seen in the previous subsection, they differ by the determinant of a boundary chiral multiplet, or equivalently a boundary Fermi multiplet. Recall that the 2d determinants in the Cardy limit are given by \eqref{eq:2d}:
\begin{align}
    \lim_{\beta \rightarrow 0} \log \mathcal Z_\text{chiral}^{2d}
    = -\lim_{\beta \rightarrow 0} \log \mathcal Z_\text{Fermi}^{2d}
    = -\frac{\pi^2}{12 \beta}.
\end{align}
They simply vanish when we glue two copies of them following the fusion rule \eqref{eq:fusion_SCI}. Thus, we need not to worry about the boundary matters when we factorize the superconformal index into two copies of the hemisphere indices in the Cardy limit; and also the distinction between the two boundary conditions, i.e., how we factorize the determinants, is not significant.
\\

Lastly, let us examine the Chern-Simons terms. In terms of the holomorphic variables, they are written as follows:
\begin{align}
    e^{i k \, \mathrm{tr} (a m)} &= e^{\frac{k}{2 \beta} \, \mathrm{tr} \left(\frac{1}{2} u^2-\frac{1}{2} \bar u^2\right)}, \\
    z_A^{m_B} z_B^{m_A} &= e^{\frac{1}{2 \beta} \, \mathrm{tr} \left(u_A u_B-\bar u_A \bar u_B\right)}
\end{align}
where the definition of $u$ is given in \eqref{eq:u}.
Thus, the classical action contribution is also factorized in the same way:
\begin{align}
    Z_\text{classical} = \mathcal Z_\text{classical} (z,t;\beta) \times \overline{\mathcal Z}_\text{classical} (\bar z,\bar t;-\beta)
\end{align}
where $\mathcal Z_\text{classical} (z,t;\beta)$ consists of two types of contributions:
\begin{align}
    e^{\frac{k}{2 \beta} \, \mathrm{tr} \left(\frac{1}{2} u^2\right)}
\end{align}
for a level $k$ Chern-Simons term and
\begin{align}
    e^{\frac{1}{2 \beta} \, \mathrm{tr} \left(u_A u_B\right)}
\end{align}
for a mixed Chern-Simons term between $U(1)_A \times U(1)_B$, which are exactly what appears in the hemisphere index.
\\

Combining those Cardy limits of the 1-loop determinants and the classical action contributions, the entire superconformal index is completely factorized into the holomorphic integral and the anti-holomorphic integral. In particular, the superconformal index has the holonomy integration as well as the magnetic flux sum where the latter is replaced by another integration in the Cardy limit \cite{Pasquetti:2019uop,Choi:2019zpz}. Combined with the original holonomy integration, it gives rise to
\begin{align}
    \sum_{m = -\infty}^\infty \oint_{|\mathfrak z| = 1} \frac{d \mathfrak z}{2 \pi i \mathfrak z} = \sum_{r \in e^{\beta \mathbb Z}} \int_0^{2 \pi} \frac{d \theta}{2 \pi} \quad \rightarrow \quad \int_\mathbb C \frac{dz d \bar{z}}{4 \pi \beta |z|^2}
\end{align}
up to $\mathcal O(\beta)$.
If we consider a $U(N)$ theory for simplicity, the Cardy limit of the superconformal index is given by
\begin{align}
\label{eq:Cardy_SCI}
    \lim_{\beta \rightarrow 0} I_{S^2} = \frac{1}{N!} \int_{\mathbb C^N} \left(\prod_{a = 1}^{N} \frac{dz_a d\bar z_a}{4 \pi \beta |z_a|^2}\right) e^{-\frac{1}{2 \beta} \left(\mathcal W(z,\mathsf t)-\overline{\mathcal W( z,\mathsf t)}\right)}
\end{align}
where
\begin{align}
    -\frac{1}{2 \beta} \mathcal W(z,\mathsf t) = \lim_{\beta \rightarrow 0} \left[\log\mathcal Z_\text{classical}(z,t;\beta)+\log\mathcal Z_\text{vector}(z;\beta)+\log\mathcal Z_\text{chiral}(z,t;\beta)\right]
\end{align}
with each component described above. As seen in the previous subsection, there is no explicit $\beta$-dependence in $\mathcal W$ as long as we use $\mathsf t = (t,-\hat x e^{-\pi i})$, on which the global weight $\sigma$ acts such that $\mathsf t^\sigma = t^\sigma \hat x^{r_\sigma-1} e^{-\pi i r_\sigma}$. According to the rule \eqref{eq:conjugate}, $\bar {\mathsf t}$ is defined by
\begin{align}
    \bar {\mathsf t} = (\bar t,-\hat x^{-1} e^{\pi i}) \,.
\end{align}

In the Cardy limit, the superconformal index can be also evaluated by the saddle point approximation.
One should note that the holomorphic part of \eqref{eq:Cardy_SCI} is nothing but the hemisphere index in the Cardy limit:
\begin{align}
    \frac{1}{N!} \int \left(\prod_{a = 1}^{N} \frac{dz_a}{2 \pi i z_a}\right) e^{-\frac{1}{2 \beta} \mathcal W(z,\mathsf t)}
\end{align}
while the anti-holomorphic part is given by its conjugate defined by \eqref{eq:conjugate}.
Naively the holomorphic variable and anti-holomorphic variable may solve the saddle point equations independently as they are formal conjugate variables rather than complex conjugate. However, regarding $\hat \beta$ as an independent variable, if $\hat \beta$ is pure imaginary, such conjugate variables are truly complex conjugate because they are related by $\hat \beta \leftrightarrow -\hat \beta$. In that case, the two saddle point equations are essentially the same; and there is no reason to solve the equations independently because it is basically the saddle point approximation of a real function on the real axis. If $\hat \beta$ is slightly away from the imaginary axis, however, $\hat \beta \rightarrow -\hat \beta$ is not compatible with the complex conjugate anymore, in which case, the two saddle point equations should be solved independently. Nevertheless, if we assume a smooth transition when $\hat \beta$ moves away from the imaginary axis, the saddle point should be determined such that $z$ and $\bar z$ at a saddle should become complex conjugate to each other as $\hat \beta$ approaches the imaginary axis; i.e., there is a natural one-to-one map between the holomorphic saddles and the anti-holomorphic saddles, and only the pairs of saddles related by this map contributes to the index.
Thus, the saddle points are determined by the same equation for the hemisphere index, and each saddle point contribution is that for the hemisphere index times its conjugate.

While we have assumed a smooth transition when $\hat \beta$ moves aways from the imaginary axis, it would be worth studying the behavior of those saddles more rigorously. Nevertheless, in spite of such a subtlety in dealing with the two complex saddle point equations, it is not very significant at the end as long as the index exponentially grows in the Cardy limit. Recall that the free energy for the hemisphere index is determined by the contribution of the dominant saddle because the contributions of the other saddles are exponentially suppressed. As the same thing happens for the superconformal index, we find a simple relation between their free energies:
\begin{align}
    F_{S^2}(\mathfrak t,n;\beta) = N \log \beta + F_{D^2}(t;\beta) + \overline F_{D^2}(\bar t;-\beta) - \log\left[N! \, \pi^N\right] +\mathcal O(\beta) \,,
\end{align}
or in terms of the Cardy block in \eqref{eq:Cardy block},
\begin{align}
\label{eq:SCI}
    F_{S^2}(\mathfrak t,n;\beta) = N \log \beta - \log \mathcal C(t;\beta) - \log \overline{\mathcal C}(\bar t;-\beta) - \log\left[N! \, \pi^N\right] +\mathcal O(\beta) \,,
\end{align}
with
\begin{align}
t = \mathfrak t \, e^{\beta n} \,, \qquad \bar t = \mathfrak t^{-1} \, e^{\beta n} \,,
\end{align}
where only the dominant saddle plays the role. While the formula is written for a $U(N)$ theory for simplicity, the generalization is straightforward.
\\

\subsection{The Refined Topologically Twisted Index}

One can also define an index for a topologically twisted theory on a circle fibered over a Riemann surface of genus $g$ \cite{Benini:2015noa,Benini:2016hjo,Closset:2016arn,Closset:2017zgf,Closset:2018ghr}. If $g = 0$, i.e., if the Riemann surface is a sphere, one can refine the index by turning on the angular momentum fugacity. In that case, the index is defined by
\begin{align}
    I_{S^2}^\text{twisted} = \mathrm{tr}_{\mathcal H_{BPS}(S^2;n,n_R = -1)} \left[(-1)^F e^{-2 \beta \mathsf J_3} e^{-F_l M_l}\right]
\end{align}
where the trace is taken over the Hilbert space on $S^2$ in the presence of the $R$-symmetry flux $n_R = -1$. Again $n$ denotes the external magnetic flux for the flavor symmetry.
Using the supersymmetric localization, we have
\begin{align}
    I_{S^2}^\text{twisted} = \frac{1}{|W_G|} \sum_{m \in \Gamma_G^\vee} \oint \left(\prod_{a = 1}^{\mathrm{rk}(G)} \frac{d \mathfrak z_a}{2 \pi i \mathfrak z_a}\right) \, Z_\text{classical}^\text{twisted} \, Z_\text{vector}^\text{twisted} \, Z_\text{chiral}^\text{twisted} \,.
\end{align}
where the integration contour is determined by the rule of the Jeffrey-Kirwan residue \cite{Benini:2015noa}. The 1-loop determinants of the vector multiplet and the chiral multiplet are given by
\begin{align}
    Z^\text{twisted}_\text{vector} &= \prod_{\alpha \in \Delta} \left(x e^{-\pi i}\right)^{-\frac{|\alpha(m)|}{2}} \left(1-\mathfrak z^\alpha x^{|\alpha(m)|}\right), \\
    Z^\text{twisted}_\text{chiral} &= \prod_{\rho \, \otimes \, \sigma \in \mathfrak R} \frac{(\mathfrak z^\rho \mathfrak t^\sigma e^{-\pi i (r_\sigma-1)})^\frac{\rho(m)+\sigma(n)-r_\sigma+1}{2}}{(\mathfrak z^\rho \mathfrak t^\sigma x^{r_\sigma-\rho(m)-\sigma(n)} e^{-\pi i r_\sigma};x^2)_{\rho(m)+\sigma(n)-r_\sigma+1}} \,.
\end{align}
The twist by the (non-superconformal) $R$-symmetry demands that the $R$-charge of a matter should be an integer. Again please note extra $e^{-\pi i r_\sigma}$ due to our choice $F = \mathsf R$. The classical action contributions are given by
\begin{align}
    e^{i k \, \mathrm{tr} (a m)}
\end{align}
for a canonical Chern-Simons term with level $k$ and
\begin{align}
    z_A^{m_B} z_B^{m_A}
\end{align}
for a mixed Chern-Simons term between $U(1)_A \times U(1)_B$, each of which is either gauge or global $U(1)$.
Again each pair of $\mathfrak z = e^{i a}$ and $m$ denote the holonomy and the magnetic flux for the corresponding symmetry.
\\

One can manipulate the determinants in a similar way to the superconformal index:
\begin{align}
    Z^\text{twisted}_\text{vector} 
    &= \prod_{\alpha} e^{-\frac{1}{8 \beta} \left(\alpha(u)^2-\alpha(\bar u)^2\right)} \frac{\left(z^{\alpha};x^2\right)}{\left(\bar z^{-\alpha} x^{2};x^2\right)} \,, \\
    Z^\text{twisted}_\text{chiral} 
    &= \prod_{\rho \, \otimes \, \sigma} e^{-\frac{r_\sigma-1}{4} (\rho(u)+\sigma(v)-\rho(\bar u)-\sigma(\bar v)-2 \pi i (r_\sigma-1))+\frac{1}{8 \beta} \left((\rho(u)+\sigma(v)-\pi i (r_\sigma-1))^2-(\rho(\bar u)+\sigma(\bar v)+\pi i (r_\sigma-1))^2\right)} \nonumber \\
    &\qquad \times \frac{\left(\bar z^{-\rho} \bar t^{-\sigma} x^{2-r_\sigma} e^{-\pi i r_\sigma};x^2\right)}{\left(z^\rho t^\sigma x^{r_\sigma} e^{-\pi i r_\sigma};x^2\right)} \,,
\end{align}
\begin{gather}
\begin{gathered}
    z^\rho = e^{\rho(u)} = \mathfrak z^\rho x^{-\rho(m)} \,, \qquad t^\sigma = e^{\sigma(v)} = \mathfrak t^\sigma x^{-\sigma(m)} \,, \\
    \bar z^\rho = e^{\rho(\bar u)} = \mathfrak z^{-\rho} x^{-\rho(m)} \,, \qquad \bar t^\sigma = e^{\sigma(\bar v)} = \mathfrak t^{-\sigma} x^{-\sigma(m)} \,,
\end{gathered}
\end{gather}
which lead to the following factorization
\begin{align}
    Z^\text{twisted}_\text{vector} &= \mathcal Z_\text{vector}(z;\beta) \times \mathcal Z_\text{vector}(\bar z^{-1};-\beta) \,, \\
    Z^\text{twisted}_\text{chiral} &= \mathcal Z_\text{chiral}(z,t;\beta) \times \mathcal Z_\text{chiral}(\bar z^{-1},\bar t^{-1};-\beta)
\end{align}
where $\mathcal Z_\text{vector}$ and $\mathcal Z_\text{chiral}$ are defined exactly in the same way as the hemisphere determinants:
\begin{align}
    \mathcal Z_\text{vector}(z;\beta) &= \prod_{\alpha} e^{-\frac{1}{8 \beta} (\alpha(u)+\pi i)^2} \left(z^{\alpha};x^2\right), \\
    \mathcal Z_\text{chiral}(z,t;\beta) &= \prod_{\rho \, \otimes \, \sigma} e^{\frac{\beta}{8} \left((r_\sigma-1)^2-\frac{1}{3}\right)-\frac{r_\sigma-1}{4} (\rho(u)+\sigma(v)-\pi i (r_\sigma-1))+\frac{1}{8 \beta} (\rho(u)+\sigma(v)-\pi i (r_\sigma-1))^2} \nonumber \\
    &\qquad \times \left(z^\rho t^\sigma x^{r_\sigma} e^{-\pi i r_\sigma};x^2\right)^{-1}.
\end{align}
The Cardy limits of those are given in \eqref{eq:Cardy}. In the Cardy limit, the boundary matters are irrelevant for the factorization of the topologically twisted index, due to the same reason as the superconformal index.
\\

Next, the classical action contribution is factorized into
\begin{align}
    Z_\text{classical}^\text{twisted} = \mathcal Z_\text{classical} (z,t;\beta) \times \mathcal Z_\text{classical} (\bar z^{-1},\bar t^{-1};-\beta)
\end{align}
where $\mathcal Z_\text{classical} (z,t;\beta)$ consists of two types of contributions:
\begin{align}
    e^{\frac{k}{2 \beta} \, \mathrm{tr} \left(\frac{1}{2} u^2\right)}
\end{align}
for a level $k$ Chern-Simons term and
\begin{align}
    e^{\frac{1}{2 \beta} \, \mathrm{tr} \left(u_A u_B\right)}
\end{align}
for a mixed Chern-Simons term between $U(1)_A \times U(1)_B$.
\\

Like the superconformal index, the entire topologically twisted index is completely factorized into the holomorphic integral and the anti-holomorphic integral given that the magnetic flux sum is replaced by an integration in the Cardy limit:
\begin{align}
\label{eq:Cardy_twisted}
    \lim_{\beta \rightarrow 0} I_{S^2}^\text{twisted} = \frac{1}{N!} \int_{\mathbb C^N} \left(\prod_{a = 1}^N \frac{dz_a d\bar z_a}{4 \pi \beta |z_a|^2}\right) e^{-\frac{1}{2 \beta} \left(\mathcal W(z,\mathsf t)-\mathcal W(\bar z^{-1},\tilde {\mathsf t}^{-1})\right)}
\end{align}
where
\begin{align}
\label{eq:TTI}
    -\frac{1}{2 \beta} \mathcal W(z,\mathsf t) = \lim_{\beta \rightarrow 0} \left[\log\mathcal Z_\text{classical}(z,t;\beta)+\log\mathcal Z_\text{vector}(z;\beta)+\log\mathcal Z_\text{chiral}(z,t;\beta)\right].
\end{align}
$\mathsf t = (t,-\hat x e^{-\pi i})$ is defined such that $\mathsf t^\sigma = t^\sigma \hat x^{r_\sigma-1} e^{-\pi i r_\sigma}$ as before, and $\tilde{\mathsf t} = (\bar t,-\hat x e^{\pi i})$.
While this form of the topologically twisted index and that of the superconformal index are very much alike, there is a crucial difference that the topologically twisted index does not grow in the small $\beta$ limit because $\beta^{-1}$ terms should be canceled out at the end. This is anticipated because the strict Cardy limit is basically the unrefined limit of the topologically twisted index, which is $\mathcal O(1)$. Thus, if one does the saddle point approximation, every saddle democratically contributes to the index, which ends up with the BAE formula for the unrefined index \cite{Benini:2015noa}. Nevertheless, if there happens to be only one saddle or if there is a dominant saddle due to other large parameters such as large $N$, one can write down the free energy in a simple manner as before:
\begin{align}
\begin{aligned}
\label{eq:twisted}
    F_{S^2}^\text{twisted}(\mathfrak t,n;\beta) = N \log \beta + F_{D^2}(t;\beta) + F_{D^2}(\bar t^{-1};-\beta) - \log\left[N! \, \pi^N\right] + \mathcal O(\beta) \,, \\
    = N \log \beta - \log \mathcal C(t;\beta) - \log \mathcal C(\bar t^{-1};-\beta) - \log\left[N! \, \pi^N\right] + \mathcal O(\beta)
\end{aligned}
\end{align}
where $t = \mathfrak t \, e^{\beta n}$, $\bar t = \mathfrak t^{-1} \, e^{\beta n}$, and $\mathcal C$ is the Cardy block in \eqref{eq:Cardy block}. Furthermore, although the strict Cardy limit is the unrefined limit of the topologically twisted index, \eqref{eq:TTI} will suggest a conjecture for finite $\beta$ in the large $N$ limit. We will discuss it with explicit examples in section \ref{sec:ex1} and \ref{sec:ex2}.
\\

\subsection{The Squashed Sphere Partition Function}

Lastly, we consider the supersymmetric partition function on the squashed sphere $S^3_b$. Again using the supersymmetric localization, it is given by \cite{Kapustin:2009kz,Hama:2010av,Hama:2011ea}
\begin{align}
    Z_{S^3_b} = \frac{1}{|W_G|} \int d^{\mathrm{rk}(G)} \hat u \, Z^{S^3_b}_\text{classical} \, Z^{S^3_b}_\text{vector} \, Z^{S^3_b}_\text{chiral}.
\end{align}
The 1-loop determinants of the vector multiplet and the chiral multiplet are
\begin{align}
    Z^{S^3_b}_\text{vector} &= \prod_{\alpha \in \Delta} s_b\left(i \frac{Q}{2}-\alpha(\hat u)\right)^{-1}, \\
    Z^{S^3_b}_\text{chiral} &= \prod_{\rho \, \otimes \, \sigma \in \mathfrak R} s_b\left(i \frac{Q}{2} (1-r_\sigma)-\rho(\hat u)-\sigma(\hat v)\right)
\end{align}
with $Q = b+\frac{1}{b}$.
$s_b(\hat u)$ is the double-sine function defined by
\begin{align}
    s_b\left(i\frac{Q}{2}-\hat u\right) &= e^{-\frac{\pi i}{2} \left(\left(i \frac{Q}{2}-\hat u\right)^2+\frac{Q^2}{12}-\frac{1}{6}\right)} \frac{(e^{-2 \pi b^{-1} \hat u+2 \pi i b^{-2}};e^{2 \pi i b^{-2}})}{(e^{-2 \pi b \hat u};e^{-2 \pi i b^2})}
\end{align}
for $\mathrm{Im}(b^2) \neq 0$, which leads to the following factorization of the determinants:
\begin{align}
\begin{aligned}
\label{eq:fusion_S3}
    Z^{S^3_b}_\text{vector} &= \mathcal Z^{S^3_b}_\text{vector} \left(e^{-2 \pi b \hat u};\pi i b Q\right) \times \mathcal Z^{S^3_b}_\text{vector} \left(e^{-2 \pi b^{-1} \hat u};\pi i b^{-1} Q\right), \\
    Z^{S^3_b}_\text{chiral} &= \mathcal Z^{S^3_b}_\text{chiral} \left(e^{-2 \pi b \hat u},e^{-2 \pi b \hat v};\pi i b Q\right) \times \mathcal Z^{S^3_b}_\text{chiral} \left(e^{-2 \pi b^{-1} \hat u},e^{-2 \pi b^{-1} \hat v};\pi i b^{-1} Q\right)
\end{aligned}
\end{align}
where
\begin{gather}
\begin{align}
    \mathcal Z^{S^3_b}_\text{vector} (z;\beta) &= \prod_{\alpha} e^{-\frac{\beta}{12}-\frac{\pi i}{24}-\frac{1}{8 \beta} \alpha(u)^2} \left(z^\alpha;x^2\right), \\
    \mathcal Z^{S^3_b}_\text{chiral} (z,t;\beta) &= \prod_{\rho \, \otimes \, \sigma} e^{\frac{\beta}{8} \left((r_\sigma-1)^2-\frac{1}{3}\right)-\frac{r_\sigma-1}{4} (\rho(u)+\sigma(v))+\frac{\pi i}{24}+\frac{1}{8 \beta} (\rho(u)+\sigma(v))^2} \frac{1}{\left(z^\rho t^\sigma x^{r_\sigma};x^2\right)}
\end{align} \\
\begin{gathered}
    u = \log z \, , \qquad v = \log t \, , \qquad \beta = -\log x \, .
\end{gathered}
\end{gather}
\\

For the sphere partition function the limit of our interest is the highly squashed limit of the sphere. More precisely, we take the limit $b^2 \rightarrow i 0^+$ with fixed $u = -2 \pi b \hat u$ and $v = -2 \pi b \hat v$. We will compare this limit with the Cardy limits of the other partition functions we have discussed. An interesting thing is that in this limit, unlike the other partition functions, the holomorphic part and the anti-holomorphic part do not contribute democratically. Indeed, even though the squashed sphere partition function has the factorized structure as in \eqref{eq:fusion_S3}, what corresponds to the hemisphere index in the Cardy limit is the entire squashed partition function rather than its holomorphic part. To see this, let us first take the highly squashed limit of each component in \eqref{eq:fusion_S3}:
\begin{align}
    & \lim_{b \rightarrow 0} \log \mathcal Z^{S^3_b}_\text{vector} \left(e^{-2 \pi b \hat u};\pi i b Q\right) = -\frac{1}{2 \pi i b^2} \sum_\alpha\mathrm{Li}_2 \left(z^\alpha e^{\pi i b^2}\right)-\frac{1}{8 \pi i} \sum_\alpha (\alpha(u)+\pi i)^2, \\
    & \lim_{b \rightarrow 0} \log \mathcal Z^{S^3_b}_\text{vector} \left(e^{-2 \pi b^{-1} \hat u};\pi i b^{-1} Q\right) \nonumber \\
    &= -\frac{1}{2 \pi i b^2} \sum_\alpha \left[\frac{\pi^2}{12}+\frac{1}{4} (\alpha(u)+\pi i)^2\right]+\frac{1}{8 \pi i} \sum_\alpha \left[\alpha(u)^2+\pi^2\right], \\
    & \lim_{b \rightarrow 0} \log \mathcal Z^{S^3_b}_\text{chiral} \left(e^{-2 \pi b \hat u},e^{-2 \pi b \hat v};\pi i b Q\right) \nonumber \\
    &= \frac{1}{2 \pi i b^2} \sum_{\rho \, \otimes \, \sigma} \mathrm{Li}_2 \left(z^\rho t^\sigma e^{-\pi i r_\sigma-\pi i b^2 (r_\sigma-1)} \right)+\frac{1}{8 \pi i} \sum_{\rho \, \otimes \, \sigma} \left[\rho(u)+\sigma(v)-\pi i (r_\sigma-1)\right]^2, \\
    & \lim_{b \rightarrow 0} \log \mathcal Z^{S^3_b}_\text{chiral} \left(e^{-2 \pi b^{-1} \hat u},e^{-2 \pi b^{-1} \hat v};\pi i b^{-1} Q\right) \nonumber \\
    &= \frac{1}{2 \pi i b^2} \sum_{\rho \, \otimes \, \sigma} \left[\frac{\pi^2}{12}+\frac{1}{4} \left(\rho(u)+\sigma(v)- \pi i (r_\sigma-1)\right)^2\right]-\frac{1}{8 \pi i} \sum_{\rho \, \otimes \, \sigma} \left[(\rho(u)+\sigma(v))^2+\pi^2 (r_\sigma-1)^2\right],
\end{align}
provided that the argument sits in a chamber: $-2 \pi < \mathrm{Im}(\rho(u)+\sigma(v)-\pi i r_\sigma) < 0$. As mentioned above, the holomorphic part and the anti-holomorphic part do not contribute democratically. Indeed, the holomorphic and anti-holomorphic parts combined give rise to
\begin{align}
    & \lim_{b \rightarrow 0} \log Z^{S^3_b}_\text{vector} = -\frac{1}{2 \pi i b^2} \sum_\alpha \left[\frac{\pi^2}{12}+\frac{1}{4} (\alpha(u)+\pi i)^2+\mathrm{Li}_2 \left(z^\alpha e^{\pi i b^2}\right)\right]-\frac{\pi i}{4} |\Delta| \,, \\
    & \lim_{b \rightarrow 0} \log Z^{S^3_b}_\text{chiral} \nonumber \\
    &= \frac{1}{2 \pi i b^2} \sum_{\rho \, \otimes \, \sigma} \left[\frac{\pi^2}{12}+\frac{1}{4} \left(\rho(u)+\sigma(v)-\pi i (1+b^2) (r_\sigma-1)\right)^2+\mathrm{Li}_2 \left(z^\rho t^\sigma e^{-\pi i r_\sigma-\pi i b^2 (r_\sigma-1)} \right)\right]
\end{align}
up to $\mathcal O(b^2)$, which are the same as the hemisphere determinants up to constant terms. Indeed, such constant terms are just a phase or remnants of the 2d boundary matter contributions, which however do not affect the factorization of the superconformal index and of the topologically twisted index. Thus, one can ignore such constant terms.
\\

Next, the classical action contribution $Z^{S^3_b}_\text{classical}$ includes two types of contributions:
\begin{align}
    e^{-\pi i k \, \mathrm{tr} (\hat u^2)}
\end{align}
for a canonical Chern-Simons term with level $k$ and
\begin{align}
    e^{-2 \pi i \, \hat u_A \hat u_B}
\end{align}
for a mixed Chern-Simons term between $U(1)_A \times U(1)_B$, which can be factorized in the same way as the determinants. Moreover, in the highly squashed limit, they become
\begin{gather}
    e^{\frac{k}{2 \pi i b^2} \, \mathrm{tr} \left(\frac{1}{2} u^2\right)}, \\
    e^{\frac{1}{2 \pi i b^2} u_A u_B}
\end{gather}
since we are keeping $u = -2 \pi b \hat u$ and $v = -2 \pi b \hat v$ finite.
\\

Therefore, the sphere partition function in the highly squashed limit is given in the following form:
\begin{align}
    \lim_{b \rightarrow 0} Z_{S^3_b} = \frac{1}{(i b)^N} \frac{1}{N!} \int \left(\prod_{a = 1}^N \frac{dz_a}{2 \pi i z_a}\right) e^{-\frac{1}{2 \pi i b^2} \mathcal W(z,\mathsf t)}
\end{align}
where
\begin{align}
    -\frac{1}{2 \pi i b^2} \mathcal W(z,\mathsf t) = \lim_{b \rightarrow 0} \left[\log\mathcal Z_\text{classical}(z,t;\pi i b^2)+\log\mathcal Z_\text{vector}(z;\pi i b^2)+\log\mathcal Z_\text{chiral}(z,t;\pi i b^2)\right]
\end{align}
As before, we have used the variable $\mathsf t = (t,-e^{-\pi i-\pi i b^2})$, which is defined such that $\mathsf t^\sigma = t^\sigma e^{-\pi i r_\sigma-\pi i b^2 (r_\sigma-1)}$.
In terms of the hemisphere index, the highly squashed sphere partition function is given by
\begin{align}
\begin{aligned}
\label{eq:sphere}
    F_{S^3_b}(\Delta;\beta) &= \frac{N}{2} \log \beta + F_{D^2}(e^{\Delta};\beta) - \frac{N}{2} \log[-\pi i] + \mathcal O(\beta) \,,\\
     &= \frac{N}{2} \log \beta - \log \mathcal C(e^{\Delta};\beta) - \frac{N}{2} \log[-\pi i] + \mathcal O(\beta) \,.
\end{aligned}
\end{align}
with $\Delta = -2 \pi b \hat v$, $\beta = \pi i b^2$ and the Cardy block $\mathcal C$ in \eqref{eq:Cardy block}.
\\

\section{The Universal Formula}
\label{sec:universal}

In this section, we discuss the universal relations in the Cardy limit among the partition functions we have discussed in the previous section. So far we have seen that those quantities can be solely written in terms of the Cardy block. We examine this relation more carefully and propose general index theorems for those quantities in the Cardy limit. In particular, such index theorems will prove very useful when we consider the large $N$ limit of those quantities, which relates to the entropic quantities of the dual gravity.
\\

In the previous section, we have already found how the hemisphere index, or the Cardy block, relates to the other three partition functions. See \eqref{eq:SCI}, \eqref{eq:twisted} and \eqref{eq:sphere}. Since the topologically twisted index needs more care, let us first focus on the generalized superconformal index and the squashed sphere partition function:
\begin{align}
\begin{aligned}
\label{eq:fusion}
    &F_{S^2}(\mathfrak t,n;\beta) = N \log \beta - \log \mathcal C(t;\beta) - \log \overline{\mathcal C}(\bar t;-\beta) + \mathcal O(\beta) \,, \\
    &F_{S^3_b}(\Delta;\beta) = \frac{N}{2} \log \beta - \log \mathcal C(e^{\Delta};\beta) \, + \mathcal O(\beta)
\end{aligned}
\end{align}
up to irrelevant numerical constants. $t = \mathfrak t \, e^{\beta n}$ and $\bar t = \mathfrak t^{-1} \, e^{\beta n}$ are understood for the superconformal index while $\Delta = -2 \pi b \hat v$ and $\beta = \pi i b^2$ are understood for the squashed sphere partition function. Combining those, we find our first index theorem:
\begin{align}
\label{eq:theorem1}
    F_{S^2}(e^\Delta,n;\beta) = F_{S^3_b}(\Delta+\beta n;\beta) + \overline F_{S^3_b}(-\Delta+\beta n;-\beta)+\mathcal O(\beta) \,.
\end{align}
One may recall that we have taken integer $r_\sigma$. This is necessary condition for the topologically twisted index while it is not for the other partition functions. Even for the other partition functions, however, the $R$-charges are not completely arbitrary and are restricted by the superpotential of the theory. Indeed, such restricted $R$-charges can be parametrized by the flavor chemical potentials. Namely, one can obtain any allowed $R$-charges, which are generically non-integer values, by shifting 
the flavor chemical potentials by $\Delta_i \rightarrow \Delta_i-(\beta+\pi i) \delta_i$. Then, the $R$-charge is deformed from $r_\sigma$ to $r_\sigma+\sigma(\delta)$. From now on, this deformed $R$-charge is understood except the topologically twisted index.
\\

One can also expand the right hand side with respect to $\beta$. Recall that $-\log \mathcal C(t;\beta) \approx F_{D^2}(t;\beta)$ has the form
\begin{align}
    -\log \mathcal C(t;\beta) = \frac{1}{\beta} G^{(0)} (\mathsf t) - \frac{N}{2} \log \beta + G^{(1)}(\mathsf t)+\mathcal O(\beta)
\end{align}
where $\mathsf t = (t,-\hat x e^{-\pi i})$ is defined such that
\begin{align}
    \mathsf t^\sigma = t^\sigma \hat x^{r_\sigma-1} e^{-\pi i r_\sigma}.
\end{align}
$t$ has a different definition for each partition function, but most generally it is defined by
\begin{align}
    t = \mathfrak t e^{\beta (n-\delta)-\pi i \delta} \,, \qquad \bar t = \mathfrak t^{-1} e^{\beta (n+\delta)+\pi i \delta}
\end{align}
where the $R$-charge deformation is taken into account. For later convenience, we include $e^{-\pi i \delta}$ in the definition of $\mathfrak t$ from now on. The small $\beta$ expansion of $-\log \mathcal C(t;\beta)$ is thus given by
\begin{align}
    -\log \mathcal C(t;\beta) &= \left[\frac{1}{\beta} G^{(0)} (\mathsf t) - \frac{N}{2} \log \beta + \frac{d \mathsf t}{d \beta} \cdot \frac{\partial}{\partial \mathsf t} G^{(0)}(\mathsf t) + G^{(1)}(\mathsf t)\right]_{\hat \beta \rightarrow 0} + \mathcal O(\beta) \,.
\end{align}
Here, the third term can be expanded as
\begin{align}
    \frac{d \mathsf t}{d \beta} \cdot \frac{\partial}{\partial \mathsf t} G^{(0)}(\mathsf t) = \sum_i (n_i-\delta_i) t_i \frac{\partial}{\partial t_i} G^{(0)}(\mathsf t)-\hat x \frac{\partial}{\partial \hat x} G^{(0)}(\mathsf t)
\end{align}
where we have used $\frac{d \hat \beta}{d \beta} = 1$.
Using this expansion, we have the following expressions for $-\log \mathcal C$ and $-\log \overline{\mathcal C}$:
\begin{align}
    & -\log \mathcal C(t;\beta) = \nonumber \\
    &\frac{1}{\beta} G^{(0)} (\mathfrak t) - \frac{N}{2} \log \beta + \sum_i (n_i-\delta_i) \mathfrak t_i \frac{\partial}{\partial \mathfrak t_i} G^{(0)}(\mathfrak t) - \left.\frac{\partial}{\partial \hat x} G^{(0)}(\mathsf t)\right|_{\hat \beta \rightarrow 0} + G^{(1)}(\mathfrak t) + \mathcal O(\beta) \,, \label{eq:free_D2} \\
    & -\log \overline{\mathcal C}(\bar t;-\beta) = \nonumber \\
    & -\frac{1}{\beta} \overline G^{(0)} (\mathfrak t^{-1}) - \frac{N}{2} \log (-\beta) - \sum_i (n_i+\delta_i) \mathfrak t_i^{-1} \frac{\partial}{\partial \mathfrak t_i^{-1}} \overline G^{(0)}(\mathfrak t^{-1}) - \left.\frac{\partial}{\partial \hat x^{-1}} \overline G^{(0)}(\bar{\mathsf t})\right|_{\hat \beta \rightarrow 0} + \overline G^{(1)}(\mathfrak t^{-1}) \nonumber \\
    & + \mathcal O(\beta)
\end{align}
where $\bar{\mathsf t} = (\bar t,-\hat x^{-1} e^{\pi i})$. In addition, we have used a shorthand expression
\begin{align}
    G^{(0,1)}(\mathfrak t) \equiv \left.G^{(0,1)}(\mathsf t)\right|_{\beta \rightarrow 0} = G^{(0,1)}(\mathfrak t,-e^{-\pi i})
\end{align}
where $\sigma$ acts on $-e^{\pi i}$ as $e^{-\pi i r_\sigma}$.
$n$ is nontrivial for the superconformal index while it is set to be zero for the hemisphere index and the squashed sphere partition function.
Combining all those expansions, in the end, we find that
\begin{align}
    &F_{S^2}(\mathfrak t e^{-\beta \delta},n;\beta) = \frac{2}{\beta} i \, \mathrm{Im}\left[G^{(0)} (\mathfrak t)\right] + 2 \sum_i n_i i \, \mathrm{Im} \left[\mathfrak t_i \frac{\partial}{\partial \mathfrak t_i} G^{(0)}(\mathfrak t)\right] \nonumber \\
    &\qquad - 2 \sum_i \delta_i \, \mathrm{Re} \left[\mathfrak t_i \frac{\partial}{\partial \mathfrak t_i} G^{(0)}(\mathfrak t)\right] - 2 \left. \mathrm{Re}\left[\frac{\partial}{\partial \hat x} G^{(0)}(\mathsf t)\right]\right|_{\hat \beta \rightarrow 0} + 2 \mathrm{Re}\left[G^{(1)}(\mathfrak t)\right] + \mathcal O(\beta) \label{eq:free_SCI} \\
    &F_{S^3_b}\left(\Delta-(\beta+\pi i) \delta;\beta\right) = \frac{1}{\beta} G^{(0)} (\mathfrak t) - \sum_i \delta_i \mathfrak t_i \frac{\partial}{\partial \mathfrak t_i} G^{(0)}(\mathfrak t) - \left.\frac{\partial}{\partial \hat x} G^{(0)}(\mathsf t)\right|_{\hat \beta \rightarrow 0} + G^{(1)}(\mathfrak t) + \mathcal O(\beta) \label{eq:free_S3}
\end{align}
up to numerical constants one can ignore. For the squashed sphere partition function, $\mathfrak t = e^{\Delta-\pi i \delta} = e^{-2 \pi b \hat v-\pi i \delta}$, $\hat x = e^{-\hat \beta} = e^{-\pi i b^2}$ is understood. Here the real part and the imaginary part should be understood with the conjugate \eqref{eq:conjugate}.
\\

On the other hand, for the topologically twisted index, we have seen that the Cardy limit is not completely determined by a dominant saddle. Nevertheless, it is worth mentioning the case that there is only one saddle or there is a dominant saddle due to other large parameters. In such situations, the topologically twisted index is written in terms of the Cardy block as
\begin{align}
    F_{S^2}^\text{twisted}(\mathfrak t,n;\beta) = N \log \beta - \log \mathcal C(t;\beta) - \log \mathcal C(\bar t^{-1};-\beta)+\mathcal O(\beta) \,.
\end{align}
The expansion of the second term is given by \eqref{eq:free_D2} with $\delta = 0$ while the third term is expanded as follows:
\begin{align}
    & -\log \mathcal C(\bar t^{-1};-\beta) = \nonumber \\
    & -\frac{1}{\beta} G^{(0)} (\mathfrak t) - \frac{N}{2} \log (-\beta) + \sum_i n_i \mathfrak t_i \frac{\partial}{\partial \mathfrak t_i} G^{(0)}(\mathfrak t) - \left.\frac{\partial}{\partial \hat x^{-1}} G^{(0)}(\tilde {\mathsf t}^{-1})\right|_{\hat \beta \rightarrow 0} + G^{(1)}(\mathfrak t) + \mathcal O(\beta)
\end{align}
where $\tilde{\mathsf t} = (\bar t,-\hat x e^{\pi i})$. Combining the two contributions, therefore, we find
\begin{align}
    F_{S^2}^\text{twisted}(\mathfrak t,n;\beta) &= 2 \sum_i n_i \mathfrak t_i \frac{\partial}{\partial \mathfrak t_i} G^{(0)}(\mathfrak t) - 2 \left.\frac{\partial}{\partial \hat x} G^{(0)}(\mathsf t)\right|_{\hat \beta \rightarrow 0} + 2 G^{(1)}(\mathfrak t) + \mathcal O(\beta), \label{eq:free_twisted}
\end{align}
which is only valid if there is only one saddle or if there is a dominant saddle due to other large parameters such as large $N$, which we discuss shortly. Also note that
\begin{align}
\label{eq:semi-theorem}
    F_{S^2}^\text{twisted}(e^\Delta,n;\beta) = F_{S^3_b}(\Delta+\beta n;\beta) + F_{S^3_b}(\Delta-\beta n;-\beta)+\mathcal O(\beta)
\end{align}
in such cases.
\\

\subsection*{The Large $N$ Limit}

So far we have seen that the Cardy limits discussed in the previous section are determined by two functions $G^{(0,1)}(\mathsf t)$. See \eqref{eq:free_D2}, \eqref{eq:free_SCI}, \eqref{eq:free_S3} and \eqref{eq:free_twisted}. On the other hand, we will see that, in the large $N$ limit, only $G^{(0)}$ plays the crucial role.

Before jumping into the large $N$ limit, let us recall that a large class of 3d supersymmetric gauge theories are known to have their gravity dual on AdS$_4$. In the large $N$ limit, for example, the superconformal and topologically twisted indices are supposed to count the microstates of the corresponding dual black holes. This has been confirmed for a wide class of theories for the topologically twisted index \cite{Benini:2015eyy,Hosseini:2016tor,Hosseini:2016ume,Benini:2016rke,Azzurli:2017kxo,Hosseini:2017fjo,Benini:2017oxt} and is also recently tested for the superconformal index using the ABJM theory and its mirror dual theory \cite{Choi:2019zpz}. Also the squashed sphere partition function relates to the supersymmetric R\'{e}nyi entropy \cite{Nishioka:2013haa}, which accounts for the Bekenstein-Hawking entropy of a charged topological black hole in AdS$_4$ \cite{Huang:2014gca,Nishioka:2014mwa}. Furthermore, although we have not discussed it so far, the round sphere partition function, which is a basic quantity counting the degrees of freedom in odd dimensions, relates to the Euclidean on-shell action in AdS$_4$ \cite{Maldacena:1997re} as well as the holographic entanglement entropy of Ryu-Takayanagi \cite{Ryu:2006bv,Dowker:2010yj,Casini:2011kv}.

As such, many observables of a field theory directly capture the entropic quantities of its gravity dual. More surprisingly, in the large $N$ limit, some of those different looking quantities are proven to be related to each other. For example, there is an index theorem between the topologically twisted index and the round sphere partition function \cite{Hosseini:2016tor}:
\begin{align}
\label{eq:Hosseini}
    F_{S^2}^\text{twisted}(\mathfrak t,n;\beta) = \frac{\pi i}{2} \sum_i \left(n_i-\frac{\Delta_i}{\pi i}\right) \frac{\partial}{\partial \Delta_i} F_{S^3}\left(-\frac{\Delta}{\pi i}\right) + F_{S^3}\left(-\frac{\Delta}{\pi i}\right) \,,
\end{align}
which is written in our notation. This index theorem relates two different entropic quantities of the dual gravity theory: the black hole entropy and the holographic entanglement entropy for a spherical entangling surface.
Here we extend this index theorem to include the superconformal index, which is supposed to capture the entropy of the rotating black hole, and also rederive the above index theorem in our factorization context.
\\

First, we note an interesting relation between the squashed sphere partition function and the round sphere partition function in the large $N$ limit \cite{Martelli:2011fu}:
\begin{align}
\label{eq:round}
   F_{S^3_b}\left(-\pi i b Q \delta;\pi i b^2\right) = \frac{Q^2}{4} F_{S^3} \left(\delta\right) \,,
\end{align}
or equivalently
\begin{align}
\label{eq:relation_S3}
   F_{S^3_b}\left(\Delta;\beta\right) = \frac{(\pi i+\beta)^2}{4 \pi i \beta} F_{S^3} \left(-\frac{\Delta}{\pi i+\beta}\right).
\end{align}
The explicit field theoretic derivation of this relation is discussed in \cite{Martelli:2011fu}, where the relation is derived for non-chiral Chern-Simons quiver gauge theories dual to M-theory on AdS$_4 \times SE_7$. On the other hand, we will see that this relation holds more generally, in particular for theories dual to massive IIA string theory as well. Note that flavor chemical potentials are turned off while the deformations of the $R$-charges are parametrized by $\delta$. Putting this back into the index theorem \eqref{eq:theorem1}, and \eqref{eq:semi-theorem}, we find our second index theorem in the large $N$ limit:
\begin{align}
\begin{aligned}
\label{eq:theorem2}
    &F_{S^2}(\Delta,n;\beta) = \frac{(\pi i+\beta)^2}{4 \pi i \beta} F_{S^3} \left(-\frac{\Delta+\beta n}{\pi i+\beta}\right) + \frac{(\pi i+\beta)^2}{4 \pi i \beta} \overline F_{S^3} \left(-\frac{\Delta-\beta n}{\pi i+\beta}\right)+\mathcal O(\beta) \,, \\
    \\
    &F_{S^2}^\text{twisted}(\Delta,n;\beta) = \frac{(\pi i+\beta)^2}{4 \pi i \beta} F_{S^3} \left(-\frac{\Delta+\beta n}{\pi i+\beta}\right) - \frac{(\pi i-\beta)^2}{4 \pi i \beta} F_{S^3} \left(-\frac{\Delta-\beta n}{\pi i-\beta}\right)+\mathcal O(\beta) \,,
\end{aligned}
\end{align}
which relates two types of black hole entropies with the holographic entanglement entropy for a spherical entangling surface.
Note that, for parity invariant theories, $F_{S^3}$ is a real function. Furthermore although we have derived \eqref{eq:theorem2} in the Cardy limit, many examples suggest that they are exact for arbitrary $\beta$ in the large $N$ limit, at least for the known large $N$ saddle point capturing the black hole microstates. In particular, recall that the Cardy block is given by
\begin{align}
-\log \mathcal C(t;\beta) = \frac{1}{\beta} G^{(0)} (\mathsf t) - \frac{N}{2} \log \beta + G^{(1)}(\mathsf t)+\mathcal O(\beta) \,.
\end{align}
In the next section, we will show that $G^{(1)}$ is subdominant in $N$ compared to $G^{(0)}$ at the large $N$ saddle point for the M2-brane example. In the same manner, we expect that the $\mathcal O(\beta)$ corrections are also all subdominant at the large $N$ saddle, which implies that our formula \eqref{eq:theorem2} is indeed exact for arbitrary $\beta$. We will come back to this point in section \ref{sec:ex1} and \ref{sec:ex2}. 
Also note that recently similar relations are found in terms of gravitational blocks on the dual gravity side \cite{Hosseini:2019iad}, which are also inspired by the holomorphic blocks. It will be interesting to obtain those relations using the equivariant localization in supergravity \cite{Hristov:2018lod,Hristov:2019xku}.

In particular, one can consider the ordinary superconformal index by turning off all the magnetic flux for the flavor symmetry, $n = 0$. In that case, we have
\begin{align}
    F_{S^2}(\Delta,n = 0;\beta) = \frac{(\pi i+\beta)^2}{2 \pi i \beta} \mathrm{Re}\left[F_{S^3} \left(-\frac{\Delta}{\pi i+\beta}\right)\right]
\end{align}
where $\mathrm{Re}[\dots]$ should be understood with the conjugation defined by \eqref{eq:conjugate}. Indeed, one can further unrefine the index by turning off the flavor chemical potentials as well. Restoring the superconformal $R$-charge by setting $\Delta = -(\pi i+\beta) \delta_* \equiv -\Delta_R \delta_*$, we obtain a simple formula
\begin{align}
\label{eq:unrefined}
    F_{S^2}(\beta) = \frac{\Delta_R^2}{2 \pi i \beta} \mathrm{Re} \left[F_{S^3} \left(\delta_*\right)\right]
\end{align}
where $\Delta_R = \pi i+ \beta$ is the chemical potential for the superconformal $R$-symmetry and $\delta_*$ is the shift of $R$-charges restoring the superconformal values, which are determined by the $F$-maximization. This relates the unrefined superconformal index and the round sphere partition function at the superconformal point in a very simple manner, where the latter accounts for the degrees of freedom of 3d SCFTs \cite{Jafferis:2010un}. This is reminiscent of the Cardy formula for 2d CFTs \cite{Cardy:1986ie}\footnote{Recently, it is shown that the 2d Cardy formula can be derived in a rigorous way using Tauberian theorems \cite{Mukhametzhanov:2019pzy,Pal:2019zzr}.} or 4d $\mathcal N = 1$ SCFTs \cite{Kim:2019yrz} in the large $N$ limit, where the real part of the round sphere free energy, $\mathrm{Re} \left[F_{S^3} \left(\delta_*\right)\right]$, plays the role of central charges in 2d and in 4d. Recently the same relation has been observed for 3d SCFTs arising from M5-branes wrapped on hyperbolic three-manifolds \cite{Bobev:2019zmz,Benini:2019dyp}, which show $N^3$ degrees of freedom in the large $N$ limit. Our result shows that this relation holds much more generally. Also we would like to mention that similar universal relations among quantities in different dimensions are discussed in \cite{Benini:2015bwz,Bobev:2017uzs}.
\\

Again one can expand the right hand side with respect to $\beta$. First, comparing \eqref{eq:round} with our formula \eqref{eq:free_S3}, we obtain the following expression for the round sphere:
\begin{align}
    F_{S^3}(\delta) = \frac{4}{\pi i} G^{(0)}(e^{-\pi i \delta}) = \frac{2}{\pi i} \sum_i \delta_i \frac{\partial}{\partial \delta_i} G^{(0)}(e^{-\pi i \delta}) - 2 \left.\frac{\partial}{\partial \hat x} G^{(0)}(\mathsf t)\right|_{t \rightarrow e^{-\pi i \delta}, \hat \beta \rightarrow 0} + 2 G^{(1)}(e^{-\pi i \delta}) \,,
\end{align}
which implies that
\begin{align}
    \frac{2}{\pi i} G^{(0)}(\mathfrak t) = \frac{1}{\pi i} \sum_i (\Delta_i-\pi i \delta_i) \mathfrak t_i \frac{\partial}{\partial \mathfrak t_i} G^{(0)}(\mathfrak t) - \left.\frac{\partial}{\partial \hat x} G^{(0)}(\mathsf t)\right|_{\hat \beta \rightarrow 0} + G^{(1)}(\mathfrak t)
\end{align}
where $\mathfrak t = e^{\Delta-\pi i \delta}$.
Putting this back into the general formula above, we obtain
\begin{align}
    &F_{D^2}(\mathfrak t e^{-\beta \delta};\beta) = \frac{1}{\beta} G^{(0)} (\mathfrak t) - \frac{N}{2} \log \beta - \frac{1}{\pi i} \sum_i \Delta_i \mathfrak t_i \frac{\partial}{\partial \mathfrak t_i} G^{(0)}(\mathfrak t) + \frac{2}{\pi i} G^{(0)}(\mathfrak t) + \mathcal O(\beta) \,, \\
    &F_{S^2}(\mathfrak t e^{-\beta \delta},n;\beta) = \nonumber \\
    &\frac{2}{\beta} i \, \mathrm{Im}\left[G^{(0)} (\mathfrak t)\right] + 2 \sum_i n_i i \, \mathrm{Im} \left[\mathfrak t_i \frac{\partial}{\partial \mathfrak t_i} G^{(0)}(\mathfrak t)\right] - 2 \sum_i \mathrm{Re}\left[\frac{\Delta_i}{\pi i} \mathfrak t_i \frac{\partial}{\partial \mathfrak t_i} G^{(0)}(\mathfrak t)\right] + \frac{4}{\pi} \mathrm{Im}\left[G^{(0)}(\mathfrak t)\right] \nonumber \\
    & + \mathcal O(\beta) \,, \\
    &F_{S^2}^\text{twisted}(\mathfrak t,n;\beta) = 2 \sum_i \left(n_i-\frac{\Delta_i}{\pi i}\right) \mathfrak t_i \frac{\partial}{\partial \mathfrak t_i} G^{(0)}(\mathfrak t) + \frac{4}{\pi i} G^{(0)}(\mathfrak t) + \mathcal O(\beta)
\end{align}
for the three indices and
\begin{align}
     F_{S^3_b}\left(-2 \pi b \hat v-\pi i b Q \delta;\pi i b^2\right) &= \frac{1}{\pi i b^2} G^{(0)} (e^{-2 \pi b \hat v-\pi i \delta}) - \frac{1}{\pi i} \sum_i \hat v_i \frac{\partial}{\partial \hat v_i} G^{(0)}(e^{-2 \pi b \hat v-\pi i \delta}) \nonumber \\
    &\quad + \frac{2}{\pi i} G^{(0)}(e^{-2 \pi b \hat v-\pi i \delta}) + \mathcal O(\beta) \,, \\
    F_{S^3}(\delta) &= \frac{4}{\pi i} G^{(0)}(e^{-\pi i \delta})
\end{align}
for the sphere partition functions. $\mathfrak t$ is defined by $\mathfrak t = e^{\Delta-\pi i \delta}$; $n_i$ vanishes for the hemisphere index; $\delta_i$ vanishes for the topologically twisted index; and finite $b \hat v$ is understood for the squashed sphere partition function.

Especially, if we only look at the leading term in each partition function,
\begin{align}
    &F_{D^2}(\mathfrak t e^{-\beta \delta};\beta) = \frac{1}{\beta} G^{(0)} (\mathfrak t) \,, \\
    &F_{S^2}(\mathfrak t e^{-\beta \delta},n;\beta) = \frac{2}{\beta} i \mathrm{Im}\left[G^{(0)} (\mathfrak t)\right], \\
    &F_{S^2}^\text{twisted}(\mathfrak t,n;\beta) = 2 \sum_i \left(n_i-\frac{\Delta_i}{\pi i}\right) \mathfrak t_i \frac{\partial}{\partial \mathfrak t_i} G^{(0)}(\mathfrak t) + \frac{4}{\pi i} G^{(0)}(\mathfrak t) \,, \\
    &F_{S^3_b}\left(-2 \pi b \hat v-\pi i b Q \delta;\pi i b^2\right) = \frac{1}{\pi i b^2} G^{(0)} (e^{-2 \pi b \hat v-\pi i \delta}) \,, \\
    &F_{S^3}(\delta) = \frac{4}{\pi i} G^{(0)}(e^{-\pi i \delta}) \,,
\end{align}
from which one can read off various relations among those partition functions in the strict Cardy limit. For example, one can write down the superconformal index and the topologically twisted index in terms of the round sphere partition function as follows:
\begin{align}
\begin{aligned}
\label{eq:leading}
    &F_{S^2}(\mathfrak t e^{-\beta \delta},n;\beta) \approx -\frac{\pi}{2 \beta} \mathrm{Re}\left[F_{S^3}\left(-\frac{\Delta}{\pi i}\right)\right], \\
    \\
    &F_{S^2}^\text{twisted}(\mathfrak t,n;\beta) \approx \frac{\pi i}{2} \sum_i \left(n_i-\frac{\Delta_i}{\pi i}\right) \frac{\partial}{\partial \Delta_i} F_{S^3}\left(-\frac{\Delta}{\pi i}\right) + F_{S^3}\left(-\frac{\Delta}{\pi i}\right)
\end{aligned}
\end{align}
where the former provides a new universal formula for the large $N$ superconformal indices of 3d $\mathcal N = 2$ supersymmetric non-chiral Chern-Simons quiver theories dual to M-theory on AdS$_4 \times SE_7$ while the latter reproduces the known index theorem for the unrefined topologically twisted index \cite{Hosseini:2016tor}.

Indeed, the relations in \eqref{eq:leading}, and \eqref{eq:theorem2}, hold more generally, in particular, for theories dual to massive IIA string theory. Recall that the second relation in \cite{Hosseini:2016tor} is derived not only for theories dual to M-theory but also for those dual to massive IIA string theory. Thus, by reversing the logic, combining the index theorem of \cite{Hosseini:2016tor} and our formula \eqref{eq:semi-theorem}, which relates the topologically twisted index and the squashed sphere partition function, one can find that the relation \eqref{eq:relation_S3} should hold for massive IIA duals as well as M-theory duals. Therefore, the resulting relations in \eqref{eq:theorem2} and \eqref{eq:leading} also hold for massive IIA duals. We will encounter such a massive type IIA example in section \ref{sec:ex2}. Furthermore, the Cardy limit formula \eqref{eq:theorem1} without the large $N$ limit is even more general as it does not have any such restrictions.
\\

\section{Example: $\mathcal{N}=4$ $U(N)$ SYM with One Fundamental and One Adjoint Hypermultiplets}
\label{sec:ex1}
In this section, we examine the $\mathcal N = 4$ $U(N)$ SYM theory with one fundamental and one adjoint hypermultiplets as an explicit example. It lives on $N$ D2-branes and 1 D6-brane, and flows in IR to $\mathcal{N}=8$ SCFT on M2-branes. Its holographic dual is the 11d SUGRA, or M-theory, on AdS$_4 \times S^7$. The partition functions of this theory are already discussed in the literature. For example, the Cardy limits of the superconformal and hemisphere indices are discussed in \cite{Choi:2019zpz}, and the (unrefined) topologically twisted index in the large $N$ limit is discussed in \cite{Hosseini:2016ume}. Here, we reexamine those partition functions in the perspective of the Cardy factoriation, which sheds more light on their relations and the universal structure such as \eqref{eq:theorem2} or \eqref{eq:leading}. For simplicity, we will only consider the large $N$ limit. The finite $N$ superconformal index in the Cardy limit can be found in \cite{Choi:2019zpz}.
\\

\subsection{The Hemisphere Index}
We first examine the hemisphere index. Following \cite{Yoshida:2014ssa}, the hemisphere index of the theory is given by
\begin{align} \label{D2}
    &I_{D^2} (\hat{Q}, z, t^{\frac{1}{2}};\beta) = \textrm{tr}_{\mathcal{H}(D^2;\alpha)} \left[(-1)^{\mathsf R} x^{\mathsf  R+2\mathsf  J_3} t^{l-r} z^{2m} \hat{Q}^T \right]  \nonumber \\
    &= \frac{1}{N!} \left(\frac{\theta(z t;x^2)}{\theta(z x^2;x^2)}\right)^{\frac{N^2-N}{2}} \left(\frac{\theta(z t^\frac{1}{2};x^2) \theta(t^\frac{1}{2} e^{-\pi i} x;x^2)}{\theta(z e^{-\pi i} x;x^2)}\right)^{N} \nonumber \\
    &\quad \times \oint \left(\prod_{a = 1}^N \frac{d s_a}{2 \pi i s_a}\right) \left(\prod_{a = 1}^N \frac{\theta(s_a \hat{Q} t^\frac{1}{2} ;x^2)}{\theta(s_a e^{-\pi i} x;x^2) \theta(\hat{Q} t^\frac{1}{2} ;x^2)}\right) \left(\prod_{1 \leq a \neq b \leq N} (s_a s_b^{-1};x^2)\right) \nonumber \\
    &\qquad \times \left(\prod_{a = 1}^N \frac{(s_a t^{-\frac{1}{2}} x^2;x^2)}{(s_a t^\frac{1}{2};x^2)}\right)
    \left(\prod_{a,b = 1}^N \frac{(s_a s_b^{-1} z t^{-\frac{1}{2}} x^2;x^2)}{(s_a s_b^{-1} t^{-1} x^2;x^2) (s_a s_b^{-1} z t^\frac{1}{2};x^2)}\right) \,,
\end{align}
where $x = e^{-\beta}$. $s_a$'s denote the gauge holonomies on $S^1$, which were originally denoted by $z_a$'s in section \ref{sec:fact}. Our theory has $\mathcal{N}=4$ SUSY in UV, which is associated with $SO(4) \cong SU(2)_{l} \times SU(2)_{r}$ $R$-symmetry. Also, the adjoint hypermultiplet can be decomposed into two half-hypermultiplets being a doublet of $SU(2)_m$ flavor symmetry. Finally, there is a topological $U(1)_T$ symmetry coming from the conserved current of the $U(N)$ gauge symmetry $j^\mu = \textrm{tr} (\star F)^\mu$. $l,r,m$ denote the Cartan charges of $SU(2)_l \times SU(2)_r \times SU(2)_m$, and $T$ is the $U(1)_T$ charge. From the $\mathcal{N}=2$ viewpoint, $l+r$ corresponds to the $SO(2) \cong U(1)$ superconformal $R$-charge while $l-r$ is a flavor charge. As explained in section \ref{sec:fact}, we will take the integer quantized $R$-charge to define the index with $(-1)^{\mathsf R}$, $\mathsf R = 2 r$ in this case, instead of the superconformal one. Those two merely differ by a shift of the chemical potentials. Then, the adjoint chiral multiplet in the $\mathcal{N}=4$ vector multiplet has $R$-charge 2, while the other chiral multiplets have vanishing $R$-charges. This can be achieved by successive replacing $t \to t x^{-1}$ and $x \to e^{-\pi i} x$ for the index in \cite{Choi:2019zpz}, which made use of the superconformal $R$-charge and $F= 2 \mathsf J_3$, where $\mathsf  J_3$ is the angular momentum on $D^2$. Further note that $\hat{Q}=e^{-\hat{\xi}}, z=e^f, t^{\frac{1}{2}}=e^{\frac{T}{2}}$ are the fugacities conjugate to the integer quantized flavor charges. Lastly we have introduced 2d degrees of freedom to cancel the boundary mixed/flavor anomalies such that there is no UV mixed/flavor Chern-Simons term except the one between the diagonal gauge $U(1)$ and the topological $U(1)$. In particular, the $\theta$ functions in front of the integral are due to the flavor anomalies while the $\theta$ functions inside the integral are due to the mixed anomalies between the gauge and the flavor symmetries. 

For this particular model, one may avoid 2d degrees of freedom by turning on appropriate mixed/flavor Chern-Simons terms, as done in \cite{Choi:2019zpz}. On the other hand, here we keep 2d degrees of freedom and assume no extra mixed/flavor Chern-Simons terms. In particular, if $Q = e^{-\xi} \equiv \hat{Q} t^\frac{1}{2} e^{\pi i} x^{-1}$ has discrete value $x^{2n}$ with an integer $n$, the $\theta$ functions inside the integral are simplified to
\begin{align}
\left.\prod_{a = 1}^N \frac{\theta(s_a \hat{Q} t^\frac{1}{2} ;x^2)}{\theta(s_a e^{-\pi i} x;x^2) \theta(\hat{Q} t^\frac{1}{2} ;x^2)}\right|_{\hat{Q} t^\frac{1}{2} e^{\pi i} x^{-1} = x^{2n}} = \frac{1}{(e^{-\pi i} x;x^2)^2} \prod_{a=1}^N s_a^{-\frac{\xi}{2\beta}}\ ,
\end{align}
which can be regarded as the classical contribution of the Fayet-Iliopoulos action, whose coupling constant should have a quantized real value on $D^2$, up to a gauge holonomy independent factor. In this regard, the above $\theta$ functions incorporate a generalized FI parameter $\hat \xi = \log \hat Q$, which can take any complex value. This is because, in 3d, the Fayet-Iliopoulos action can be realized as a mixed Chern-Simons action between the gauge $U(1)$ and the topological $U(1)$. For computational convenience, we will temporarily use the parameter $\xi = \hat \xi - \frac{T}{2} -\pi i -\beta$ rather than $\hat \xi$. At the final answer, we will convert it back to $\hat \xi$.

In the Cardy limit $\beta \to 0$, all the other chemical potentials $\xi, f, T$ may be taken as pure imaginary and finite \cite{Choi:2018hmj}. We will restrict the parameter region as
\begin{gather}
\begin{gathered}
\label{para-reg}
-2\pi < \textrm{Im} (\xi)<0 \,, \qquad -2\pi<\textrm{Im} (T)< 0 \,, \\
0 < \textrm{Im} \left(f-\frac{T}{2}\right) < 2\pi \,, \qquad 0 < \textrm{Im} \left(-f-\frac{T}{2}\right) < 2\pi \,.
\end{gathered}
\end{gather}
After gluing two hemisphere indices to make the superconformal index, or the topologically twisted index, the resulting index at the other parameter regions can be easily generated by periodic shifts of the chemical potentials and complex conjugation of the index \cite{Choi:2019zpz}. So, it suffices to consider the above case only to cover the whole parameter region. Then, recall that the Cardy limit $\beta \to 0$ of the hemisphere index \eqref{D2} can be evaluated by the saddle point method as
\begin{equation} \label{asyint}
I_{D^2} = \left( \frac{\beta}{\pi}\right)^{\frac{N}{2}} \exp \left(-\frac{1}{2\beta} \mathcal{W}^* \right) \left( \det ( - \partial_s^2 \mathcal{W})^*\right)^{-\frac{1}{2}} \left(\prod_{a=1}^N \frac{1}{s_a^*} +\mathcal{O}(\beta) \right)\ ,
\end{equation}
with an effective twisted superpotential \cite{Yoshida:2014ssa, Choi:2019zpz}
\begin{equation} \label{eff-action}
\begin{aligned}
&\mathcal{W} = \\
&\frac{N^2-N}{2} \big(T-2\pi i (p_1-p_2) + 2\beta \big) \left(-f-\frac{T}{2}+\pi i(p_1+p_2+1) \right) \\
&+\frac{N}{2}(T +2\pi i +2\beta) \left(-f-\frac{T}{2}\right)\\
&+N\left(-\frac{3}{2}\pi^2+2\pi^2\big((p_3-p_4)(p_3+p_4+2)-p_5 (p_5+2)\big)-2\pi i (p_3-p_5) \xi  -\frac{1}{2}\beta (2\pi i +\beta)\right) \\
&+ N \left( - \textrm{Li}_2(t^{-1}x) + \textrm{Li}_2(z t^{-\frac{1}{2}}x) - \textrm{Li}_2(z t^{\frac{1}{2}}x^{-1}) \right) \\
&+ \sum_{a=1}^N \left( \big(\xi+2\pi i (p_3-p_4)\big) \log s_a + \textrm{Li}_2(s_a t^{-\frac{1}{2}}x) - \textrm{Li}_2(s_a t^{\frac{1}{2}} x^{-1}) \right) \\
&+ \sum_{1 \leq a \neq b \leq N} \left(\textrm{Li}_2(s_a {s_b}^{-1}x^{-1}) - \textrm{Li}_2(s_a {s_b}^{-1} t^{-1}x) + \textrm{Li}_2(s_a {s_b}^{-1} z t^{-\frac{1}{2}}x) - \textrm{Li}_2(s_a {s_b}^{-1} z t^{\frac{1}{2}}x^{-1}) \right) \\
&+ \mathcal{O}(\beta^2)\ ,
\end{aligned}
\end{equation}
where
\begin{gather}
\begin{gathered}
2\pi p_1 < \textrm{Im} \left(f+T \right) < 2\pi (p_1+1) \,, \\
2\pi p_2 < \textrm{Im} \left(f \right) < 2\pi (p_2+1) \,, \\
2\pi p_3 < \textrm{Im} \left(\log s_a - \xi -\pi i\right) < 2\pi (p_3+1) \,, \\
2\pi p_4 < \textrm{Im} \left(\log s_a - \pi i\right) < 2\pi (p_4+1) \,, \\
2\pi p_5 < \textrm{Im} \left(-\xi-\pi i \right)< 2\pi (p_5+1) \,.
\end{gathered}
\end{gather}
Here, we used the asymptotic formulae \eqref{poch}, \eqref{theta}. Also, we assumed that the eigenvalue distribution $s_a$ does not pass across the branch cuts, and $\mathcal{O}(\beta)$ corrections do not change the branches of the arguments. Due to the conditions \eqref{para-reg},
\begin{gather}
\begin{gathered}
(p_1,p_2) = (-2,-1) \,, (-1,-1) \,, (-1,0) \,, (0,0) \,, \\
p_3-p_4 = 0, \, 1 \,, \qquad p_5 = -1, \, 0 \,.
\end{gathered}
\end{gather}

Focusing on the gauge holonomy dependent parts of \eqref{eff-action}, $2\pi i (p_3-p_4)$ effectively shifts the range of $\textrm{Im} (\xi)$ to $\left(2\pi (p_3-p_4-1),2\pi (p_3-p_4)\right)$. However, according to \cite{Choi:2019zpz}, the known Cardy saddle point of the hemisphere index \eqref{D2} only exists when $-2\pi < \textrm{Im} (\xi)<0$. Therefore, we will set $p_3-p_4=0$. This restricts the possible range of the argument of $s_a$'s. The known Cardy saddle point belongs to that range. When $p_3-p_4=1$, there is no known saddle point. Also, as the hemisphere index \eqref{D2} is invariant under $s_a \to e^{2\pi i}s_a$, due to the large gauge invariance of our QFT, we can freely tune the value of $p_4$. We shall set $p_3=p_4=p_5 \equiv p$, which can be either $-1$ or $0$. Then, $\mathcal{W}$ becomes
\begin{equation}
\begin{aligned}
&\mathcal{W} = \\
&\frac{N^2-N}{2} \big(T-2\pi i (p_1-p_2) + 2\beta \big) \left(-f-\frac{T}{2}+\pi i(p_1+p_2+1) \right) \\
&+\frac{N}{2}(T +2\pi i +2\beta) \left(-f-\frac{T}{2}\right)  \\
&+ N \left( -\frac{4p+3}{2}\pi^2  -\frac{1}{2}\beta (2\pi i +\beta) - \textrm{Li}_2(t^{-1}x) + \textrm{Li}_2(z t^{-\frac{1}{2}}x) - \textrm{Li}_2(z t^{\frac{1}{2}}x^{-1}) \right) \\
&+ \sum_{a=1}^N \left( \xi \log s_a + \textrm{Li}_2(s_a t^{-\frac{1}{2}}x) - \textrm{Li}_2(s_a t^{\frac{1}{2}} x^{-1}) \right) \\
&+ \sum_{1 \leq a \neq b \leq N} \left(\textrm{Li}_2(s_a {s_b}^{-1}x^{-1}) - \textrm{Li}_2(s_a {s_b}^{-1} t^{-1}x) + \textrm{Li}_2(s_a {s_b}^{-1} z t^{-\frac{1}{2}}x) - \textrm{Li}_2(s_a {s_b}^{-1} z t^{\frac{1}{2}}x^{-1}) \right) \\
&+ \mathcal{O}(\beta^2)\ .
\end{aligned}
\end{equation}
The first and second lines become $\frac{N^2}{2}(T+2\pi i +2\beta) \left(-f-\frac{T}{2}\right)$ when $(p_1,p_2)=(-1,0)$, which is proportional to $N^2$. This exactly cancels $\mathcal{W}_0$ in \cite{Choi:2019zpz}, which is the term proportional to $N^2$ of $\mathcal{W}^*$ in the large $N$ limit. Namely, by introducing appropriate boundary degrees of freedom, we can get rid of the term proportional to $N^2$ in the large $N$ Cardy free energy, which does not come from the degrees of freedom of $N$ M2-branes. However, one may suspect what happens in the other branches as it will not exactly cancels $\mathcal{W}_0$ in those cases. First note that it does not depend on $s_a$'s, so it does not affect the saddle point. Furthermore, as explained in \cite{Choi:2019zpz}, this term cancels out when we glue two hemisphere indices to make the superconformal index or the topologically twisted index. Therefore, this term does not affect the resulting indices at all so that we can effectively ignore it in the hemisphere index even in the other branches.

To analytically compute $\mathcal{W}^*$, let us consider the large $N$ limit. It was basically studied in \cite{Choi:2019zpz}. Carefully following the section 4.1 of \cite{Choi:2019zpz}, it is not hard to keep $\mathcal{O}(\beta)$ terms in $\mathcal{W}$. We obtain
\begin{align} \label{W}
-\frac{\mathcal{W}^*}{2\beta}=&- i \, \frac{\sqrt{2} N^{\frac{3}{2}}}{3\beta} \sqrt{\left(-\hat{\xi}+\frac{T}{2} + \pi i+\beta \right)\left(\hat{\xi}+\frac{T}{2} + \pi i +\beta\right)\left(f-\frac{T}{2} \right) \left(-f-\frac{T}{2}  \right)} \nonumber \\
&+ \frac{o(N^{\frac{3}{2}})}{\beta}+o(N^{\frac{3}{2}}) \beta^0 + \mathcal{O}(\beta)\ ,
\end{align}
where $\hat{\xi}=\xi+\frac{T}{2}+\pi i+\beta$.\footnote{Indeed, the subleading correction in $N$ turns out to be not $\mathcal{O} (N)$ but $\mathcal{O} (N^{\frac{1}{2}})$, according to the numerical analysis \cite{Choi:2019zpz}. So, one may replace $o(N^{\frac{3}{2}}) \to \mathcal{O}(N^{\frac{1}{2}})$ in \eqref{W}. Nevertheless, we shall keep using $o(N^{\frac{3}{2}})$ as we do not have its analytic proof yet.} This large $N$ Cardy saddle point value exists only when we further restrict the parameters as
\begin{equation}
0<\textrm{Im} \left(\hat{\xi}+\frac{T}{2}+\pi i + \beta \right)<2\pi\ , \quad \textrm{Im} (\beta)<0\ .
\end{equation}
Note that we shifted $T \to T+\beta$ and removed $\mathcal{W}_0$ in \cite{Choi:2019zpz} as explained before.

According to \eqref{asyint}, the following also contributes to the hemisphere index at $\mathcal{O}(\beta^0)$:
\begin{equation}
\left(\frac{\beta}{\pi}\right)^{\frac{N}{2}} \left(\det ( - \partial_s^2 \mathcal{W})^*\right)^{-\frac{1}{2}} \prod_{a=1}^N \frac{1}{s_a^*}\ .
\end{equation}
Also, from \cite{Choi:2019zpz}, one can find that the Cardy saddle point satisfies $\textrm{Im} (s_a)=0, \, \textrm{Re} (s_a)>0$ for our parameter region. Then, we may sort the eigenvalues in the ascending order as follows:
\begin{equation}
0<s_1<s_2<\cdots < s_N\ .
\end{equation}
Assuming $\textrm{Im} (\beta)<0$ as before, which is the relevant region for our microstate counting, the Hessian of $\mathcal{W}$ at $\beta \to 0$ is given by
\begin{equation}
\begin{aligned}
& \frac{\partial^2 \mathcal{W}}{\partial s_a \partial s_b} = \frac{1}{s_a s_b} \left[ 1 + \sum_\mu s_\mu  \left(\frac{1}{1- e^{-\mu} s_a s_b^{-1}} + \frac{1}{1- e^{-\mu} s_a^{-1} s_b} \right) \right]\ , \qquad (a \neq b)\ , \\
& \frac{\partial^2 \mathcal{W}}{\partial s_a^2} = \\
& \frac{1}{s_a^2} \left[ -(N-1) - (N-2a+1) \pi i - \xi + \textrm{Li}_1 (t^{1/2} s_a) - \textrm{Li}_1 (t^{-1/2} s_a) + \frac{1}{1-t^{-1/2}s_a^{-1}} - \frac{1}{1-t^{1/2}s_a^{-1}} \right. \\
& \left. + \sum_{\substack{b=1 \\ b \neq a}}^N \left\{  \log (s_a s_b^{-1}) + \sum_\mu s_\mu \left( - \textrm{Li}_1 (e^\mu s_a s_b^{-1}) + \textrm{Li}_1 (e^\mu s_a^{-1} s_b) - \frac{1}{1-e^{-\mu} s_a s_b^{-1}} - \frac{1}{1-e^{-\mu} s_a^{-1} s_b} \right) \right\} \right]\ ,
\end{aligned}
\end{equation}
where $\mu = -T, \, f - \frac{T}{2}, \, f + \frac{T}{2}$ and $s_{-T}=-1, \, s_{f-\frac{T}{2}}=+1, \, s_{f+\frac{T}{2}}=-1$. At the large $N$ Cardy saddle point \cite{Choi:2019zpz}
\begin{equation}
s_a^*=s_0 e^{N^{1/2}x_a}\ ,\quad s_0=\frac{\sinh(\xi/2)}{\sinh(\xi/2+T/2)}\ ,
\end{equation}
the above Hessian becomes
\begin{equation}
\begin{aligned}
\frac{\partial^2 \mathcal{W}}{\partial s_a \partial s_b} = & - \frac{1}{s_a^* s_b^*} \left[ \left( \sum_\mu 2 s_\mu \sinh \mu \right) e^{-N^{1/2}|x_a-x_b|} + \mathcal{O}\left(\left(e^{-N^{1/2}}\right)^2\right) \right]\ , \quad (a \neq b)\ , \\
\frac{\partial^2 \mathcal{W}}{\partial s_a^2} 
= & - \frac{1}{{s_a^*}^2} \left[ \xi + \theta (x_a) (T+2\pi i) - \delta_{x_a, 0} \left(\xi + \frac{1}{1-t^{-1/2}s_0^{-1}} - \frac{1}{1-t^{1/2}s_0^{-1}} \right) + \mathcal{O}\left(e^{-N^{1/2}}\right) \right]\ ,
\end{aligned}
\end{equation}
where $\theta(x>0)=1, \, \theta(x \leq 0)=0$, and $\delta_{x_a,0}$ is the Kronecker delta. Then, the Hessian determinant is given by
\begin{equation}
\det ( - \partial_s^2 \mathcal{W})^* = \prod_{a=1}^N \frac{1}{{s_a^*}^2} \; \Delta^N \left(1 + \mathcal{O}\left(e^{-N^{1/2}}\right) \beta^0 + \mathcal{O}(\beta)\right)\ ,
\end{equation}
where $\Delta^N \equiv \prod_{a=1}^N \left[\xi + \theta (x_a) (T+2\pi i) - \delta_{x_a, 0} \left(\xi + \frac{1}{1-t^{-1/2}s_0^{-1}} - \frac{1}{1-t^{1/2}s_0^{-1}} \right)\right]$, i.e. $\Delta$ is an $\mathcal{O}(1)$ constant. Finally, we obtain
\begin{equation}
\log \left[\left(\frac{\beta}{\pi}\right)^{\frac{N}{2}}\left(\det (-\mathcal{W}''_{ss})\right)^{-\frac{1}{2}} \prod_{a=1}^N \frac{1}{s_a^*} \right] = \frac{N}{2} \log \left(\frac{\beta}{\pi \Delta}\right) + \mathcal{O}\left(e^{-N^{1/2}}\right) \beta^0 + \mathcal{O}(\beta)\ ,
\end{equation}
which contributes as $\mathcal{O}(N)$ to the large $N$ Cardy free energy. Therefore, together with \eqref{W}, the large $N$ Cardy free energy of the hemisphere index is given by
\begin{equation} \label{D2-Cardy-N}
\begin{aligned}
& \log I_{D^2} (e^{-\hat{\xi}}, e^{f}, e^{\frac{T}{2}};\beta) \\
&= \log \mathcal C(e^{-\hat{\xi}}, e^{f}, e^{\frac{T}{2}};\beta) \\
&= - i \, \frac{\sqrt{2} N^{\frac{3}{2}}}{3\beta} \sqrt{\left(-\hat{\xi}+\frac{T}{2} + \pi i +\beta \right)\left(\hat{\xi}+\frac{T}{2} + \pi i +\beta\right)\left(f-\frac{T}{2} \right) \left(-f-\frac{T}{2}  \right)} \\
&\quad + \frac{N}{2} \log \left(\frac{\beta}{\pi \Delta}\right) + \frac{o(N^{\frac{3}{2}})}{\beta}+o(N^{\frac{3}{2}}) \beta^0 + \mathcal{O}(\beta) \ ,
\end{aligned}
\end{equation}
when
\begin{equation}
0< \textrm{Im} \left(-\hat{\xi}+\frac{T}{2} + \pi i+\beta\right), \; \textrm{Im} \left(\hat{\xi}+\frac{T}{2}+\pi i +\beta \right), \; \textrm{Im} \left(f-\frac{T}{2}\right), \;  \textrm{Im} \left(-f-\frac{T}{2} \right) < 2\pi\ ,
\end{equation}
with
\begin{equation}
\textrm{Im} (\beta) <0\ .
\end{equation}
Note that the $\log \beta$ term is correct even at finite $N$ as explained in section \ref{sec:fact}. One can also restore the superconformal $R$-charge by shifting $T \rightarrow T-\pi i-\beta$.
\\


\subsection{The Generalized Superconformal Index}
In this subsection, we construct the large $N$ Cardy limit of the generalized superconformal index from \eqref{D2-Cardy-N}. This index should statistically account for the microstates of the electrically charged \cite{Cvetic:2005zi} or dyonic \cite{Hristov:2019mqp} rotating BPS black holes with vanishing magnetic charge for the $R$-symmetry and large angular momentum $J/N^{\frac{3}{2}} = \mathcal{O}(\beta^{-2})$ in AdS$_4 \times S^7$ \cite{Choi:2019zpz}. In order to preserve SUSY, the black holes should have both electric charges and angular momentum while it is free to turn off the magnetic charges for the (non-$R$) flavor symmetries \cite{Cvetic:2005zi, Choi:2018fdc, Hristov:2019mqp}.

Recall that, in the Cardy limit, the generalized superconformal index is given in terms of the Cardy block \eqref{eq:SCI} as follows:
\begin{align}
    F_{S^2}(\mathfrak t,n;\beta) =& N \log \beta - \log \mathcal C(e^{-\hat{\xi}+\beta n_\xi}, e^{f+\beta n_f}, e^{\frac{T}{2} + \beta n_T};\beta) \nonumber \\
    &- \log \overline{\mathcal C}(e^{\hat{\xi}+\beta n_\xi}, e^{-f+\beta n_f}, e^{-\frac{T}{2} + \beta n_T};-\beta) - \log\left[N! \, \pi^N\right] +\mathcal O(\beta)
\end{align}
Applying \eqref{D2-Cardy-N} to the above factorization formula, we obtain
\begin{equation} \label{SCI}
\begin{aligned}
\log I_{S^2} (\Delta_I, n_I ; \beta) =& - i \frac{2\sqrt{2} N^{\frac{3}{2}}}{3} \frac{\sqrt{\left(\Delta_1+n_1 \beta \right)\left(\Delta_2 +n_2 \beta\right)\left(\Delta_3 +n_3 \beta \right) \left(\Delta_4 +n_4 \beta \right)}}{2\beta} \\
& - i \frac{2\sqrt{2} N^{\frac{3}{2}}}{3} \frac{\sqrt{\left(\Delta_1-n_1 \beta \right)\left(\Delta_2 -n_2 \beta\right)\left(\Delta_3 -n_3 \beta \right) \left(\Delta_4 -n_4 \beta \right)}}{2\beta} +o(N^{\frac{3}{2}}) \\
& +\mathcal{O} (\beta)\ ,
\end{aligned}
\end{equation}
where
\begin{equation}
\begin{gathered}
\Delta_1 \equiv - \hat{\xi}+ \frac{T}{2} + \pi i + \beta\ , \; \Delta_2 \equiv \hat{\xi}+ \frac{T}{2} + \pi i +\beta\ ,  \;\Delta_3 \equiv f- \frac{T}{2}\ , \; \Delta_4 \equiv -f- \frac{T}{2}\ , \\
n_1 \equiv n_\xi+n_T\ , \; n_2 \equiv -n_\xi+n_T\ , \; n_3 \equiv n_f-n_T\ , \; n_4 \equiv -n_f -n_T\ ,
\end{gathered}
\end{equation}
which satisfy
\begin{equation} \label{SCI-const}
\sum_{I=1}^4 \Delta_I - 2\beta = 2\pi i\ ,\quad \sum_{I=1}^4 n_I =0\ , \quad 0<\textrm{Im} (\Delta_I) <2\pi\ , \quad \textrm{Im} (\beta) <0\ .
\end{equation}
Here, $\Delta_I$'s are the four Cartan chemical potentials for $SO(8)$ $R$-symmetry of the $\mathcal{N}=8$ SCFT in IR, and $n_I$'s denote the magnetic fluxes for the Cartan subgroup. Then, the first constraint of \eqref{SCI-const} can be considered as the index-like condition implied on a partition function defined without $(-1)^F$ \cite{Choi:2018hmj, Choi:2019zpz}. Also, the second constraint reveals that we do not turn on magnetic flux on $S^2$ for the $R$-symmetry; i.e., there is no topological twist.

The above large $N$ Cardy free energy \eqref{SCI} statistically accounts for the microstates of dyonic rotating BPS black holes in AdS$_4 \times S^7$ \cite{Hristov:2019mqp, Hosseini:2019iad}. In particular, turning off all the magnetic fluxes for the flavor symmetries, we get
\begin{equation} \label{M2-large}
\log I_{S^2} (\Delta_I ; \beta) =  - i \frac{4\sqrt{2} N^{\frac{3}{2}}}{3} \frac{\sqrt{\Delta_1 \Delta_2 \Delta_3 \Delta_4}}{2\beta} +o(N^{\frac{3}{2}}) +\mathcal{O} (\beta)\ .
\end{equation}
This large $N$ Cardy free energy of the superconformal index accounts for the microstates of electrically charged rotating BPS black holes in AdS$_4 \times S^7$ \cite{Choi:2018fdc}. Namely, performing the Legendre transformation of the above free energies with respect to $\Delta_I$'s and $2\beta$ under the constraints \eqref{SCI-const}, one obtains the Bekenstein-Hawking entropy of the corresponding BPS black holes in AdS$_4 \times S^7$. Further note that although we derive these large $N$ free energies in the Cardy limit $\beta \to 0$, they in fact perfectly capture the entropy of the BPS black holes even at finite $\beta$.
\\

\subsection{The Refined Topologically Twisted Index}
In this subsection, we construct the large $N$ Cardy limit of the refined topologically twisted index from \eqref{D2-Cardy-N}. This index should statistically account for the microstates of the magnetically charged or dyonic, static \cite{Cacciatori:2009iz,Katmadas:2014faa,Halmagyi:2014qza} or rotating \cite{Hristov:2018spe} BPS black holes with the non-vanishing magnetic flux for the gauged $R$-symmetry in mAdS$_4 \times S^7$ \cite{Benini:2015eyy}, which is a particular asymptotically locally AdS spacetime \cite{Hristov:2011ye}. For these black holes, one can freely turn off the electric charges, angular momentum, and magnetic charges for the (non-$R$) flavor symmetries. The magnetic charge for the $R$-symmetry should be properly tuned to preserve SUSY by a topological twist.

Recall that, in the Cardy limit, the refined topologically twisted index is given in terms of the Cardy block \eqref{eq:twisted} as follows:
\begin{align}
    F_{S^2}^\text{twisted}(\mathfrak t,n;\beta) = & N \log \beta - \log \mathcal C(e^{-\hat{\xi}+\beta n_\xi}, e^{f+\beta n_f}, e^{\frac{T}{2} + \beta n_T};\beta) \nonumber \\
    & - \log \mathcal C(e^{-\hat{\xi}-\beta n_\xi}, e^{f-\beta n_f}, e^{\frac{T}{2} - \beta n_T};-\beta) - \log\left[N! \, \pi^N\right] + \mathcal O(\beta) \,.
\end{align}
Applying \eqref{D2-Cardy-N} to the above factorization formula, we obtain
\begin{equation} \label{TTI}
\begin{aligned}
\log I_{S^2}^{\textrm{twisted}} (\Delta_I, n_I ; \beta) =   & - i \frac{2\sqrt{2} N^{\frac{3}{2}}}{3} \frac{\sqrt{\left(\Delta_1+n_1 \beta \right)\left(\Delta_2 +n_2 \beta\right)\left(\Delta_3 +n_3 \beta \right) \left(\Delta_4 +n_4 \beta \right)}}{2\beta} \\
& + i \frac{2\sqrt{2} N^{\frac{3}{2}}}{3} \frac{\sqrt{\left(\Delta_1-n_1 \beta \right)\left(\Delta_2 -n_2 \beta\right)\left(\Delta_3 -n_3 \beta \right) \left(\Delta_4 -n_4 \beta \right)}}{2\beta} \\
& +o(N^{\frac{3}{2}}) +\mathcal{O} (\beta)\ ,
\end{aligned}
\end{equation}
where
\begin{equation}
\begin{aligned}
&\Delta_1 \equiv - \hat{\xi}+ \frac{T}{2} + \pi i\ , \; \Delta_2 \equiv \hat{\xi}+ \frac{T}{2} + \pi i\ ,  \;\Delta_3 \equiv f- \frac{T}{2}\ , \; \Delta_4 \equiv -f- \frac{T}{2}\ , \\
&n_1 \equiv 1+n_\xi+n_T\ , \; n_2 \equiv 1-n_\xi+n_T\ , \; n_3 \equiv n_f-n_T\ , \; n_4 \equiv -n_f -n_T\ ,
\end{aligned}
\end{equation}
which satisfy
\begin{equation} \label{TTI-const}
\sum_{I=1}^4 \Delta_I= 2\pi i\ ,\quad \sum_{I=1}^4 n_I =2\ , \quad 0<\textrm{Im} (\Delta_I) <2\pi\ , \quad \textrm{Im} (\beta) <0\ .
\end{equation}
Here, $\Delta_I$'s are again the four Cartan chemical potentials for $SO(8)$ $R$-symmetry of the $\mathcal{N}=8$ SCFT in IR, and $n_I$'s denote the magnetic fluxes for each Cartan subgroup. The first constraint of \eqref{TTI-const} comes from the index-like condition for the partition function without $(-1)^F$ while the second constraint implies the topological twist by the magnetic flux for the $R$-symmetry on $S^2$.

The above large $N$ Cardy free energy \eqref{TTI} statistically accounts for the microstates of magnetically charged or dyonic rotating BPS black holes in mAdS$_4 \times S^7$ \cite{Hristov:2018spe, Hosseini:2019iad}. As before, this large $N$ free energy gives the correct entropy of the BPS black holes at arbitrary $\beta$. Also, expanding the free energy in $\beta$, one can find that it is regular at $\beta=0$. Then, setting $\beta=0$, we get
\begin{equation}
\log I_{S^2}^{\textrm{twisted}} (\Delta_I, n_I) =  - i \frac{\sqrt{2} N^{\frac{3}{2}}}{3} \sqrt{\Delta_1 \Delta_2 \Delta_3 \Delta_4} \sum_{I=1}^4 \frac{n_I}{\Delta_I}+o(N^{\frac{3}{2}})\ ,
\end{equation}
whose Legendre transformation in $\Delta_I$'s yields the entropy of the magnetically charge or dyonic static BPS black holes in mAdS$_4 \times S^7$ \cite{Benini:2015eyy}.
\\

\subsection{The Squashed Sphere Partition Function}
In this subsection, we construct the large $N$ Cardy limit of the squashed sphere partition function from \eqref{D2-Cardy-N}. This partition function should be related to the round sphere partition function by \eqref{eq:round} \cite{Martelli:2011fu}. Then, the free energy of the round sphere partition function \cite{Herzog:2010hf} is supposed to be identified with the regularized Euclidean on-shell action on AdS$_4 \times S^7$ \cite{Drukker:2010nc, Emparan:1999pm}.

Recall that, in the Cardy limit, the squashed sphere partition function is given in terms of the Cardy block \eqref{eq:sphere} as follows:
\begin{align}
    F_{S^3_b}(-\pi i b Q \delta;\pi i b^2) = \frac{N}{2} \log \left[-b^2\right] - \log \mathcal C(e^{-\pi i bQ \delta_\xi}, e^{-\pi i b Q \delta_f}, e^{-\pi i b Q \delta_T};\pi i b^2) + \mathcal O(b^2)
\end{align}
where $Q=b+\frac{1}{b}$; $\delta_\mu$'s are the $R$-charge deformations by $U(1)_\mu$ flavor charges; and we turned off all the mass parameters on $S_b^3$ for simplicity. Applying \eqref{D2-Cardy-N} to the above formula, we obtain
\begin{equation} \label{Sb3}
\log Z_{S_b^3}(\Delta_I;\pi i b^2)  =- \frac{1}{4} \left(b+\frac{1}{b}\right)^2 \frac{4\sqrt{2} \pi N^{\frac{3}{2}}}{3}\sqrt{\Delta_1 \Delta_2 \Delta_3 \Delta_4}  +o(N^{\frac{3}{2}}) + \mathcal{O} (b^2)\ ,
\end{equation}
where
\begin{equation}
\Delta_1 \equiv 1-\delta_\xi - \delta_T\ , \; \Delta_2 \equiv 1+\delta_\xi-\delta_T\ ,  \;\Delta_3 \equiv -\delta_f+ \delta_T\ , \; \Delta_4 \equiv \delta_f+\delta_T\ ,
\end{equation}
which satisfy
\begin{equation}
\sum_{I=1}^4 \Delta_I =2\ .
\end{equation}
Here, $\Delta_I$'s parametrize trial $R$-charges of the theory, which are constrained as above due to the condition that the superpotential should have $R$-charge 2. As before, this large $N$ free energy \eqref{Sb3} is indeed exact at arbitrary $b$.

Using the relation between the squashed and round sphere partition function \eqref{eq:round}, we get the following large $N$ limit of the round sphere partition function with arbitrary $R$-charges:
\begin{equation}
\log Z_{S^3} (\Delta_I) = 4 \left(b+\frac{1}{b}\right)^{-2} \log Z_{S_b^3} (\Delta_I;\pi i b^2) = - \frac{4\sqrt{2} \pi N^{\frac{3}{2}}}{3}\sqrt{\Delta_1 \Delta_2 \Delta_3 \Delta_4}  +o(N^{\frac{3}{2}})\ ,
\end{equation}
which exactly agrees with the known field theory result \cite{Jafferis:2011zi} and the confirmation on the gravity side \cite{Freedman:2013ryh}. In particular, setting $\delta_T = \frac{1}{2}$ and $\delta_\xi=\delta_f=0$, one can restore the superconformal $R$-charge. In that case, we find $\Delta_1=\Delta_2=\Delta_3=\Delta_4=\frac{1}{2}$, which indeed maximize $F=-\log Z_{S^3} (\Delta_I)$ \cite{Jafferis:2010un}. Then, the large $N$ limit of the round sphere partition function at the superconformal $R$-charge is given by
\begin{equation}
\log Z_{S^3} = - \frac{\sqrt{2} \pi N^{\frac{3}{2}}}{3}  +o(N^{\frac{3}{2}})\ ,
\end{equation}
which precisely matches the regularized Euclidean on-shell action on AdS$_4 \times S^7$ \cite{Herzog:2010hf}. In addition, one can easily check that our second index theorem \eqref{eq:theorem2} indeed holds for the results obtained in this section.
\\

\section{Other Examples}
\label{sec:ex2}

In this section, we study the Cardy limit of the superconformal index of various $\mathcal{N} \geq 2$ SCFTs applying \eqref{eq:fusion} or \eqref{eq:theorem2}. For the definiteness, we shall only consider the SCFTs which can be obtained from M-theory or string theory.

Before moving on to explicit examples, we first make a comment on the Hessian determinant of the hemisphere index in the large $N$ limit. While computing the Hessian determinant in section \ref{sec:ex1}, the crucial point was that the eigenvalues spread as $s_a \sim e^{N^{1/2} x_a}$. This eigenvalue spreading is a common feature of 3d SCFTs while the precise factor $N^\alpha$ depends on a specific model. For SCFTs with M-theory dual, $\alpha=\frac{1}{2}$; for SCFTs with massive IIA dual, $\alpha=\frac{1}{3}$. As one can see in section \ref{sec:ex1}, the precise factor is not important in the computation. Thus, we expect that the log of the Hessian determinant is $\mathcal{O}(N)$ in general. Namely, for generic $\mathcal{N}=2$ SCFTs, we expect that
\begin{equation}
G^{(1)} = \mathcal{O} (N) + \mathcal{O} (\beta)\ .
\end{equation}
It is well known that $G^{(0)} = \mathcal{O} (N^{\frac{3}{2}})$ for the M-theory dual while $G^{(0)} = \mathcal{O} (N^{\frac{5}{3}})$ for massive IIA dual. Therefore, we expect that $G^{(1)}$ is negligible in the large $N$ limit.

There are lots of examples whose round sphere partition functions \cite{Drukker:2010nc, Herzog:2010hf, Jafferis:2011zi, Fluder:2015eoa} or topologically twisted indices on $S^2 \times S^1$ \cite{Hosseini:2016ume,Hosseini:2017fjo,Benini:2017oxt,PandoZayas:2019hdb} are known in the large $N$ limit. Using \eqref{eq:theorem2}, one can easily read off the large $N$ Cardy limit of the superconformal index from the round sphere partition function. Also, in the literature, the topologically twisted index was computed via its index theorem \eqref{eq:Hosseini}, which was expressed in terms of the Bethe potential $\overline{\mathcal{V}}$. In our notation, the Bethe potential is translated as
\begin{equation}
\overline{\mathcal{V}} (\Delta) = \frac{2}{i} G^{(0)}(e^{-i\Delta})\ .
\end{equation}
Since $G^{(1)}$ can be ignored in the large $N$ limit, reading off $\overline{\mathcal{V}}$ from the topologically twisted index suffices to compute the superconformal index in the large $N$ Cardy limit by \eqref{eq:fusion}. We will illustrate such examples whose large $N$ Cardy limit of the superconformal indices can be obtained either from the round sphere partition functions or the topologically twisted indices.

Furthermore, we will also provide finite $N$ Cardy results of the superconformal indices for a few examples, with rank less than 3 for simplicity, which are severed as nontrivial tests of known dualities of those examples.
\\

\subsection{M2-Branes Probing a CY 4-Fold Singularity}
In this subsection, we consider quiver gauge theories, which describe the low energy dynamics of M2-branes probing a conical Calabi-Yau 4-fold singularity. For this class of theories, the sum of the CS levels for each gauge group vanishes, i.e. $\sum_{g} k_g=0$. Also, those theories are parity invariant so that the round sphere free energy $F_{S^3}$ is real. Then, from \eqref{eq:theorem2}, the generalized superconformal index can be expressed as
\begin{equation}
F_{S^2} (\Delta, n ; \beta ) = \frac{F_{S^2} (\Delta+n\beta,0; \beta )+F_{S^2} (\Delta- n\beta,0; \beta )}{2}\ .
\end{equation}
Keeping the above formula in mind, we shall turn off all the magnetic fluxes for the flavor symmetries. The generalization can be easily done by the above formula. There are two relevant regimes for these theories: M-theory regime, and type IIA string theory regime.

\subsubsection{M-Theory Regime}
One can take the large $N$ limit with fixed CS levels $k \sim \mathcal{O}(1)$. Then, the field theory is supposed to be dual to the M-theory or 11d SUGRA on AdS$_4 \times \textrm{SE}_7$ where SE$_7$ is the Sasaki-Einstein 7-manifold serving as the base of a conical CY 4-fold. The characteristic large $N$ behavior of the free energy in this M-theory regime is
\begin{equation}
    F \sim k^{\frac{1}{2}} N^{\frac{3}{2}}\ .
\end{equation}
First, the round sphere free energy at the superconformal $R$-charge is given by \cite{Herzog:2010hf}
\begin{equation}
F_{S^3} = N^{\frac{3}{2}} \sqrt{\frac{2\pi^6}{27 \textrm{Vol} (\textrm{SE}_7)}}\ .
\end{equation}
Applying \eqref{eq:unrefined}, the (unrefined) superconformal index in the large $N$ Cardy limit, for the generic $\mathcal{N} \geq 2$ SCFTs describing the low energy dynamics of M2-branes, is given by
\begin{equation}
F_{S^2}  = \frac{\Delta_R^2}{2\pi i \beta} N^{\frac{3}{2}} \sqrt{\frac{2\pi^6}{27 \textrm{Vol} (\textrm{SE}_7)}}\ ,
\end{equation}
which precisely matches the result from the dual supergravity analysis on universal spinning black holes in AdS$_4$ \cite{Bobev:2019zmz}, even in the non-Cardy regime.

\paragraph{Example 1: ABJM theory}~\\
The most common example in this class is the ABJM$_k$ theory, which describes the low energy dynamics of $N$ M2-branes probing $\mathbb{C}_4/\mathbb{Z}_k$ singularity \cite{Aharony:2008ug}. Its holographic dual is given by 11d SUGRA on AdS$_4 \times S^7/\mathbb{Z}_k$. The round sphere free energy with generic $R$-charge assignment is given by \cite{Jafferis:2011zi}
\begin{equation}
F_{S^3} =  \frac{4\sqrt{2} \pi k^{\frac{1}{2}} N^{\frac{3}{2}}}{3}\sqrt{\Delta_1 \Delta_2 \Delta_3 \Delta_4}\ .
\end{equation}
The large $N$ Cardy free energy of the superconformal index is given by \cite{Choi:2019zpz}
\begin{equation} \label{ABJM-large}
F_{S^2} =   i \frac{4\sqrt{2} k^{\frac{1}{2}} N^{\frac{3}{2}}}{3} \frac{\sqrt{\Delta_1 \Delta_2 \Delta_3 \Delta_4}}{2\beta}\ .
\end{equation}
Indeed, they satisfy our large $N$ Cardy formula \eqref{eq:theorem2}:
\begin{equation}
F_{S^2}(\Delta;\beta) = \frac{(\pi i+\beta)^2}{2 \pi i \beta} F_{S^3} \left(-\frac{\Delta}{\pi i+\beta}\right)\ .
\end{equation}
Note that the above large $N$ Cardy free energy of the superconformal index, in fact, precisely captures the entropy of dual BPS black holes in AdS$_4 \times S^7/\mathbb{Z}_k$ \cite{Cvetic:2005zi, Hristov:2019mqp, Choi:2018fdc}, even in the non-Cardy regime.

\subparagraph{$SL(2,\mathbb{Z})$ duality}
When $k=1$, the ABJM theory is supposed to be dual to $\mathcal{N}=4$ $U(N)$ SYM with one fundamental and one adjoint hypermultiplets in section \ref{sec:ex1} by the $SL(2,\mathbb{Z})$ duality \cite{Aharony:2008ug}. Indeed, in the large $N$ Cardy limit, two free energies \eqref{M2-large}, \eqref{ABJM-large} coincide. One can also test this duality in the finite $N$ Cardy limit using \eqref{eq:fusion}. When $N=1$, the Cardy free energy of the ABJM$_1$ theory is given by
\begin{equation}
\label{eq:ABJM}
F_{S^2} \sim \frac{8iG}{2\beta} \approx \frac{7.33i}{2\beta}\ ,
\end{equation}
at $\Delta_1=\Delta_2=\Delta_3=\Delta_4=\frac{\pi i}{2}$. $G$ is Catalan's constant, which is defined by
\begin{align}
G = \frac{\mathrm{Li}_2(i)-\mathrm{Li}_2(-i)}{2 i} \approx 0.915966 \,.
\end{align}
It turns out that \eqref{eq:ABJM} is identical to the Cardy free energy of the dual theory when $N=1$ \cite{Choi:2019zpz}. One would test this duality at $N>1$ by numerical analysis.
\\

Now we discuss a few more examples belonging to this class. The topologically twisted indices of those examples are examined in \cite{Hosseini:2016ume}. We examine their superconformal indices in the Cardy limit, mainly with large $N$ but also with finite $N$ for a few theories with rank less than 3.

\paragraph{Example 2: $\mathcal{N}=4$ $U(N)$ SYM with $N_f$ fundamental and one adjoint hypermultiplets} ~\\
This theory is a natural generalization of the theory we have considered in section \ref{sec:ex1}. This theory describes $N$ M2-branes probing $\mathbb{C}^2 \times \mathbb{C}^2/\mathbb{Z}_{N_f}$ singularity. Referring to the topologically twisted index in \cite{Hosseini:2016ume}, we obtain the following large $N$ Cardy free energy of the superconformal index from \eqref{eq:fusion}:
\begin{equation} \label{ex1-large}
F_{S^2}=  i \frac{4\sqrt{2} N_f^{\frac{1}{2}} N^{\frac{3}{2}}}{3} \frac{\sqrt{\Delta_1 \Delta_2 \Delta_3 \Delta_4}}{2\beta}\ .
\end{equation}
Also, at $N_f=2$, we give the Cardy free energy at finite $N$. When $N=1$,
\begin{equation} \label{ex1-1}
\begin{aligned}
F_{S^2} & \sim \frac{i}{2\beta} \left[4G+8\textrm{Im} \left\{ \textrm{Li}_2 ((-1+\sqrt{2})i)\right\} - \pi \log (-1+\sqrt{2})\right] \\
& \approx	\frac{9.68691i}{2\beta}\ ,
\end{aligned}
\end{equation}
and when $N=2$,
\begin{equation}
F_{S^2} \approx	\frac{22.6365i}{2\beta}\ ,
\end{equation}
at $\Delta_1=\Delta_2=\Delta_3=\Delta_4=\frac{\pi i}{2}$. Note that the above free energies comes from the dominant one among several saddle points.

\paragraph{Example 3: $\mathcal{N}=4$ $U(N)^{g}$ necklace quiver SYM with $N_f$ fundamental hypermultiplets for the $g$-th gauge group} ~\\
This theory contains $g > 1$ bifundamental hypermultiplets, each of which connects adjacent gauge nodes, as well as $N_f$ fundamental hypermultiplets attached to the last gauge node. The theory lives on $N$ M2-branes probing $\mathbb{C}^2/\mathbb{Z}_{g} \times \mathbb{C}^2/\mathbb{Z}_{N_f}$ singularity. The large $N$ Cardy free energy of the superconformal index is given by
\begin{equation} \label{ex2-large}
F_{S^2} = i \frac{4\sqrt{2} (gN_f)^{\frac{1}{2}} N^{\frac{3}{2}}}{3} \frac{\sqrt{\Delta_1 \Delta_2 \Delta_3 \Delta_4}}{2\beta}\ .
\end{equation}

\subparagraph{Mirror symmetry}
This theory at $g=2,N_f=1$ is dual to the former theory at $N_f=2$ by the mirror symmetry \cite{Intriligator:1996ex}. Indeed, the Cardy free energies of two theories, \eqref{ex1-large} and \eqref{ex2-large}, agree at large $N$ limit. Also, when $g=2, N_f=1$ and $N=1$,
\begin{equation} \label{ex2-1}
\begin{aligned}
F_{S^2} & \sim \frac{i}{2\beta} \left[4G+4\textrm{Im} \left\{ \textrm{Li}_2 (e^{\frac{\pi i}{4}}) + \textrm{Li}_2 (e^{\frac{3\pi i}{4}})\right\} \right] \\
& \approx	\frac{9.68691i}{2\beta}\ ,
\end{aligned}
\end{equation}
at $\Delta_1=\Delta_2=\Delta_3=\Delta_4=\frac{\pi i}{2}$. Hence, two Cardy free energies, \eqref{ex1-1} and \eqref{ex2-1}, also agree at $N=1$.

\paragraph{Example 4: $\mathcal{N}=3$ $U(N)^{g}$ necklace quiver CS matter theory with CS levels $(+k,-k,0,0,\ldots,0)$}~\\
This theory lives on $N$ M2-branes probing $(\mathbb{C}^2\times \mathbb{C}^2/\mathbb{Z}_{g-1})/\mathbb{Z}_k$ singularity. The large $N$ Cardy free energy is given by
\begin{equation} \label{ex4-large}
F_{S^2} = i \frac{4\sqrt{2} [(g-1)k]^{\frac{1}{2}} N^{\frac{3}{2}}}{3} \frac{\sqrt{\Delta_1 \Delta_2 \Delta_3 \Delta_4}}{2\beta}\ .
\end{equation}
This theory with $g$ gauge nodes at $k=1$ is dual to the third example with $g-1$ gauge nodes at $N_f=1$ by the $SL(2,\mathbb{Z})$ duality \cite{Aharony:1997ju}. Indeed, the large $N$ Cardy free energies of two theories, \eqref{ex2-large} and \eqref{ex4-large}, are identical.

\paragraph{Example 5: $\mathcal{N}=3$ $U(N)^{2g}$ necklace quiver CS matter theory with alternating CS levels $\pm k$}~\\
This theory describes the low energy dynamics of $N$ M2-branes probing $(\mathbb{C}^2/\mathbb{Z}_{g} \times \mathbb{C}^2/\mathbb{Z}_{g})/\mathbb{Z}_k$ singularity. The large $N$ Cardy free energy is
\begin{equation}
F_{S^2} = i \frac{4\sqrt{2} gk^{\frac{1}{2}} N^{\frac{3}{2}}}{3} \frac{\sqrt{\Delta_1 \Delta_2 \Delta_3 \Delta_4}}{2\beta}\ .
\end{equation}

\subsubsection{Type IIA String Theory Regime}
One can take the large $N$ limit with large but fixed 't Hooft couplings $\lambda = \frac{N}{k} \gg 1 \sim \mathcal{O}(N^0)$. Then, the field theory in this 't Hooft limit is dual to type IIA string theory. The characteristic large $N, \, \lambda$ behavior of the free energy in this type IIA string theory regime is
\begin{equation}
    F \sim \frac{N^2}{\sqrt{\lambda}}\ .
\end{equation}
Our example is the ABJM$_k$ theory. In the above 't Hooft limit, the ABJM$_k$ theory is dual to type IIA SUGRA on AdS$_4 \times \mathbb{CP}^3$ \cite{Aharony:2008ug}. This can be understood as the 10d reduction of the M-theory on AdS$_4 \times S^7/\mathbb{Z}_k$ where $S^7$ is a $U(1)$ bundle over $\mathbb{CP}^3$. Using \eqref{eq:fusion}, the Cardy free energy of the superconformal index in the large $N, \, \lambda$ limit can be read off from the topologically twisted index in \cite{PandoZayas:2019hdb} as follows:
\begin{equation}
F_{S^2} =   i \frac{4\sqrt{2} N^2}{3\sqrt{\lambda}} \frac{\sqrt{\Delta_1 \Delta_2 \Delta_3 \Delta_4}}{2\beta}\ .
\end{equation}
The above large $N$ Cardy free energy perfectly captures the entropy of dual BPS black holes in AdS$_4 \times \mathbb{CP}^3$ \cite{Choi:2018fdc,PandoZayas:2019hdb}.
\\

\subsection{D2-Branes Probing a CY 3-Fold Singularity in Massive IIA String Theory}
In this subsection, we consider Chern-Simons matter gauge theories describing the low energy dynamics of D2-branes probing a conical CY 3-fold singularity in the presence of a non-vanishing quantized Romans mass, i.e. in massive IIA string theory \cite{Guarino:2015jca,Fluder:2015eoa}. For this class of theories, sum of the CS levels for each gauge group does not vanish, i.e. $\sum_{g} k_g \neq 0$. Hence, those theories are not parity invariant so that the round sphere free energy $F_{S^3}$ is complex.

We shall consider the large $N$ limit with fixed CS levels $k \sim \mathcal{O}(1)$. The holographic dual is given by massive IIA SUGRA on AdS$_4 \times M_6$ where $M_6 = \mathcal{S}Y_5$ is the suspension of a Sasaki-Einstein 5-manifold $Y_5$, which serves as the base of a conical CY 3-fold. The characteristic large $N$ behavior of the free energy of this class is given by \cite{Guarino:2015jca}
\begin{equation}
    F \sim n^{\frac{1}{3}} N^{\frac{5}{3}}\ ,
\end{equation}
where $n=\sum_g k_g$, sum of the CS levels for each gauge node. The round sphere free energy at the superconformal $R$-charge is given by \cite{Fluder:2015eoa}
\begin{equation}
\textrm{Re}[F_{S^3}] = \frac{2^{1/3} 3^{1/6}\pi^3}{5 \textrm{Vol} (Y_5)^{2/3}} n^{\frac{1}{3}}N^{\frac{5}{3}} = \frac{2^{5/3} 3^{1/6}\pi}{5} (nN)^{\frac{1}{3}} (a_{4d})^{\frac{2}{3}}\ ,
\end{equation}
where $a_{4d}$ is the $a$-anomaly coefficient of the parent 4d SCFT, which lives on D3-branes probing the same CY 3-fold singularity. Applying \eqref{eq:unrefined}, the (unrefined) superconformal index in the large $N$ Cardy limit, for the generic $\mathcal{N} \geq 2$ SCFTs describing the low energy dynamics of D2-branes in massive IIA string theory, is given by
\begin{equation}
F_{S^2}  = \frac{\Delta_R^2}{2\pi i \beta} \frac{2^{1/3} 3^{1/6}\pi^3}{5 \textrm{Vol} (Y_5)^{2/3}} n^{\frac{1}{3}}N^{\frac{5}{3}} =  \frac{\Delta_R^2}{2\pi i \beta} \frac{2^{5/3} 3^{1/6}\pi}{5} (nN)^{\frac{1}{3}} (a_{4d})^{\frac{2}{3}}\ ,
\end{equation}
as expected from the dual gravity side in the non-Cardy regime \cite{Bobev:2019zmz}. 

For a generic $\mathcal{N}=2$ $U(N)^g$ quiver theory with equal CS level $k$ and bifundamental and adjoint matters, the large $N$ Cardy free energy of the generalized superconformal index can be read off from the round sphere partition function \cite{Fluder:2015eoa} using \eqref{eq:theorem2} as follows:
\begin{equation}
\begin{aligned}
&F_{S^2} = \\
& \frac{ 2^{1/3}3^{3/2}}{80i\beta} \left(1-\frac{i}{\sqrt{3}} \right) \left[\sum_{I \in \textrm{matters}} (\Delta_I+n_I \beta) (\Delta_I +n_I \beta +\pi i  +\beta) (\Delta_I +n_I \beta +2\pi i + 2\beta)\right]^{\frac{2}{3}} n^{\frac{1}{3}} N^{\frac{5}{3}} \\
&+\frac{ 2^{1/3}3^{3/2}}{80i\beta} \left(1+\frac{i}{\sqrt{3}} \right) \left[\sum_{I \in \textrm{matters}} (\Delta_I-n_I \beta) (\Delta_I -n_I \beta +\pi i  +\beta) (\Delta_I -n_I \beta +2\pi i + 2\beta)\right]^{\frac{2}{3}} n^{\frac{1}{3}} N^{\frac{5}{3}}\ .
\end{aligned}
\end{equation}

One particular example is an $\mathcal{N}=2$ CS deformation of the maximal SYM. This SCFT is dual to massive IIA SUGRA on AdS$_4 \times S^6$. The large $N$ Cardy free energy of the generalized superconformal index can be read off from the topologically twisted index \cite{Hosseini:2017fjo,Benini:2017oxt} using \eqref{eq:fusion} as
\begin{equation}
\begin{aligned}
F_{S^2}=\frac{2^{1/3}3^{13/6}}{80i\beta}k^{\frac{1}{3}} N^{\frac{5}{3}}  & \left[ \left(1-\frac{i}{\sqrt{3}} \right)  [(\Delta_1+n_1 \beta) (\Delta_2+n_2 \beta) (\Delta_3+n_3 \beta)]^{\frac{2}{3}} \right. \\
& \left.  + \left(1+\frac{i}{\sqrt{3}} \right)  [(\Delta_1-n_1 \beta) (\Delta_2-n_2 \beta) (\Delta_3-n_3 \beta)]^{\frac{2}{3}} \right]\ ,
\end{aligned}
\end{equation}
which is consistent with the above general formula. This large $N$ Cardy free energy is supposed to account for the microstates of the dyonic rotating BPS black holes with vanishing magnetic charge for the $R$-symmetry in the massive IIA SUGRA background AdS$_4 \times S^6$.
\\

\section{Concluding Remarks}

In this paper, we have examined the Cardy limit of 3d supersymmetric partition functions using their factorization into the Cardy block, which is defined as the dominant saddle point contribution to the hemisphere index in the Cardy limit. The Cardy block plays the role of a building block of other 3d partition functions such as the generalized superconformal index, the refined topologically twisted index and the squashed sphere partition function. The factorization to the Cardy block allows us to find universal relations among those partition functions in the Cardy limit.

Furthermore, our analysis can be applied to holographic SCFTs in 3d, which are dual to AdS$_4$ gravity in the large $N$ limit. In the large $N$ limit, such universal relations extend to include the round sphere partition function, which is known to count the degrees of freedom of a SCFT in odd dimensions and is also dual to the holographic entanglement entropy in dual AdS$_4$ for a spherical entangling surface. In addition, the two supersymmetric indices we have examined correspond to the entropy functions of BPS black holes in AdS$_4$; the generalized superconformal index, in the presence of the magnetic flux for the flavor symmetry, captures the microstates of rotating dyonic BPS black holes in AdS$_4$ while the refined topologically twisted index captures the microstates of rotating dyonic BPS black holes in mAdS$_4$, an asymptotically locally AdS$_4$ spacetime. Therefore, our analysis provides a field theoretic derivation of universal relations among the black hole entropies and the holographic entanglement entropy in AdS$_4$. We have also provided explicit examples, which confirm the universal relations we have found.
\\

We would like to remark a few interesting points and future directions.
\begin{itemize}

\item \textbf{Black hole microstate counting in the non-Cardy regime}

In section \ref{sec:ex1}, we have seen that our Cardy formulae for the M2-brane theory, which are derived in the $\beta \rightarrow 0$ limit, exactly account for the microstates of various BPS black holes in AdS$_4 \times S^7$ even at finite $\beta$. Those BPS black holes are supposed to be realized as the local large N saddle points on the dual field theory side. Indeed, there are a lot of examples showing that the Cardy formula is exact at finite $\beta$: from the pioneering work of Strominger and Vafa counting the microstates of the D1-D5-P black holes \cite{Strominger:1996sh} to recent works counting the microstates of the electrically charged rotating BPS black holes in AdS$_5$ \cite{Choi:2018hmj, Kim:2019yrz}, AdS$_7$ \cite{Choi:2018hmj, Nahmgoong:2019hko}, AdS$_6$ \cite{Choi:2019miv}, AdS$_4$ \cite{Choi:2019zpz}. Remarkably, the resulting Cardy free energy of the index at large $N$ perfectly captures the Bekenstein-Hawking entropy of the dual BPS black holes even in the non-Cardy regime.

Accordingly, we expect that our Cardy formulae such as \eqref{eq:theorem2} are exact in the non-Cardy regime as long as we consider the large $N$ saddle point corresponding to the dual BPS black holes in AdS$_4$. For example, let us consider the superconformal index without magnetic flux for the flavor symmetry. The form of the entropy function of generic rotating electric BPS black holes in AdS$_4$ can be found in \cite{Hristov:2019mqp}, which is the same as our Cardy free energy once we identify $\mathrm{Im}\left[G^{(0)}\right]$ with the supergravity prepotential $\mathcal{F}$ up to some multiplicative constant factors. We expect that $G^{(1)}$ and all $\mathcal O(\beta)$ corrections are subdominant in $N$ at the large $N$ saddle point capturing the dual black hole microstates. Also the unrefined superconformal index leads to the Cardy formula \eqref{eq:unrefined}, which is proven exact by the supergravity analysis for the universal spinning black holes in AdS$_4$ \cite{Bobev:2019zmz}. It is worth studying such exactness of our Cardy formulae further, both on the field theory side and on the gravity side.

\item \textbf{Cardy limit for finite $N$}

In section \ref{sec:ex2}, we have examined the superconformal indices in the Cardy limit of some examples for finite $N$. In \cite{Choi:2019zpz}, the finite $N$ Cardy limit of the superconformal index was examined for $\mathcal N = 4$ SYM with one fundamental and one adjoint matters both analytically and numerically. In particular, for $N = 2$, they obtained the exact coefficient of the free energy:
\begin{align}
&\log I_{S^2} \nonumber \\
&\sim \frac{i}{2 \beta} \left[-8 G-2 \, \mathrm{Im} \left\{2 \mathrm{Li}_2(ix)+ 2 \mathrm{Li}_2\left(\frac{i}{x}\right)+2 \mathrm{Li}_2(i x^2)+ 2 \mathrm{Li}_2\left(\frac{i}{x^2}\right)+\mathrm{Li}_2\left(\frac{1}{x}\right)\right\}\right] \nonumber \\
&\approx \frac{-17.4771i}{2 \beta}
\end{align}
with
\begin{gather}
\begin{gathered}
G = \frac{\mathrm{Li}_2(i)-\mathrm{Li}_2(-i)}{2 i} \approx 0.915966 \,, \\
x = \frac{1}{2} (1-3^\frac{1}{4} \sqrt2+\sqrt3) \approx 0.435421 \,,
\end{gathered}
\end{gather}
which is the finite $N$ version of the $N^\frac{3}{2}$ scaling of the M2-brane degrees of freedom. Our analysis provides similar results for more examples, and it would be interesting to find the physical interpretation of those numbers. Also our Cardy free energy is a simple but non-trivial observable of a theory. For example, its matching can be regarded as a non-trivial test of a duality for 3d SCFTs. We have illustrated such examples in section \ref{sec:ex2}, where the Cardy free energy shows perfect matches under 3d mirror symmetry or $SL(2,\mathbb Z)$ duality. While our analysis is restricted for theories of rank less than 3, it would be worth studying higher rank theories.

\item \textbf{Twisted compactification of 5d, 6d SCFTs}

While our analysis relies on the localization computation of supersymmetric partition functions, and thus on the Lagrangian description of a theory, recently similar results are obtained for class $\mathcal R$ theories, which are generically non-Lagrangian theories, realized as twisted compactification of 6d (2,0) $A_{N-1}$ theory on hyperbolic 3-manifolds \cite{Dimofte:2011ju, Cecotti:2011iy, Dimofte:2011py}. The superconformal indices of those theories can be computed as topological invariants of $SL(N,\mathbb C)$ Chern-Simons theories by the 3d-3d correspondence and are shown to capture $N^3$ degrees of freedom of $N$ M5-branes \cite{Bobev:2019zmz, Benini:2019dyp}. Remarkably, they satisfy the same relation \eqref{eq:unrefined}:
\begin{align}
    F_{S^2}(\Delta = -\Delta_R \delta_*,n = 0;\beta) = \frac{\Delta_R^2}{2 \pi i \beta} \mathrm{Re} \left[F_{S^3} \left(\delta_*\right)\right],
\end{align}
which we derive for 3d SCFTs with Lagrangian descriptions. Also the same relation is expected by the supergravity analysis \cite{Bobev:2019zmz} for 3d theories obtained as the twisted compactification of 5d SCFTs \cite{Crichigno:2018adf}. From those results, we may expect that our Cardy analysis for Lagrangian theories would hold for a broader class of 3d SCFTs, which will be interesting to clarify.
\\

\end{itemize}

\acknowledgments

We would like to thank Aprile Francesco, Dongmin Gang, Nakwoo Kim, Hee-Cheol Kim, Sara Pasquetti, Valentin Reys, Alberto Zaffaroni for helpful discussions, and especially Seok Kim for insightful comments and discussions. SC is partially supported by the National Research Foundation of Korea Grant 2018R1A2B6004914 and by NRF-2017-Global Ph.D. Fellowship Program. CH is partially supported by the ERC-STG grant 637844-HBQFTNCER and by the INFN.
\\

\appendix

\section{Asymptotic Behavior of q-Pochhammer Symbols}

The $q$-Pochhammer symbol $(|q|<1)$ is defined as
\begin{equation}
\begin{aligned}
&(a;q)_n:=\displaystyle \prod^{n-1}_{i=0} (1-aq^i), \quad n>0, \\
&(a;q)_0:=1, \\
&(a;q)_n:=\frac{1}{(aq^n;q)_{-n}}, \quad n<0.
\end{aligned}
\end{equation}
The Cardy limit ($q \to 1^-$) of the infinite $q$-Pochhammer symbol is given by
\begin{equation} \label{poch}
\begin{aligned}
(aq^m;q^2) & \equiv (aq^m;q^2)_{\infty} = \prod_{i=0}^{\infty}(1-aq^{m+2i})\ , \quad  q=e^{-\beta}\ , \\
\lim_{\beta \to 0^+}(aq^m;q^2)_\infty &= \,   (1-aq^m)^{1/2} \, \exp\Big[- \frac {1}{2\beta} \textrm{Li}_2 (aq^m)\Big]\big(1+\mathcal O(\beta)\big) \\
&=\, \exp\Big[- \frac {1}{2\beta} \textrm{Li}_2 (aq^{m-1})\Big]\big(1+\mathcal O(\beta)\big)\ , \quad  a \in \mathbb{C} \; \& \;a \notin [1,\infty)\ , \\
\lim_{\beta \to 0^+}(q^m;q^2)_{\infty}&= \frac{\sqrt{2\pi}}{\Gamma(m/2)} (2\beta)^{-(m-1)/2}  \exp \Big[-\frac{1}{2\beta} \textrm{Li}_2 (1) \Big]   \big(1+\mathcal O(\beta)\big)\ .
\end{aligned}
\end{equation}
These asymptotic formulae \eqref{poch} are well-known for $|a| \leq 1$. To extend it to whole complex plane $\mathbb{C}$, we used the modular property of the Jacobi theta function and the Dedekind eta function. Also note that these asymptotic formulae \eqref{poch} have a branch cut at $(1,\infty)$. For further details, refer to Appen. A of \cite{Choi:2019zpz}.

We can extend the definition of the infinite $q$-Pochhammer symbol to $|q|>1$ region using the Plethystic exponential as the following:
\begin{equation} \label{pope}
(a;q^2)_{\infty}=\textrm{PE}\left[-\frac{a}{1-q^2}\right]=\textrm{PE}\left[\frac{aq^{-2}}{1-q^{-2}}\right]=\textrm{PE}\left[-\frac{aq^{-2}}{1-q^{-2}}\right]^{-1}=\frac{1}{(aq^{-2};q^{-2})_{\infty}}\ .
\end{equation}
Using this formula, one can easily check that the $q \to 1^+$ limit of the infinite $q$-Pochhammer symbol is given by
\begin{equation} \label{pope2}
\lim_{\beta \to 0^-}(aq^m;q^2)_\infty =\, \exp\Big[- \frac {1}{2\beta} \textrm{Li}_2 (aq^{m-1})\Big]\big(1+\mathcal O(\beta)\big)\ , \quad  a \in \mathbb{C} \; \& \;a \notin [1,\infty)\ ,
\end{equation}
i.e. the same as the $q \to 1^-$ limit.


The polylogarithm function is defined by a power series when $|a|<1$:
\begin{equation} \label{power-polylog}
\textrm{Li}_n (a) = \sum_{k=1}^\infty \frac{a^k}{k^n}\ ,
\end{equation}
and can be extended to $|a| \geq 1$ by the process of analytic continuation. Then, the polylogarithm function satisfies the following inversion formula involving Bernoulli polynomials $B_n(x)$ ($n \in \mathbb{Z}$):
\begin{equation} \label{invm}
\begin{aligned}
&\textrm{Li}_n (a) +  (-1)^n \textrm{Li}_n (1/a) = - \frac{(2\pi i)^n}{n!} B_n \left(\frac{\log a}{2\pi i} - p\right), \quad 2\pi p<  \textrm{Im} (\log a) < 2\pi (p+1) \; \& \; a \notin (0,1) \,, \\
& B_1(x)=x-\frac{1}{2}, \quad B_2(x) = x^2 -x + \frac{1}{6}, \quad B_3(x)=x^3-\frac{3}{2}x^2 +\frac{1}{2}x, \; \ldots.
\end{aligned}
\end{equation}
Also, the polylogarithm function exhibits the following limiting behavior:
\begin{equation} \label{lim-polylog}
\lim_{|a| \to 0} \textrm{Li}_n (a) = a\ .
\end{equation}
Note that there are a branch point at $a=1$ and the branch cut on $(1, +\infty)$ as usual. Hence, the principal branch of the polylogarithm will be $(0,2\pi)$. Accordingly, we will set the branch cut of the logarithm function as $(0,-\infty)$, i.e. its principal branch is $(-\pi, +\pi)$.

We also use the theta function defined as following:
\begin{equation}
\theta(a;q^2) \equiv (a;q^2)_\infty (a^{-1}q^2;q^2)_\infty\ ,
\end{equation}
whose Cardy limit $\beta \to 0$ is given by
\begin{equation} \label{theta}
\lim_{\beta \to 0} \theta(aq^m;q^2) = \exp \left[ - \frac{1}{2\beta} \left( - \frac{1}{2} \log^2 (\hat{a}q^{m-1}) + \pi i \log (\hat{a}q^{m-1}) + \frac{\pi^2}{3} \right)\right] \left(1+\mathcal O(\beta)\right)\ ,
\end{equation}
where
\begin{equation}
\hat{a} \equiv a e^{-2\pi i p} \quad \Rightarrow \quad \log \hat{a} = \log a -2\pi i p\ .
\end{equation}
Here, we used \eqref{poch} and \eqref{invm}.
\\


\bibliography{mybib}

\providecommand{\href}[2]{#2}\begingroup\raggedright\begin{thebibliography}{10}

\bibitem{Beem:2012mb}
C.~Beem, T.~Dimofte and S.~Pasquetti, \emph{{Holomorphic Blocks in Three
  Dimensions}}, \href{http://dx.doi.org/10.1007/JHEP12(2014)177}{\emph{JHEP}
  {\bfseries 12} (2014) 177},
  [\href{https://arxiv.org/abs/1211.1986}{{\ttfamily 1211.1986}}].

\bibitem{Nieri:2015yia}
F.~Nieri and S.~Pasquetti, \emph{{Factorisation and holomorphic blocks in 4d}},
  \href{http://dx.doi.org/10.1007/JHEP11(2015)155}{\emph{JHEP} {\bfseries 11}
  (2015) 155}, [\href{https://arxiv.org/abs/1507.00261}{{\ttfamily
  1507.00261}}].

\bibitem{Pasquetti:2011fj}
S.~Pasquetti, \emph{{Factorisation of N = 2 Theories on the Squashed
  3-Sphere}}, \href{http://dx.doi.org/10.1007/JHEP04(2012)120}{\emph{JHEP}
  {\bfseries 04} (2012) 120},
  [\href{https://arxiv.org/abs/1111.6905}{{\ttfamily 1111.6905}}].

\bibitem{Hwang:2012jh}
C.~Hwang, H.-C. Kim and J.~Park, \emph{{Factorization of the 3d superconformal
  index}}, \href{http://dx.doi.org/10.1007/JHEP08(2014)018}{\emph{JHEP}
  {\bfseries 08} (2014) 018},
  [\href{https://arxiv.org/abs/1211.6023}{{\ttfamily 1211.6023}}].

\bibitem{Taki:2013opa}
M.~Taki, \emph{{Holomorphic Blocks for 3d Non-abelian Partition Functions}},
  \href{https://arxiv.org/abs/1303.5915}{{\ttfamily 1303.5915}}.

\bibitem{Alday:2009aq}
L.~F. Alday, D.~Gaiotto and Y.~Tachikawa, \emph{{Liouville Correlation
  Functions from Four-dimensional Gauge Theories}},
  \href{http://dx.doi.org/10.1007/s11005-010-0369-5}{\emph{Lett. Math. Phys.}
  {\bfseries 91} (2010) 167--197},
  [\href{https://arxiv.org/abs/0906.3219}{{\ttfamily 0906.3219}}].

\bibitem{Aganagic:2013tta}
M.~Aganagic, N.~Haouzi, C.~Kozcaz and S.~Shakirov, \emph{{Gauge/Liouville
  Triality}},  \href{https://arxiv.org/abs/1309.1687}{{\ttfamily 1309.1687}}.

\bibitem{Aganagic:2014oia}
M.~Aganagic, N.~Haouzi and S.~Shakirov, \emph{{$A_n$-Triality}},
  \href{https://arxiv.org/abs/1403.3657}{{\ttfamily 1403.3657}}.

\bibitem{Kim:2012uz}
H.-C. Kim, J.~Kim, S.~Kim and K.~Lee, \emph{{Vortices and 3 dimensional
  dualities}},  \href{https://arxiv.org/abs/1204.3895}{{\ttfamily 1204.3895}}.

\bibitem{Hwang:2015wna}
C.~Hwang and J.~Park, \emph{{Factorization of the 3d superconformal index with
  an adjoint matter}},
  \href{http://dx.doi.org/10.1007/JHEP11(2015)028}{\emph{JHEP} {\bfseries 11}
  (2015) 028}, [\href{https://arxiv.org/abs/1506.03951}{{\ttfamily
  1506.03951}}].

\bibitem{Hwang:2017kmk}
C.~Hwang, P.~Yi and Y.~Yoshida, \emph{{Fundamental Vortices, Wall-Crossing, and
  Particle-Vortex Duality}},
  \href{http://dx.doi.org/10.1007/JHEP05(2017)099}{\emph{JHEP} {\bfseries 05}
  (2017) 099}, [\href{https://arxiv.org/abs/1703.00213}{{\ttfamily
  1703.00213}}].

\bibitem{Hwang:2018uyj}
C.~Hwang, H.~Kim and J.~Park, \emph{{On 3d Seiberg‐Like Dualities with Two
  Adjoints}}, \href{http://dx.doi.org/10.1002/prop.201800064}{\emph{Fortsch.
  Phys.} {\bfseries 66} (2018) 1800064},
  [\href{https://arxiv.org/abs/1807.06198}{{\ttfamily 1807.06198}}].

\bibitem{Strominger:1996sh}
A.~Strominger and C.~Vafa, \emph{{Microscopic origin of the Bekenstein-Hawking
  entropy}}, \href{http://dx.doi.org/10.1016/0370-2693(96)00345-0}{\emph{Phys.
  Lett.} {\bfseries B379} (1996) 99--104},
  [\href{https://arxiv.org/abs/hep-th/9601029}{{\ttfamily hep-th/9601029}}].

\bibitem{Benini:2015eyy}
F.~Benini, K.~Hristov and A.~Zaffaroni, \emph{{Black hole microstates in
  AdS$_{4}$ from supersymmetric localization}},
  \href{http://dx.doi.org/10.1007/JHEP05(2016)054}{\emph{JHEP} {\bfseries 05}
  (2016) 054}, [\href{https://arxiv.org/abs/1511.04085}{{\ttfamily
  1511.04085}}].

\bibitem{Hosseini:2016tor}
S.~M. Hosseini and A.~Zaffaroni, \emph{{Large $N$ matrix models for 3d ${\cal
  N}=2$ theories: twisted index, free energy and black holes}},
  \href{http://dx.doi.org/10.1007/JHEP08(2016)064}{\emph{JHEP} {\bfseries 08}
  (2016) 064}, [\href{https://arxiv.org/abs/1604.03122}{{\ttfamily
  1604.03122}}].

\bibitem{Hosseini:2016ume}
S.~M. Hosseini and N.~Mekareeya, \emph{{Large $N$ topologically twisted index:
  necklace quivers, dualities, and Sasaki-Einstein spaces}},
  \href{http://dx.doi.org/10.1007/JHEP08(2016)089}{\emph{JHEP} {\bfseries 08}
  (2016) 089}, [\href{https://arxiv.org/abs/1604.03397}{{\ttfamily
  1604.03397}}].

\bibitem{Benini:2016rke}
F.~Benini, K.~Hristov and A.~Zaffaroni, \emph{{Exact microstate counting for
  dyonic black holes in AdS4}},
  \href{http://dx.doi.org/10.1016/j.physletb.2017.05.076}{\emph{Phys. Lett.}
  {\bfseries B771} (2017) 462--466},
  [\href{https://arxiv.org/abs/1608.07294}{{\ttfamily 1608.07294}}].

\bibitem{Azzurli:2017kxo}
F.~Azzurli, N.~Bobev, P.~M. Crichigno, V.~S. Min and A.~Zaffaroni, \emph{{A
  universal counting of black hole microstates in AdS$_{4}$}},
  \href{http://dx.doi.org/10.1007/JHEP02(2018)054}{\emph{JHEP} {\bfseries 02}
  (2018) 054}, [\href{https://arxiv.org/abs/1707.04257}{{\ttfamily
  1707.04257}}].

\bibitem{Hosseini:2017fjo}
S.~M. Hosseini, K.~Hristov and A.~Passias, \emph{{Holographic microstate
  counting for AdS$_{4}$ black holes in massive IIA supergravity}},
  \href{http://dx.doi.org/10.1007/JHEP10(2017)190}{\emph{JHEP} {\bfseries 10}
  (2017) 190}, [\href{https://arxiv.org/abs/1707.06884}{{\ttfamily
  1707.06884}}].

\bibitem{Benini:2017oxt}
F.~Benini, H.~Khachatryan and P.~Milan, \emph{{Black hole entropy in massive
  Type IIA}}, \href{http://dx.doi.org/10.1088/1361-6382/aa9f5b}{\emph{Class.
  Quant. Grav.} {\bfseries 35} (2018) 035004},
  [\href{https://arxiv.org/abs/1707.06886}{{\ttfamily 1707.06886}}].

\bibitem{Zaffaroni:2019dhb}
A.~Zaffaroni, \emph{{Lectures on AdS Black Holes, Holography and
  Localization}},  2019.
\newblock \href{https://arxiv.org/abs/1902.07176}{{\ttfamily 1902.07176}}.

\bibitem{Choi:2019zpz}
S.~Choi, C.~Hwang and S.~Kim, \emph{{Quantum vortices, M2-branes and black
  holes}},  \href{https://arxiv.org/abs/1908.02470}{{\ttfamily 1908.02470}}.

\bibitem{Nian:2019pxj}
J.~Nian and L.~A. Pando~Zayas, \emph{{Microscopic Entropy of Rotating
  Electrically Charged AdS$_4$ Black Holes from Field Theory Localization}},
  \href{https://arxiv.org/abs/1909.07943}{{\ttfamily 1909.07943}}.

\bibitem{Cabo-Bizet:2018ehj}
A.~Cabo-Bizet, D.~Cassani, D.~Martelli and S.~Murthy, \emph{{Microscopic origin
  of the Bekenstein-Hawking entropy of supersymmetric AdS$_{\bf 5}$ black
  holes}},  \href{https://arxiv.org/abs/1810.11442}{{\ttfamily 1810.11442}}.

\bibitem{Choi:2018hmj}
S.~Choi, J.~Kim, S.~Kim and J.~Nahmgoong, \emph{{Large AdS black holes from
  QFT}},  \href{https://arxiv.org/abs/1810.12067}{{\ttfamily 1810.12067}}.

\bibitem{Choi:2018vbz}
S.~Choi, J.~Kim, S.~Kim and J.~Nahmgoong, \emph{{Comments on deconfinement in
  AdS/CFT}},  \href{https://arxiv.org/abs/1811.08646}{{\ttfamily 1811.08646}}.

\bibitem{Benini:2018ywd}
F.~Benini and P.~Milan, \emph{{Black holes in 4d $\mathcal{N}=4$
  Super-Yang-Mills}},  \href{https://arxiv.org/abs/1812.09613}{{\ttfamily
  1812.09613}}.

\bibitem{Honda:2019cio}
M.~Honda, \emph{{Quantum Black Hole Entropy from 4d Supersymmetric Cardy
  formula}}, \href{http://dx.doi.org/10.1103/PhysRevD.100.026008}{\emph{Phys.
  Rev.} {\bfseries D100} (2019) 026008},
  [\href{https://arxiv.org/abs/1901.08091}{{\ttfamily 1901.08091}}].

\bibitem{ArabiArdehali:2019tdm}
A.~Arabi~Ardehali, \emph{{Cardy-like asymptotics of the 4d $ \mathcal{N}=4 $
  index and AdS$_{5}$ blackholes}},
  \href{http://dx.doi.org/10.1007/JHEP06(2019)134}{\emph{JHEP} {\bfseries 06}
  (2019) 134}, [\href{https://arxiv.org/abs/1902.06619}{{\ttfamily
  1902.06619}}].

\bibitem{Kim:2019yrz}
J.~Kim, S.~Kim and J.~Song, \emph{{A 4d $N=1$ Cardy Formula}},
  \href{https://arxiv.org/abs/1904.03455}{{\ttfamily 1904.03455}}.

\bibitem{Cabo-Bizet:2019osg}
A.~Cabo-Bizet, D.~Cassani, D.~Martelli and S.~Murthy, \emph{{The asymptotic
  growth of states of the 4d $ \mathcal{N}=1 $ superconformal index}},
  \href{http://dx.doi.org/10.1007/JHEP08(2019)120}{\emph{JHEP} {\bfseries 08}
  (2019) 120}, [\href{https://arxiv.org/abs/1904.05865}{{\ttfamily
  1904.05865}}].

\bibitem{Larsen:2019oll}
F.~Larsen, J.~Nian and Y.~Zeng, \emph{{AdS$_5$ Black Hole Entropy near the BPS
  Limit}},  \href{https://arxiv.org/abs/1907.02505}{{\ttfamily 1907.02505}}.

\bibitem{Lezcano:2019pae}
A.~G. Lezcano and L.~A. Pando~Zayas, \emph{{Microstate Counting via Bethe
  Ans\"{a}tze in the 4d ${\cal N}=1$ Superconformal Index}},
  \href{https://arxiv.org/abs/1907.12841}{{\ttfamily 1907.12841}}.

\bibitem{Choi:2019miv}
S.~Choi and S.~Kim, \emph{{Large AdS$_6$ black holes from CFT$_5$}},
  \href{https://arxiv.org/abs/1904.01164}{{\ttfamily 1904.01164}}.

\bibitem{Nahmgoong:2019hko}
J.~Nahmgoong, \emph{{6d superconformal Cardy formulas}},
  \href{https://arxiv.org/abs/1907.12582}{{\ttfamily 1907.12582}}.

\bibitem{Choi:2018fdc}
S.~Choi, C.~Hwang, S.~Kim and J.~Nahmgoong, \emph{{Entropy functions of BPS
  black holes in AdS$_4$ and AdS$_6$}},
  \href{https://arxiv.org/abs/1811.02158}{{\ttfamily 1811.02158}}.

\bibitem{Aharony:2008ug}
O.~Aharony, O.~Bergman, D.~L. Jafferis and J.~Maldacena, \emph{{N=6
  superconformal Chern-Simons-matter theories, M2-branes and their gravity
  duals}}, \href{http://dx.doi.org/10.1088/1126-6708/2008/10/091}{\emph{JHEP}
  {\bfseries 10} (2008) 091},
  [\href{https://arxiv.org/abs/0806.1218}{{\ttfamily 0806.1218}}].

\bibitem{Kapustin:2010xq}
A.~Kapustin, B.~Willett and I.~Yaakov, \emph{{Nonperturbative Tests of
  Three-Dimensional Dualities}},
  \href{http://dx.doi.org/10.1007/JHEP10(2010)013}{\emph{JHEP} {\bfseries 10}
  (2010) 013}, [\href{https://arxiv.org/abs/1003.5694}{{\ttfamily 1003.5694}}].

\bibitem{Benini:2015noa}
F.~Benini and A.~Zaffaroni, \emph{{A topologically twisted index for
  three-dimensional supersymmetric theories}},
  \href{http://dx.doi.org/10.1007/JHEP07(2015)127}{\emph{JHEP} {\bfseries 07}
  (2015) 127}, [\href{https://arxiv.org/abs/1504.03698}{{\ttfamily
  1504.03698}}].

\bibitem{Benini:2016hjo}
F.~Benini and A.~Zaffaroni, \emph{{Supersymmetric partition functions on
  Riemann surfaces}},  \href{https://arxiv.org/abs/1605.06120}{{\ttfamily
  1605.06120}}.

\bibitem{Closset:2016arn}
C.~Closset and H.~Kim, \emph{{Comments on twisted indices in 3d supersymmetric
  gauge theories}},
  \href{http://dx.doi.org/10.1007/JHEP08(2016)059}{\emph{JHEP} {\bfseries 08}
  (2016) 059}, [\href{https://arxiv.org/abs/1605.06531}{{\ttfamily
  1605.06531}}].

\bibitem{Martelli:2011fu}
D.~Martelli, A.~Passias and J.~Sparks, \emph{{The gravity dual of
  supersymmetric gauge theories on a squashed three-sphere}},
  \href{http://dx.doi.org/10.1016/j.nuclphysb.2012.07.019}{\emph{Nucl. Phys.}
  {\bfseries B864} (2012) 840--868},
  [\href{https://arxiv.org/abs/1110.6400}{{\ttfamily 1110.6400}}].

\bibitem{Hristov:2011ye}
K.~Hristov, C.~Toldo and S.~Vandoren, \emph{{On BPS bounds in D=4 N=2 gauged
  supergravity}}, \href{http://dx.doi.org/10.1007/JHEP12(2011)014}{\emph{JHEP}
  {\bfseries 12} (2011) 014},
  [\href{https://arxiv.org/abs/1110.2688}{{\ttfamily 1110.2688}}].

\bibitem{Nishioka:2013haa}
T.~Nishioka and I.~Yaakov, \emph{{Supersymmetric Renyi Entropy}},
  \href{http://dx.doi.org/10.1007/JHEP10(2013)155}{\emph{JHEP} {\bfseries 10}
  (2013) 155}, [\href{https://arxiv.org/abs/1306.2958}{{\ttfamily 1306.2958}}].

\bibitem{Huang:2014gca}
X.~Huang, S.-J. Rey and Y.~Zhou, \emph{{Three-dimensional SCFT on conic space
  as hologram of charged topological black hole}},
  \href{http://dx.doi.org/10.1007/JHEP03(2014)127}{\emph{JHEP} {\bfseries 03}
  (2014) 127}, [\href{https://arxiv.org/abs/1401.5421}{{\ttfamily 1401.5421}}].

\bibitem{Nishioka:2014mwa}
T.~Nishioka, \emph{{The Gravity Dual of Supersymmetric Renyi Entropy}},
  \href{http://dx.doi.org/10.1007/JHEP07(2014)061}{\emph{JHEP} {\bfseries 07}
  (2014) 061}, [\href{https://arxiv.org/abs/1401.6764}{{\ttfamily 1401.6764}}].

\bibitem{Dowker:2010yj}
J.~S. Dowker, \emph{{Entanglement entropy for odd spheres}},
  \href{https://arxiv.org/abs/1012.1548}{{\ttfamily 1012.1548}}.

\bibitem{Casini:2011kv}
H.~Casini, M.~Huerta and R.~C. Myers, \emph{{Towards a derivation of
  holographic entanglement entropy}},
  \href{http://dx.doi.org/10.1007/JHEP05(2011)036}{\emph{JHEP} {\bfseries 05}
  (2011) 036}, [\href{https://arxiv.org/abs/1102.0440}{{\ttfamily 1102.0440}}].

\bibitem{Maldacena:1997re}
J.~M. Maldacena, \emph{{The Large N limit of superconformal field theories and
  supergravity}}, \href{http://dx.doi.org/10.1023/A:1026654312961,
  10.4310/ATMP.1998.v2.n2.a1}{\emph{Int. J. Theor. Phys.} {\bfseries 38} (1999)
  1113--1133}, [\href{https://arxiv.org/abs/hep-th/9711200}{{\ttfamily
  hep-th/9711200}}].

\bibitem{Hosseini:2019iad}
S.~M. Hosseini, K.~Hristov and A.~Zaffaroni, \emph{{Gluing gravitational blocks
  for AdS black holes}},  \href{https://arxiv.org/abs/1909.10550}{{\ttfamily
  1909.10550}}.

\bibitem{Jafferis:2010un}
D.~L. Jafferis, \emph{{The Exact Superconformal R-Symmetry Extremizes Z}},
  \href{http://dx.doi.org/10.1007/JHEP05(2012)159}{\emph{JHEP} {\bfseries 05}
  (2012) 159}, [\href{https://arxiv.org/abs/1012.3210}{{\ttfamily 1012.3210}}].

\bibitem{Cardy:1986ie}
J.~L. Cardy, \emph{{Operator Content of Two-Dimensional Conformally Invariant
  Theories}}, \href{http://dx.doi.org/10.1016/0550-3213(86)90552-3}{\emph{Nucl.
  Phys.} {\bfseries B270} (1986) 186--204}.

\bibitem{Bobev:2019zmz}
N.~Bobev and P.~M. Crichigno, \emph{{Universal Spinning Black Holes and
  Theories of Class $\mathcal R$}},
  \href{https://arxiv.org/abs/1909.05873}{{\ttfamily 1909.05873}}.

\bibitem{Benini:2019dyp}
F.~Benini, D.~Gang and L.~A. Pando~Zayas, \emph{{Rotating Black Hole Entropy
  from M5 Branes}},  \href{https://arxiv.org/abs/1909.11612}{{\ttfamily
  1909.11612}}.

\bibitem{Bobev:2017uzs}
N.~Bobev and P.~M. Crichigno, \emph{{Universal RG Flows Across Dimensions and
  Holography}}, \href{http://dx.doi.org/10.1007/JHEP12(2017)065}{\emph{JHEP}
  {\bfseries 12} (2017) 065},
  [\href{https://arxiv.org/abs/1708.05052}{{\ttfamily 1708.05052}}].

\bibitem{Fujitsuka:2013fga}
M.~Fujitsuka, M.~Honda and Y.~Yoshida, \emph{{Higgs branch localization of 3d
  $\mathcal{N}=2$ theories}},
  \href{http://dx.doi.org/10.1093/ptep/ptu158}{\emph{PTEP} {\bfseries 2014}
  (2014) 123B02}, [\href{https://arxiv.org/abs/1312.3627}{{\ttfamily
  1312.3627}}].

\bibitem{Benini:2013yva}
F.~Benini and W.~Peelaers, \emph{{Higgs branch localization in three
  dimensions}}, \href{http://dx.doi.org/10.1007/JHEP05(2014)030}{\emph{JHEP}
  {\bfseries 05} (2014) 030},
  [\href{https://arxiv.org/abs/1312.6078}{{\ttfamily 1312.6078}}].

\bibitem{Yoshida:2014ssa}
Y.~Yoshida and K.~Sugiyama, \emph{{Localization of 3d $\mathcal{N}=2$
  Supersymmetric Theories on $S^1 \times D^2$}},
  \href{https://arxiv.org/abs/1409.6713}{{\ttfamily 1409.6713}}.

\bibitem{Bhattacharya:2008zy}
J.~Bhattacharya, S.~Bhattacharyya, S.~Minwalla and S.~Raju, \emph{{Indices for
  Superconformal Field Theories in 3,5 and 6 Dimensions}},
  \href{http://dx.doi.org/10.1088/1126-6708/2008/02/064}{\emph{JHEP} {\bfseries
  02} (2008) 064}, [\href{https://arxiv.org/abs/0801.1435}{{\ttfamily
  0801.1435}}].

\bibitem{Bhattacharya:2008bja}
J.~Bhattacharya and S.~Minwalla, \emph{{Superconformal Indices for N = 6 Chern
  Simons Theories}},
  \href{http://dx.doi.org/10.1088/1126-6708/2009/01/014}{\emph{JHEP} {\bfseries
  01} (2009) 014}, [\href{https://arxiv.org/abs/0806.3251}{{\ttfamily
  0806.3251}}].

\bibitem{Kapustin:2011jm}
A.~Kapustin and B.~Willett, \emph{{Generalized Superconformal Index for Three
  Dimensional Field Theories}},
  \href{https://arxiv.org/abs/1106.2484}{{\ttfamily 1106.2484}}.

\bibitem{Kim:2009wb}
S.~Kim, \emph{{The Complete superconformal index for N=6 Chern-Simons theory}},
  \href{http://dx.doi.org/10.1016/j.nuclphysb.2012.07.015,
  10.1016/j.nuclphysb.2009.06.025}{\emph{Nucl. Phys.} {\bfseries B821} (2009)
  241--284}, [\href{https://arxiv.org/abs/0903.4172}{{\ttfamily 0903.4172}}].

\bibitem{Imamura:2011su}
Y.~Imamura and S.~Yokoyama, \emph{{Index for three dimensional superconformal
  field theories with general R-charge assignments}},
  \href{http://dx.doi.org/10.1007/JHEP04(2011)007}{\emph{JHEP} {\bfseries 04}
  (2011) 007}, [\href{https://arxiv.org/abs/1101.0557}{{\ttfamily 1101.0557}}].

\bibitem{Dimofte:2011py}
T.~Dimofte, D.~Gaiotto and S.~Gukov, \emph{{3-Manifolds and 3d Indices}},
  \href{http://dx.doi.org/10.4310/ATMP.2013.v17.n5.a3}{\emph{Adv. Theor. Math.
  Phys.} {\bfseries 17} (2013) 975--1076},
  [\href{https://arxiv.org/abs/1112.5179}{{\ttfamily 1112.5179}}].

\bibitem{Pasquetti:2019uop}
S.~Pasquetti and M.~Sacchi, \emph{{From 3$d$ dualities to 2$d$ free field
  correlators and back}},  \href{https://arxiv.org/abs/1903.10817}{{\ttfamily
  1903.10817}}.

\bibitem{Closset:2017zgf}
C.~Closset, H.~Kim and B.~Willett, \emph{{Supersymmetric partition functions
  and the three-dimensional A-twist}},
  \href{http://dx.doi.org/10.1007/JHEP03(2017)074}{\emph{JHEP} {\bfseries 03}
  (2017) 074}, [\href{https://arxiv.org/abs/1701.03171}{{\ttfamily
  1701.03171}}].

\bibitem{Closset:2018ghr}
C.~Closset, H.~Kim and B.~Willett, \emph{{Seifert fibering operators in 3d
  $\mathcal{N}=2$ theories}},
  \href{http://dx.doi.org/10.1007/JHEP11(2018)004}{\emph{JHEP} {\bfseries 11}
  (2018) 004}, [\href{https://arxiv.org/abs/1807.02328}{{\ttfamily
  1807.02328}}].

\bibitem{Kapustin:2009kz}
A.~Kapustin, B.~Willett and I.~Yaakov, \emph{{Exact Results for Wilson Loops in
  Superconformal Chern-Simons Theories with Matter}},
  \href{http://dx.doi.org/10.1007/JHEP03(2010)089}{\emph{JHEP} {\bfseries 03}
  (2010) 089}, [\href{https://arxiv.org/abs/0909.4559}{{\ttfamily 0909.4559}}].

\bibitem{Hama:2010av}
N.~Hama, K.~Hosomichi and S.~Lee, \emph{{Notes on SUSY Gauge Theories on
  Three-Sphere}}, \href{http://dx.doi.org/10.1007/JHEP03(2011)127}{\emph{JHEP}
  {\bfseries 03} (2011) 127},
  [\href{https://arxiv.org/abs/1012.3512}{{\ttfamily 1012.3512}}].

\bibitem{Hama:2011ea}
N.~Hama, K.~Hosomichi and S.~Lee, \emph{{SUSY Gauge Theories on Squashed
  Three-Spheres}}, \href{http://dx.doi.org/10.1007/JHEP05(2011)014}{\emph{JHEP}
  {\bfseries 05} (2011) 014},
  [\href{https://arxiv.org/abs/1102.4716}{{\ttfamily 1102.4716}}].

\bibitem{Ryu:2006bv}
S.~Ryu and T.~Takayanagi, \emph{{Holographic derivation of entanglement entropy
  from AdS/CFT}},
  \href{http://dx.doi.org/10.1103/PhysRevLett.96.181602}{\emph{Phys. Rev.
  Lett.} {\bfseries 96} (2006) 181602},
  [\href{https://arxiv.org/abs/hep-th/0603001}{{\ttfamily hep-th/0603001}}].

\bibitem{Hristov:2018lod}
K.~Hristov, I.~Lodato and V.~Reys, \emph{{On the quantum entropy function in 4d
  gauged supergravity}},
  \href{http://dx.doi.org/10.1007/JHEP07(2018)072}{\emph{JHEP} {\bfseries 07}
  (2018) 072}, [\href{https://arxiv.org/abs/1803.05920}{{\ttfamily
  1803.05920}}].

\bibitem{Hristov:2019xku}
K.~Hristov, I.~Lodato and V.~Reys, \emph{{One-loop determinants for black holes
  in 4d gauged supergravity}},
  \href{http://dx.doi.org/10.1007/JHEP11(2019)105}{\emph{JHEP} {\bfseries 11}
  (2019) 105}, [\href{https://arxiv.org/abs/1908.05696}{{\ttfamily
  1908.05696}}].

\bibitem{Mukhametzhanov:2019pzy}
B.~Mukhametzhanov and A.~Zhiboedov, \emph{{Modular invariance, tauberian
  theorems and microcanonical entropy}},
  \href{http://dx.doi.org/10.1007/JHEP10(2019)261}{\emph{JHEP} {\bfseries 10}
  (2019) 261}, [\href{https://arxiv.org/abs/1904.06359}{{\ttfamily
  1904.06359}}].

\bibitem{Pal:2019zzr}
S.~Pal and Z.~Sun, \emph{{Tauberian-Cardy formula with spin}},
  \href{https://arxiv.org/abs/1910.07727}{{\ttfamily 1910.07727}}.

\bibitem{Benini:2015bwz}
F.~Benini, N.~Bobev and P.~M. Crichigno, \emph{{Two-dimensional SCFTs from
  D3-branes}}, \href{http://dx.doi.org/10.1007/JHEP07(2016)020}{\emph{JHEP}
  {\bfseries 07} (2016) 020},
  [\href{https://arxiv.org/abs/1511.09462}{{\ttfamily 1511.09462}}].

\bibitem{Cvetic:2005zi}
M.~Cvetic, G.~W. Gibbons, H.~Lu and C.~N. Pope, \emph{{Rotating black holes in
  gauged supergravities: Thermodynamics, supersymmetric limits, topological
  solitons and time machines}},
  \href{https://arxiv.org/abs/hep-th/0504080}{{\ttfamily hep-th/0504080}}.

\bibitem{Hristov:2019mqp}
K.~Hristov, S.~Katmadas and C.~Toldo, \emph{{Matter-coupled supersymmetric
  Kerr-Newman-AdS$_4$ black holes}},
  \href{http://dx.doi.org/10.1103/PhysRevD.100.066016}{\emph{Phys. Rev.}
  {\bfseries D100} (2019) 066016},
  [\href{https://arxiv.org/abs/1907.05192}{{\ttfamily 1907.05192}}].

\bibitem{Cacciatori:2009iz}
S.~L. Cacciatori and D.~Klemm, \emph{{Supersymmetric AdS(4) black holes and
  attractors}}, \href{http://dx.doi.org/10.1007/JHEP01(2010)085}{\emph{JHEP}
  {\bfseries 01} (2010) 085},
  [\href{https://arxiv.org/abs/0911.4926}{{\ttfamily 0911.4926}}].

\bibitem{Katmadas:2014faa}
S.~Katmadas, \emph{{Static BPS black holes in U(1) gauged supergravity}},
  \href{http://dx.doi.org/10.1007/JHEP09(2014)027}{\emph{JHEP} {\bfseries 09}
  (2014) 027}, [\href{https://arxiv.org/abs/1405.4901}{{\ttfamily 1405.4901}}].

\bibitem{Halmagyi:2014qza}
N.~Halmagyi, \emph{{Static BPS black holes in AdS$_{4}$ with general dyonic
  charges}}, \href{http://dx.doi.org/10.1007/JHEP03(2015)032}{\emph{JHEP}
  {\bfseries 03} (2015) 032},
  [\href{https://arxiv.org/abs/1408.2831}{{\ttfamily 1408.2831}}].

\bibitem{Hristov:2018spe}
K.~Hristov, S.~Katmadas and C.~Toldo, \emph{{Rotating attractors and BPS black
  holes in $AdS_4$}},
  \href{http://dx.doi.org/10.1007/JHEP01(2019)199}{\emph{JHEP} {\bfseries 01}
  (2019) 199}, [\href{https://arxiv.org/abs/1811.00292}{{\ttfamily
  1811.00292}}].

\bibitem{Herzog:2010hf}
C.~P. Herzog, I.~R. Klebanov, S.~S. Pufu and T.~Tesileanu, \emph{{Multi-Matrix
  Models and Tri-Sasaki Einstein Spaces}},
  \href{http://dx.doi.org/10.1103/PhysRevD.83.046001}{\emph{Phys. Rev.}
  {\bfseries D83} (2011) 046001},
  [\href{https://arxiv.org/abs/1011.5487}{{\ttfamily 1011.5487}}].

\bibitem{Drukker:2010nc}
N.~Drukker, M.~Marino and P.~Putrov, \emph{{From weak to strong coupling in
  ABJM theory}},
  \href{http://dx.doi.org/10.1007/s00220-011-1253-6}{\emph{Commun. Math. Phys.}
  {\bfseries 306} (2011) 511--563},
  [\href{https://arxiv.org/abs/1007.3837}{{\ttfamily 1007.3837}}].

\bibitem{Emparan:1999pm}
R.~Emparan, C.~V. Johnson and R.~C. Myers, \emph{{Surface terms as counterterms
  in the AdS / CFT correspondence}},
  \href{http://dx.doi.org/10.1103/PhysRevD.60.104001}{\emph{Phys. Rev.}
  {\bfseries D60} (1999) 104001},
  [\href{https://arxiv.org/abs/hep-th/9903238}{{\ttfamily hep-th/9903238}}].

\bibitem{Jafferis:2011zi}
D.~L. Jafferis, I.~R. Klebanov, S.~S. Pufu and B.~R. Safdi, \emph{{Towards the
  F-Theorem: N=2 Field Theories on the Three-Sphere}},
  \href{http://dx.doi.org/10.1007/JHEP06(2011)102}{\emph{JHEP} {\bfseries 06}
  (2011) 102}, [\href{https://arxiv.org/abs/1103.1181}{{\ttfamily 1103.1181}}].

\bibitem{Freedman:2013ryh}
D.~Z. Freedman and S.~S. Pufu, \emph{{The holography of $F$-maximization}},
  \href{http://dx.doi.org/10.1007/JHEP03(2014)135}{\emph{JHEP} {\bfseries 03}
  (2014) 135}, [\href{https://arxiv.org/abs/1302.7310}{{\ttfamily 1302.7310}}].

\bibitem{Fluder:2015eoa}
M.~Fluder and J.~Sparks, \emph{{D2-brane Chern-Simons theories: F-maximization
  = a-maximization}},
  \href{http://dx.doi.org/10.1007/JHEP01(2016)048}{\emph{JHEP} {\bfseries 01}
  (2016) 048}, [\href{https://arxiv.org/abs/1507.05817}{{\ttfamily
  1507.05817}}].

\bibitem{PandoZayas:2019hdb}
L.~A. Pando~Zayas and Y.~Xin, \emph{{The Topologically Twisted Index in the 't
  Hooft Limit and the Dual AdS$_4$ Black Hole Entropy}},
  \href{https://arxiv.org/abs/1908.01194}{{\ttfamily 1908.01194}}.

\bibitem{Intriligator:1996ex}
K.~A. Intriligator and N.~Seiberg, \emph{{Mirror symmetry in three-dimensional
  gauge theories}},
  \href{http://dx.doi.org/10.1016/0370-2693(96)01088-X}{\emph{Phys. Lett.}
  {\bfseries B387} (1996) 513--519},
  [\href{https://arxiv.org/abs/hep-th/9607207}{{\ttfamily hep-th/9607207}}].

\bibitem{Aharony:1997ju}
O.~Aharony and A.~Hanany, \emph{{Branes, superpotentials and superconformal
  fixed points}},
  \href{http://dx.doi.org/10.1016/S0550-3213(97)00472-0}{\emph{Nucl. Phys.}
  {\bfseries B504} (1997) 239--271},
  [\href{https://arxiv.org/abs/hep-th/9704170}{{\ttfamily hep-th/9704170}}].

\bibitem{Guarino:2015jca}
A.~Guarino, D.~L. Jafferis and O.~Varela, \emph{{String Theory Origin of Dyonic
  N=8 Supergravity and Its Chern-Simons Duals}},
  \href{http://dx.doi.org/10.1103/PhysRevLett.115.091601}{\emph{Phys. Rev.
  Lett.} {\bfseries 115} (2015) 091601},
  [\href{https://arxiv.org/abs/1504.08009}{{\ttfamily 1504.08009}}].

\bibitem{Dimofte:2011ju}
T.~Dimofte, D.~Gaiotto and S.~Gukov, \emph{{Gauge Theories Labelled by
  Three-Manifolds}},
  \href{http://dx.doi.org/10.1007/s00220-013-1863-2}{\emph{Commun. Math. Phys.}
  {\bfseries 325} (2014) 367--419},
  [\href{https://arxiv.org/abs/1108.4389}{{\ttfamily 1108.4389}}].

\bibitem{Cecotti:2011iy}
S.~Cecotti, C.~Cordova and C.~Vafa, \emph{{Braids, Walls, and Mirrors}},
  \href{https://arxiv.org/abs/1110.2115}{{\ttfamily 1110.2115}}.

\bibitem{Crichigno:2018adf}
P.~M. Crichigno, D.~Jain and B.~Willett, \emph{{5d Partition Functions with A
  Twist}}, \href{http://dx.doi.org/10.1007/JHEP11(2018)058}{\emph{JHEP}
  {\bfseries 11} (2018) 058},
  [\href{https://arxiv.org/abs/1808.06744}{{\ttfamily 1808.06744}}].

\end{thebibliography}\endgroup
\bibliographystyle{JHEP}
\end{document}